\documentclass[notitlepage,pre,superscriptaddress,onecolumn]{revtex4-1}
\usepackage[T1]{fontenc}
\usepackage{amsfonts}
\usepackage{amsmath}

\usepackage{amssymb}
\usepackage[dvips,pdftex]{graphicx}
\usepackage[latin1]{inputenc}
\usepackage[margin=1 in]{geometry}
\usepackage[dvips,pdftex]{graphicx}
\usepackage{lipsum}
\usepackage{hyperref}
\usepackage{color}

\hypersetup{
backref=true, 
pagebackref=true,
hyperindex=true, 
colorlinks=true, 
breaklinks=true, 
urlcolor= blue, 
linkcolor= blue, 
}

\usepackage{caption}
\usepackage{array}
\usepackage{multirow}
\usepackage{hhline}
\usepackage{booktabs}
\usepackage{subfig}
\usepackage{float}

\begin{document}

\title{Inference of compressed Potts graphical models}

\author{Francesca Rizzato\footnote{Now at SISSA Medialab Via Bonomea, 265, 34136 Trieste, Italy}}
\affiliation{Laboratoire de Physique, Ecole Normale Sup\'{e}rieure and CNRS-UMR8023, PSL Research University, 24 Rue Lhomond, 75005 Paris, France}

\author{Alice Coucke\footnote{Now at Snips, 18 rue Saint Marc, 75002 Paris, France}}
\affiliation{Laboratoire de Physique, Ecole Normale Sup\'{e}rieure and CNRS-UMR8023, PSL Research University, 24 Rue Lhomond, 75005 Paris, France}

\author{Eleonora de Leonardis\footnote{Now at Groupe Galeries Lafayette, 44 rue de Chateaudun, 75009 Paris, France}}
\affiliation{Laboratoire de Physique, Ecole Normale Sup\'{e}rieure and CNRS-UMR8023, PSL Research University, 24 Rue Lhomond, 75005 Paris, France}

\author{John P. Barton}
\affiliation{Department of Physics and Astronomy, University of California, Riverside, 900 University Ave, Riverside, CA, United States}

\author{J\'er\^ome Tubiana\footnote{Now at Blavatnik School of Computer Science, Tel Aviv University, Israel}}
\affiliation{Laboratoire de Physique, Ecole Normale Sup\'{e}rieure and CNRS-UMR8023, PSL Research University, 24 Rue Lhomond, 75005 Paris, France}

\author{R\'emi Monasson}
\affiliation{Laboratoire de Physique, Ecole Normale Sup\'{e}rieure and CNRS-UMR8023, PSL Research University, 24 Rue Lhomond, 75005 Paris, France}

\author{Simona Cocco}
\affiliation{Laboratoire de Physique, Ecole Normale Sup\'{e}rieure and CNRS-UMR8023, PSL Research University, 24 Rue Lhomond, 75005 Paris, France}

\begin{abstract}
We consider the problem of inferring a graphical Potts model on a population of variables. This inverse Potts problem generally involves the inference of a large number of parameters, often larger than the number of available data, and, hence, requires the introduction of regularization. 
We study here a double regularization scheme, in which the number of  Potts states (colors) available to each variable is reduced, and interaction networks are made sparse.
To achieve the color compression, only Potts states with large empirical frequency (exceeding some threshold) are explicitly modeled on each site, while the others are grouped into a single state. We benchmark the performances of this mixed regularization approach, with two inference algorithms, Adaptive Cluster Expansion (ACE) and PseudoLikelihood Maximization (PLM) on synthetic data obtained by sampling disordered Potts models on an Erd\H{o}s-R\'enyi random graphs. 
We show in particular that
color compression does not affect the quality of reconstruction of the parameters corresponding to high-frequency symbols, while drastically reducing the number of the other parameters and thus the computational time. Our procedure is also applied to multi-sequence alignments of protein families, with similar results.
\end{abstract}

\maketitle


\section{Introduction}

Graphical models are an important tool to model dependencies and infer effective interactions among a population of variables from data.
We hereafter refer to this approach as the inverse Ising problem \cite{ackley1985learning,cocco2012adaptive,nguyen2017inverse} in the case of binary variables, and as the inverse Potts problem in the more general case of of multi-categorial variables\cite{wu1982potts}.
Applications include inferring functional couplings among a set of neurons from the recording of their activity \cite{cocco2017functional,schneidman2006weak,cocco2009neuronal},  between birds in  flocks \cite{bialek2012statistical} and among amino acids in sequences that belong to the same  protein family \cite{ekeberg2013improved}. 
Over the last decade, it was shown that describing these protein families by Potts models, whose parameters were learned from the corresponding sequence alignment could provide information on the protein structure 
\cite{burger2010disentangling,weigt2009identification,marks2011protein,sulkowska2012genomics,hopf2012three,ekeberg2013improved,sulkowska2012genomics,nugent2012accurate,hopf2012three,de2013emerging,jones2012psicov,morcos2014coevolutionary}, help predict fitness variations following mutations \cite{nguyen2017inverse,ferguson2013translating,mann2014fitness,figliuzzi2016coevolutionary,hopf2017mutation} and design new working proteins of the same family \cite{cocco2018inverse,socolich2005evolutionary,russ2019}.
Given the computational untractability of achieving exact solutions, different effective methods have been proposed to infer the Potts parameters from sequence data, including Gaussian approximation with different priors \cite{jones2012psicov,baldassi2014fast,burger2010disentangling,pensar2019high}, message passing \cite{weigt2009identification}, PseudoLikelihood Maximization (PLM)\cite{ekeberg2013improved,ekeberg2014fast,vuffray2019efficient}, minimum probability flow \cite{sohl2011new}, and the Adaptive Cluster Expansion (ACE) method \cite{cocco2011adaptive,barton2016adaptive}.

Even if Potts models only include pairwise interactions, the number of parameters to be inferred is huge. For an $N$-site protein, where each site can be one out of the 20 natural amino acids or an extra symbol standing for a site insertion or deletion, the number of independent parameters is $20\cdot N +20^2\cdot N(N-1)/2$. This gives about $10^6$ parameters for $N=100$ and almost $10^8$ parameters for $N=500$, while protein sequence alignments typically include few thousands to few tens thousand sequences. Moreover, amino-acid frequencies, and, hence, sampling quality may vary substantially from site to site, making it impossible to accurately reconstruct the complete set of Potts parameters.
To overcome the  problem of undersampling regularization terms are generally included. Standard $L_2$-regularization helps constraining parameter values, but does not change their number. $L_1$ and $L_0$-based regularization, on the contrary, may effectively remove many interaction parameters associated to low (in absolute value) connected correlations. 

In this paper a simple procedure to reduce the number $q$ of Potts parameters is described and analyzed.   In  physics Potts states are often referred to as colors, so we call this state reduction procedure \textit{color compression}. Our goal is to infer a compressed Potts model, where the number of states  $q_i\leq q$ depends on the site $i$. 
The basic idea is to group together rarely observed states on each site, defined as those below a given frequency threshold $f_0$. This way, the number of Potts states $q_i$ on each site $i$ is variable, leading to the reduced number of parameters $M_{cc}=\sum_i^N q_i +\sum_{i<j}^N q_i\,q_j$. 
Color compression is therefore equivalent to an $L_0$ regularization on the number of inferred parameters associated to the Potts states.
Slightly different schemes are based on grouping colors according to their entropy contributions to the site variability \cite{barton2016adaptive} or to their mutual information \cite{haldane2018coevolutionary} or compression to a fixed number of colors \cite{anton2018radi}, and comes to a similar outcome.
 As expected color compression  can help in limiting the computational time, in avoiding overfitting and, in a more theoretical framework, in understanding the intrinsic dimensionality of the problem, by distinguishing the parameters that can be reliably inferred from those fixed through regularization only.
Color compression was already used by some of us in the Adaptive Cluster Expansion (ACE) algorithm \cite{barton2016adaptive} in order to reduce the computational time and have a simpler inferred model for the analysis of protein sequence data, but its performance has not been systematically tested up to now. 

 The aim of  the present study is to benchmark the color compression procedure in a systematic way as a function of the  compression frequency threshold $f_0$ and of the sampling depth $B$:
Correlations between variables that are grouped in the color compression procedure are lost. We  expect therefore  that compression  is useful in the inference at low $B$ and low $f_0$  to discard correlations between poorly sampled colors, affected by large statistical fluctuations. Conversely,  information on the correlated structure of the data will be lost for  large $B$ and $f_0$.   
Consequences of compression on the performances of inferred models  are here investigated for different regularization types and strengths, for two particular inference procedures:  PLM with large or small $L_2$ regularization parameters, and the ACE procedure with  a fully connected or sparse interaction graph.    
 We then introduce a procedure to recover, after inference of the compressed Potts model, a full Potts model over all possible $q$ states, which we refer to as color decompression.  Decompression  is necessary to compare the inferred parameters to the ground truth  (when available), to compare the quality of the inferences for different color compression strengths (and therefore different $q_i$ values), and, more generally, when the model is used to predict the behavior of poorly sampled variables in the original data set.

The first part of the paper briefly sketches the methodological background and the inference algorithms (Section~\ref{sec:background}).  In Section~\ref{sec:compression} the procedures of color compression and decompression are introduced. We then assess the performances of the procedures on synthetic data generated from Potts model on random graphs in Section~\ref{sec:results}. 
In Section~\ref{sec:real} we show an illustrative example on fitness prediction for real proteins, to verify that the results obtained on synthetic data model translate to real cases. Section~\ref{sec:times} shows the gain our color compression procedure provides in terms of computational time. Some conclusion and perspectives are presented in Sec.~\ref{sec:conclusion}.

\section{Reminder on Inference and Algorithms}\label{sec:background}
\subsection{Inverse Potts Problem}
The Potts model describes a system of $N$ interacting sites, each assuming one of $q$ possible Potts states (or colors). The probability distribution of each color on each site is controlled by a set of parameters that can be divided into local fields $h_i(a_i)$, depending only on one site $i$ and its color $a_i$, and pairwise couplings $J_{ij}(a_i,a_j)$, depending on the pair of sites $i,j$ and the two Potts states $a_i$, $a_j$. An energy value is associated to each system configuration ${\mathbf{a}}=a_1,\ldots a_N$,
\begin{equation}
E({\mathbf{a}}|{\bf J}) = -\sum_{i=1}^{N} h_i(a_i) - \sum_{i=1}^{N-1}\sum_{j=i+1}^{N} J_{ij}(a_i, a_j)\,
\end{equation}
and, consequently, a probability
\begin{equation}
P({\mathbf{a}}|{\bf J}) = \frac{\exp\left(-E({\mathbf{a}}|{\bf J})\right)}{Z({\bf J}) }\,,
\end{equation}
where $Z({\bf J}) =\sum_{{\mathbf{a}}}\exp\left(-E({\mathbf{a}}|{\bf J})\right)$ is the partition function and ensures that all probabilities sum to one. For simplicity, here we label the set of fields and couplings as ${\bf J}$.

Given a sample of configurations, one may be interested in inferring back the model from which these samples were generated, or at least a model reproducing the statistical properties of such configurations, such as the one- and two-site frequencies, $f_i(a)$ and $f_{ij}(a,b)$.
In general the Potts model defined above is the simplest, or maximum entropy \citep{jaynes1982rationale}, probabilistic model capable of reproducing the observed frequencies. In the present case we know by construction that the Potts model is not only the simplest model to fit the data, but also the real model from which the sample was generated.
To reproduce the statistics of the data, the parameters $h_i(a)$ and $J_{ij}(a,b)$ must be chosen such that site averages and correlations in the model match those in the data, i.e.,

\begin{align}
\begin{aligned} \label{eq:constraint}
\sum_{a_1,\ldots a_N} \delta(a_i,a) P( a_1 \ldots, a_N|{\bf J}) &= f_i(a)\,,\\
\sum_{a_1,\ldots a_N} \delta(a_i,a)\delta(a_j,b) P( a_1 \ldots, a_N|{\bf J}) &= f_{ij}(a,b)\,,
\end{aligned}
\end{align}
where $\delta(a_i,a)$ is the Kronecker delta function, which is one if the symbol $a_i$ at site $i$ is equal to $a$ and zero otherwise. Finding the parameters $h_i(a)$, $J_{ij}(a,b)$ that satisfy Eq.~\ref{eq:constraint} constitutes the inverse Potts problem.

\subsection{Cross-entropy and regularization}
Formally, the solution to the inverse Potts problem is the set of fields and couplings that maximize the average log-likelihood or, equivalently,  that minimize the cross-entropy between the data and the model.
This cross-entropy can be written as
\begin{align}
\begin{aligned} \label{eq:crossentropy}
S({\bf J}| {\bf f }) &= \log{Z}({\bf J}) - \sum_{i=1}^{N}\sum_{a=1}^{q} h_i(a) f^s_i(a) - \sum_{i=1}^{N-1}\sum_{j=i+1}^{N}\sum_{a=1}^{q}\sum_{b=1}^{q} J_{ij}(a,b) f^s_{ij}(a,b)\,,
\end{aligned}
\end{align}
 where, for simplicity, we indicate the set of single and pairwise frequencies as ${\bf f}$ and the set of fields and couplings as ${\bf J}$.

To guarantee that the minimization of the cross-entropy is a well defined problem, a regularization term $\Delta S$ is added, which,
n the Bayesian formulation, corresponds to prior knowledge over the parameters $\bf J$.
A Gaussian prior distribution, also referred to as $L_2$-regularization, is a usual choice:
\begin{equation}\label{eq:L2}
\Delta S=\gamma_h \sum_{i=1}^{N}\sum_{a=1}^{q} h_i(a)^2 + \gamma_J \sum_{i=1}^{N-1}\sum_{j=i+1}^{N}\sum_{a=1}^{q}\sum_{b=1}^{q} J_{ij}(a,b)^2 .
\end{equation}
The regularization parameters $\gamma_J$ and $\gamma_h$ are related to the prior variances of fields ($\sigma^2_h$) and couplings ($\sigma^2_J$) through $\gamma_h= 1/(B \sigma^2_h)$, and $\gamma_J = 1/(B \sigma^2_J)$, where $B$ is the number of configurations in the sample. When the penalties are relatively weak ($\gamma \sim \mathcal{O}(1/B)$), this regularization can be thought of as a weakly informative prior \cite{gelman2008weakly} whose main purpose is to prevent pathologies in the inference.

\subsection{Gauge invariance}\label{subsec:gauge}
The $N\cdot q$ frequencies $f_i(a)$ and $\frac 12 N(N-1) q^2$ correlations $f_{ij}(a,b)$, with $i<j$, are related to each other: the former sum up to 1, while the latter have the frequencies as marginals. Therefore, not all constraints in Eq.~\ref{eq:constraint} are independent and multiple sets of parameters give the same probability distribution.
In the language of physics this over-parameterization of the model is referred to as \textit{gauge invariance} and the choice of one particular parameter set among the equivalent ones as \textit{gauge choice}. This gauge invariance reduces the number of free parameters in the Potts model to $q-1$ fields for each site and $(q-1)^2$ couplings for each pair of sites.

In particular, we can reparameterize the model without changing the probabilities by an arbitrary transformation of the form:
\begin{eqnarray*}
h_i(a) &\rightarrow &h_i(a)+ H_i +\sum_{j(\neq i)}K_{ij}(a)\\
J_{ij}(a,b)&\rightarrow & J_{ij}(a,b)-K_{ij}(a)-K_{ji}(b)+\kappa_{ij}
\end{eqnarray*}
for any  $K_{ij}(a)$, $H_i$ and $\kappa_{ij}$. 
In the so called lattice-gas gauge, this freedom is used to define a \textit{gauge state} $c_i$ at each site such that
\begin{equation}
J_{ij}(a,c_j)=J_{ij}(c_i,b)=h_i(c_i)=0 \,,
\label{eq:consensus}
\end{equation}
for all states $a,b$ and sites $i,j$. The couplings and fields are transformed as follows:
\begin{equation}
\begin{aligned}
h_i(a) &\to h_i(a) - h_i(c_i)+\sum_{j \neq i} \left ( J_{ij}(a,c_j) - J_{ij}(c_i,c_j) \right )\,,\\
J_{ij}(a,b) &\to J_{ij}(a,b) - J_{ij}(c_i,b) -J_{ij}(a,c_j)+ J_{ij}(c_i,c_j) \,.
\end{aligned}
\label{eq:consensus_transform}
\end{equation}

Two common gauge states are the most and the least frequent states of each site, defining respectively the \textit{consensus gauge} and the \textit{least-frequent gauge}. In protein analysis, the gauge state is often fixed to the amino acid present at site $i$ in a reference sequence, called \textit{wild-type} sequence.
An alternative choice is the so-called \textit{zero-sum gauge}, in which
\begin{equation}
\sum_{c=1}^qJ_{ij}(a,c)=\sum_{c=1}^qJ_{ij}(c,a)=\sum_{c=1}^qh_i(c)=0 \ ,
\label{eq:zerosum}
\end{equation}
for all states $a$ and all variables $i,j$. Fields and couplings can also be put in the so-called zero-sum gauge through
\begin{equation}
\begin{aligned}
h_i(a) &\to h_i(a)-{h_{i}(\cdot)}+\sum_{j(\neq i)}\left[ J_{ij}(a,\cdot )-J_{ij}(\cdot,\cdot ) \right]\ , \\
J_{ij}(a,b) &\to J_{ij}(a,b) - J_{ij}(\cdot,b) -J_{ij}(a,\cdot )+ J_{ij}(\cdot, \cdot) \ ,
\end{aligned}
\label{eq:zerosum_transform}
\end{equation}
where $g(\cdot)$ denotes the uniform average of $g(a)$ over all states $a$ at fixed position.

Note that, while all observables such as the moments of the distribution are invariant with respect to the gauge choice, the fields and the couplings are not. Arbitrary functions of the couplings and fields, such as the commonly-used Frobenius norm of the couplings, are also not generally gauge invariant.
If not explicitly stated, the comparisons shown in this paper are performed in the consensus gauge, but the choice of the gauge for the inference and for the analysis of the inferred network can be different. The gauge chosen during the inference will be further discussed in section \ref{sec:alg} in the descriptions of ACE and PLM.

\subsection{Algorithms \label{sec:alg}}
The presence of the partition function $Z$ in Eq.~\ref{eq:crossentropy} precludes direct numerical minimization of the cross-entropy when the system size is large, since this requires summing over all $\prod_{i=1}^{N} q_i$ possible configurations of the system. However many approximate solutions have been proposed to tackle this issue.
We briefly recall two of these methods to respectively approximate the cross-entropy or the log-likelihood: the Adaptive Cluster Expansion (ACE) and PseudoLikelihood Maximization (PLM).

\subsubsection{Adaptive cluster expansion (ACE)}
\label{sec:ACE}

The cross-entropy~(Eq.~\ref{eq:crossentropy}) can be exactly decomposed as a sum of cross-entropy contributions, calculated recursively (see Appendix~\ref{ace:appendix} and \ref{ace:pseudocode:appendix}).
The adaptive cluster expansion \cite{cocco2011adaptive, cocco2012adaptive, barton2016adaptive} is based on the idea of summing up cluster contributions based on their importance as quantified by their absolute contribution to the cross entropy.
To this end an inclusion threshold parameter $t$ is introduced and only clusters with cross-entropy contributions larger than the threshold $t$ are included. The inclusion threshold $t$ is then progressively decreased to include more and more clusters in the summation. 
The expansion is usually stopped when the frequencies and correlations of the inferred model reproduce the empirical ones to within the statistical error bars due to finite sampling. 
The inference routine which has been used in this paper is publicly available at \url{https://github.com/johnbarton/ACE} .
For an input sample of size $B$, the regularization parameters are set to $\gamma_J=1/B$ and $\gamma_h=0.01/B$, corresponding to a variance of the prior distribution of couplings of order 1 and a variance of fields of order 100.

\subsubsection{Pseudo-likelihood maximization (PLM)}

The idea behind Pseudo-Likelihood Maximization is to approximate the full likelihood of the data given the model (or equivalently the full cross-entropy (Eq.~\ref{eq:crossentropy}) by the site-by-site maximization of the conditional probability of observing one state at a site, given the observed states on the other sites. This approximation makes the problem tractable, and it also makes possible to parallelize the computation for the different sites. 
Pseudolikelihood is a consistent estimator of the likelihood in the limit of infinite input data.
For this study, a version of the asymmetric pseudolikelihood maximization \cite{ekeberg2013improved, ekeberg2014fast} capable of working with a site-dependent number of Potts states has been implemented adapting the public code by M.~Ekenberg and E.~Aurell at \url{https://github.com/magnusekeberg/plmDCA}.

Unlike ACE, the networks inferred by PLM with $L_2$-regularization  are always fully connected. As empirically shown in protein sequence analysis \cite{ekeberg2013improved, hopf2017mutation, cocco2018inverse} and in theoretical analyses \cite{barton2014large},
large regularization is needed in the presence of fully connected networks to avoid overfitting and thus to improve contact and fitness predictions.
We have tested different regularization strengths, see Sec.~\ref{appendix:PLM:reg:table}, and fixed $\gamma_J=N/B$, $\gamma_h=0.1/B$ for system with $N$ variables and input sampling of size $B$.

With PLM  gauge invariance is automatically broken. The inference is performed in the gauge that minimizes the $L_2$-regularization:
\begin{equation}
\gamma_J\sum_{b=1}^{q_j}J_{ij}(a,b)=\gamma_hh_i(a)\ , \quad
\gamma_J\sum_{a=1}^{q_i}J_{ij}(a,b)=\gamma_hh_j(b)\ , \quad
\sum_{a=1}^{q_i}h_i(a)=0 \ .
\end{equation}
As for ACE, the PLM fields and couplings are subsequently transformed to the consensus gauge for comparison.


\section{Regularizations}
\label{sec:compression}

\subsection{Removing variable states}

\subsubsection{Color compression}
So far we have described (Eq.~\ref{eq:crossentropy} and \ref{eq:L2}) how to infer the parameters of a Potts model where the number of states $q$ is the same at all sites, but it is easy to generalize this procedure to Potts models in which the number of states $q_i$ depends on the site $i$.
This situation naturally arises due to sampling: states with very small probabilities are rarely observed. For instance, in multiple sequence alignments of protein families, for the large majority of sites. only a subset of the full $q=21$ possible amino acids are observed. 
In the color compression procedure,  for each site $i$, we model explicitly only the $q_i$ states observed with a frequency $f_i(a)$ larger than a frequency cutoff threshold $f_0$
\begin{equation}
	f_i(a)>f_0 \ ,
	\label{eq:pcut}
\end{equation}
and we group together the remaining $q-q_i$ low frequency states into a single one.
The frequency of the grouped/compressed Potts state $q_i+1$ is then the total frequency of the states that have been grouped together: $f_i(q_i+1) \equiv \sum_{a'=q_i+1}^q f_i(a')$. 

\subsubsection{Artificial data sets on Erd\H{o}s-R\'enyi random graphs}\label{sec:ER}

To benchmark color compressed inference we  have generated synthetic data from a Potts model with  $N=50$ variables carrying $q=10$ Potts states and interacting on an Erd\H{o}s-R\'enyi random graph. To build the interaction graph, each edge in the network is included with probability $0.05$, giving a mean connectivity 2.5,  with a maximum connectivity equal to 7. An exemple of contact map is shown in Supplementary Fig.~\ref{fig:cmapACEandPLM}. The field and couplings parametes on interacting sites are selected from Gaussian distributions of mean $\mu=0$ and standard deviations $\sigma_J^2=1$ and $\sigma_h^2=5.$ 
 Therefore if $i$ and $j$ interact $J_{ij}$ is a $10 \times 10$ matrix whose elements are chosen independently according to the above distributions, and each element of the matrix is zero when the sites do not interact.

 We  have generated 10 independent realizations of such ER models (networks of interactions and sets of fields and couplings). For each realization, $B=5\;10^2,10^3,10^4$, or $10^5$ configurations are generated by Markov Chain Monte-Carlo sampling. The number of available data, $B\times N$, can be compared to the number of parameters to be inferred, $M_{tot}=qN+q^2 N(N-1)/2\simeq 1.2\; 10^5$. We have, for the previously listed values of $B$,  $B\times N=2.5\; 10^4, 5\; 10^4, 5\; 10^5$, or $5\; 10^6$, the first two being in heavy undersampling conditions, the third in scarce sampling, and only the last one being relatively well sampled.
  \begin{figure}
	{\includegraphics[width=0.3\linewidth]{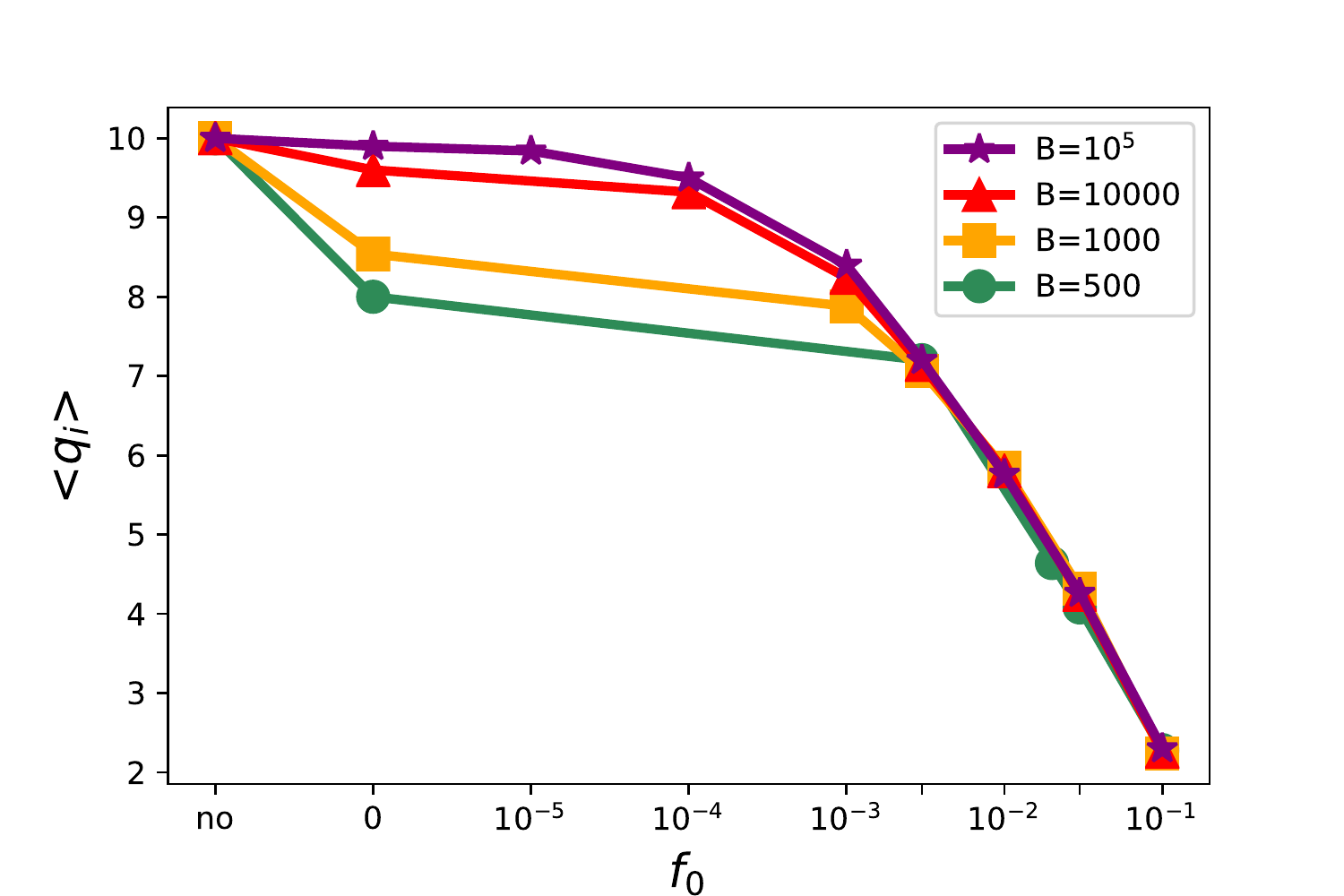}} 
	{ \includegraphics[width=0.3\linewidth]{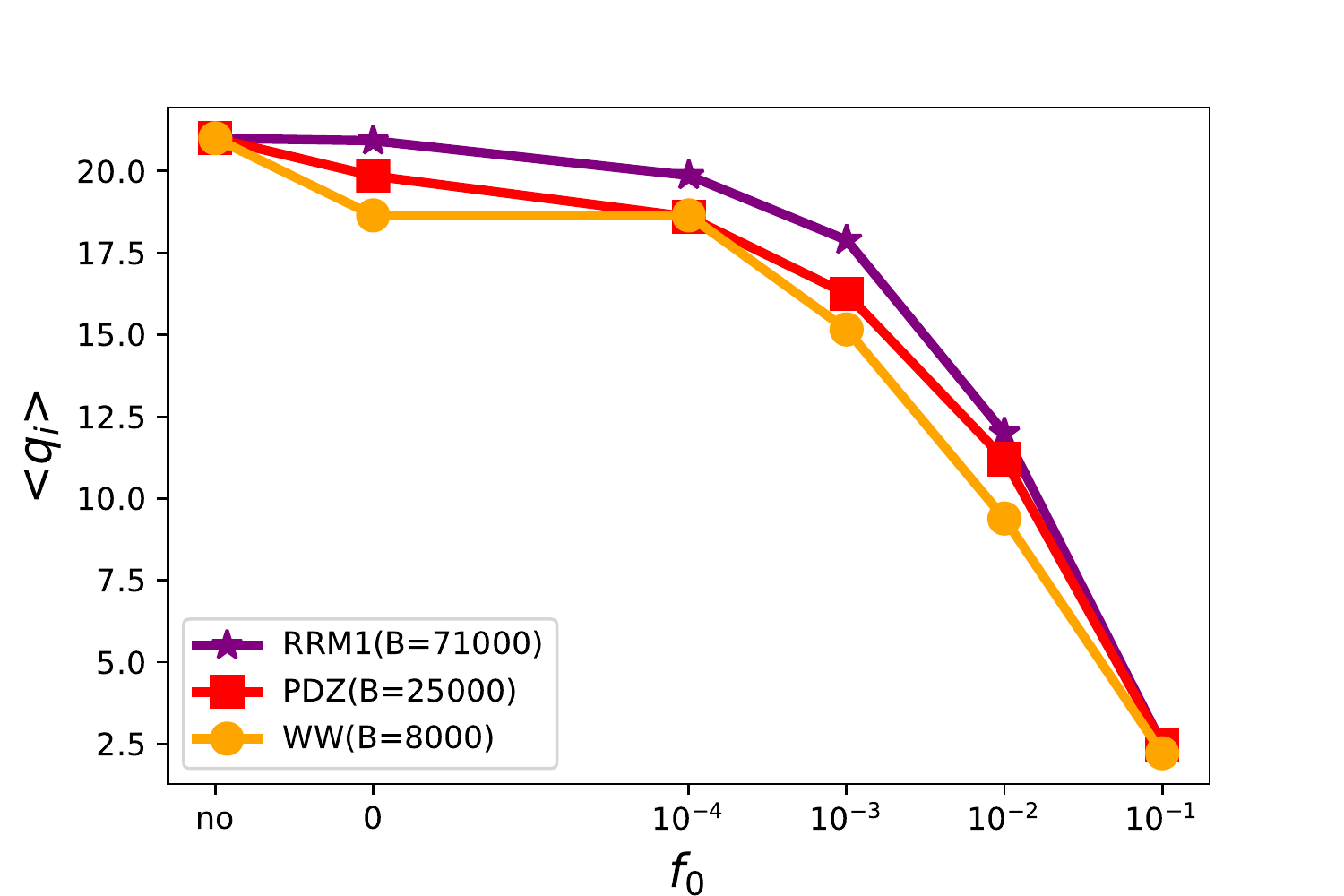} } \\
	\subfloat [Synthetic Data]
	{ \includegraphics[width=0.3\linewidth]{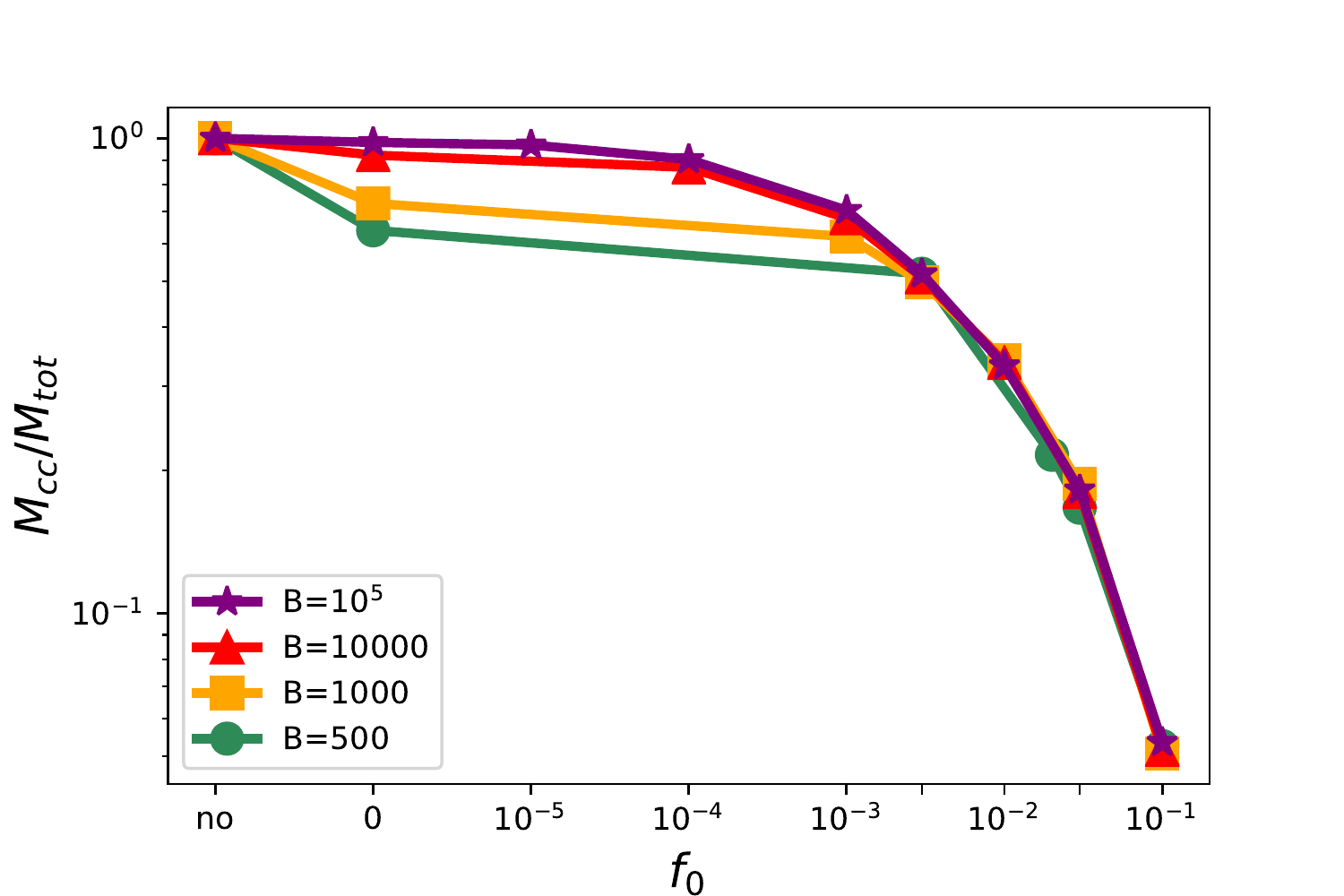} }
	 \subfloat [Protein Sequence Data]
	 { \includegraphics[width=0.3\linewidth]{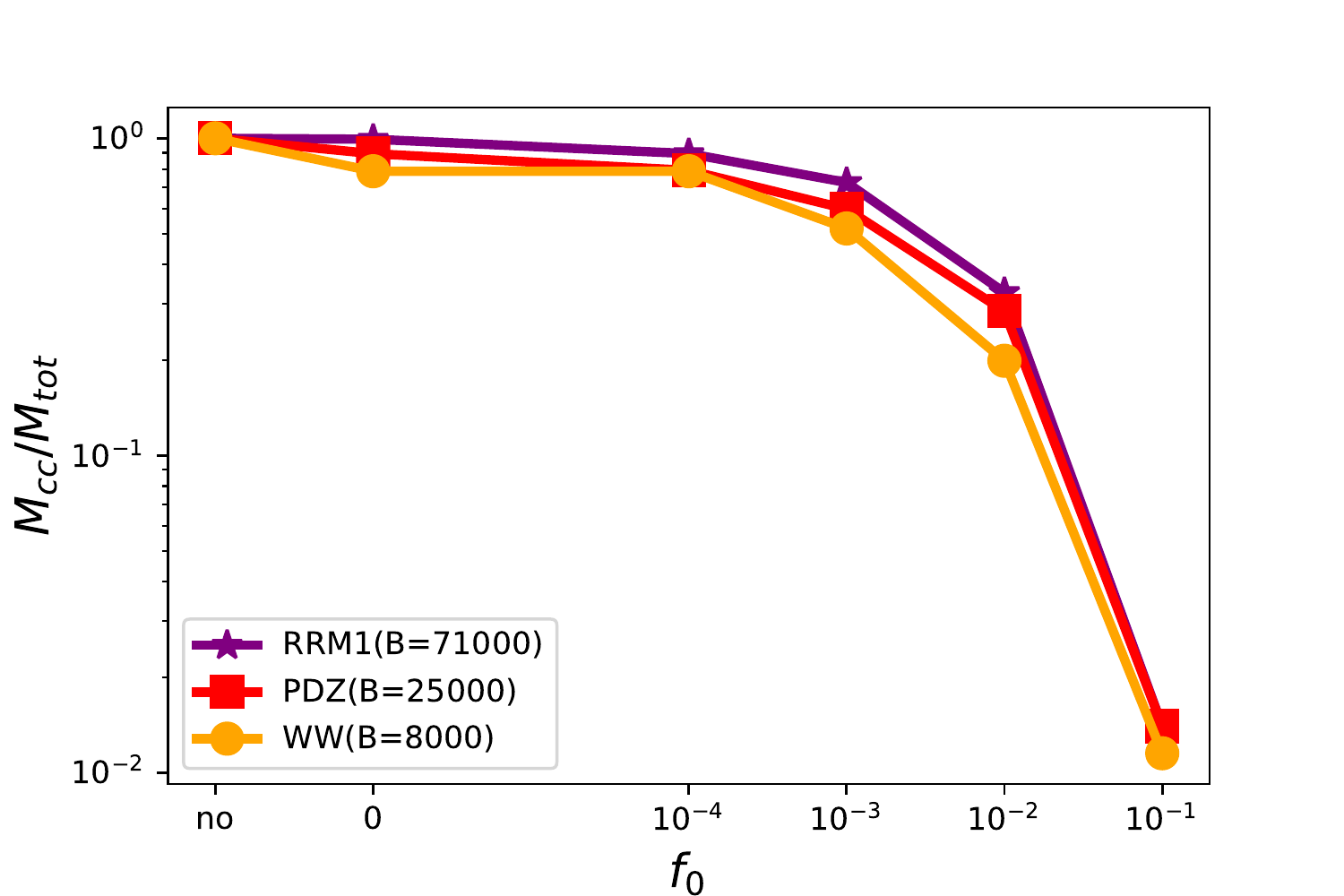} }
	 \caption{ Mean number of colors per site <$q_i$>   (top) and relative reduction in the number of parameters to be inferred  $M_{cc}/M_{tot}$ (bottom),
	 as a function of the  $f_0$: no compression, $0$ (only unseen symbols are removed from inference) and $1/B<f_0<0.1$. 
	Data from a single ER realization (left), 
	 and sequence data for 3 protein families
	 WW ($N=31$), PDZ ($N=84$), RRM ($N=82$) (right) at  different sample depths $B$.}
	\label{fig:Nparam}
\end{figure}
\subsubsection{Reduction in the number of parameters in compressed data: comparison between ER data with $q=10$ and sequence data with $q=21$ }\label{sec:redpar}
Tanks to color compression the number of colors per site, $q_i$, and thus the number of parameters   $M_{cc}=\sum_i^N q_i +\sum_{i<j}^N q_i\,q_j$ are reduced   with respect to the 
full number of colors per site, $q$, and the  total number of parameters to be inferred $M_{tot}$.
Fig.~\ref{fig:Nparam}(left)  illustrates how such reduction is achieved as a function of the compression threshold $f_0$ and for different sampling depths $B$ for the artificial data. 
Increasing $f_0$  the inferred model goes from the full-color  $\langle q_i\rangle =q$  Potts model to the Ising model with only two colors $\langle q_i\rangle =2$ per sites indicating only the presence or absence of the most probable color on each site. The  number of parameters to be inferred is divided by 50 at high compression.    \\
Fig.~\ref{fig:Nparam} (right) shows a similar behavior  for parameter reduction as a function of $f_0$ and sampling size $B$ in protein sequence data, for three families that will be further described  and analyzed in Sec.~\ref{sec:real}. In the latter case  the number of Potts states is the number of possible amino-acid types (plus the gap symbol),  $q=21$,  
and the number of  parameters is decreased by 100 times at high compression.
While protein sequence data are the direct application of color compression that we will consider in Sec.~\ref{sec:real}, we have generated  synthetic data from a Potts model with  $q=10$ instead  of $q=21$ states to speed up the numerical analysis performed varying  sampling depth, compression threshold and inference methods;  especially for the ACE algorithm, indeed, as it will be shown in Sec.~\ref{sec:times},  the computational cost of inferring a complete  $q=21$ Potts model may  become very large. Moreover  
 the aim of the present study is to investigate the interplay between sampling and  effect of compression, and we do not expect qualitative changes between $q=10$ and $q=21$: we expect that the compression threshold $f_0$, at which  the performances of the inferred model  worsen with respect to the full Potts model does not  depend on the maximal numbers of Potts state in the original model but rather on their occurrence in the sampling. At large compression threshold, we expect that model performance deteriorates because highly frequent, well-sampled colors  are grouped  and their correlations are lost.
 
\subsubsection{Color  decompression}
Once the restricted Potts model is inferred, we need to recover the complete model with $q$ states at each site in order to compare it to the ground truth. To this aim, we have to determine the parameters for the  states that were \textit{grouped} on site $i$. We use the following procedure: For each grouped state $a'$,  the fields and the couplings are estimated through
\begin{equation}
	h_i(a')=h_i(q_i+1)+\log \left(\frac{f_i(a')}{f_i(q_i+1)}\right) \ ,\quad J_{ij}(a',b)=J_{ij}(q_i+1,b) \ ,
\label{eq:decompress}
\end{equation}
where $q_i+1$ refers to the  unique Potts state for the grouped symbols.  
This procedure allows us to correctly recover the local fields parameters reproducing the  frequencies $f_i(a')$  for the  grouped symbols, while a common coupling parameters is assigned to all the grouped symbols.
Then, we associate fields and couplings  to states that are never observed in the sampling, hereafter referred to as \textit{unseen} states, through a natural extension of 
the procedure described above. We assign a pseudocount frequency $f_i(a")=\alpha/ B$ to the unseen symbols $(a")$. Herafter, we have fixed $\alpha=0.1$, in the expected range $0<\alpha<1$. 
When the grouped state is not present while unseens are present,  we  use the least probable state on the given site ($l_{i}$) instead of the grouped state in the denominator of Eq.~\ref{eq:decompress}, by replacing $q_i+1$  by $l_{i}.$

The choice of associating the unseen states to the grouped state or the least probable state is both simple and effective. Indeed, it follows the gauge choice, and yields fields with lower values than for the observed states.
A detailed comparison of the effects of the pseudo-count above and of the standard $L_2$ regularization in the simple case of an independent model is presented in Appendix~\ref{unseen:appendix}.

\subsubsection{Gauge used in the ACE inference}

ACE inference has been implemented in the lattice-gas gauge (Eq.~\ref{eq:consensus}) \cite{barton2016adaptive}. 
It is important to notice that the choice of the  gauge symbol  may have some effect on the performance of the inference procedure because the regularization term is not gauge invariant. For abundant data or in the limit of large compression, Potts states are well sampled and the choice of the  gauge symbol is largely irrelevant. However, for few data or in the absence of color compression, or at small compression, the best performance is obtained by choosing as gauge symbol  the least-probable Potts state on each site. In this way, all the fields and couplings corresponding to at least one poorly sampled state are fixed to zero, and have therefore null statistical variances by construction.
Fields and couplings are then translated to the gauge in which the most probable  symbol is chosen as gauge-symbol (consensus gauge) to perform the comparisons described in the next sections using Eq.~\ref{eq:consensus_transform}. This consensus gauge is best for comparison because the statistics of the consensus symbols (on all sites) are the easiest ones to measure accurately.

\subsection{Removing interactions}

We also regularize the inference procedure by limiting the number of non-zero interactions. Sparsification of the interaction network is sometimes achieved through $L_1$ regularization of the couplings. Hereafter, we show that the inclusion threshold of the ACE inference procedure defined in Section \ref{sec:ACE} plays a similar role, while not affecting the amplitude of non-zero couplings.

\subsubsection{Role of ACE inclusion threshold: sparse versus dense inferred graphs} 
 
\begin{figure}
	\includegraphics[width=0.6\linewidth]{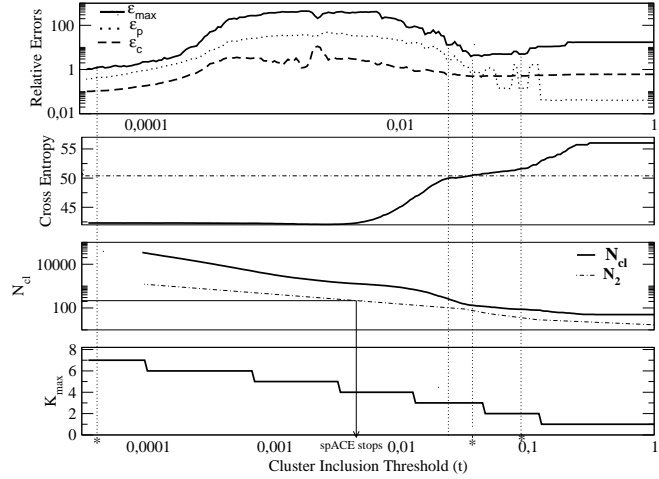}
	\caption{Cluster expansion as function of the cluster inclusion threshold $t$ for one data set, obtained by sampling one realization of an Erd\H{o}s-R\'enyi random graph with $B=1000$ configurations and applying ACE to compressed data with $f_0=0.01$. From top to bottom: {\em i)} relative reconstruction errors versus $t$ on frequencies $\epsilon_p$, connected correlations $\epsilon_c$ and maximal relative error $\epsilon_{max}$. Stars indicate possible solutions of the inverse models, reproducing well the empirical 1- and 2-point correlations  ($\epsilon_{\rm max}<5 $), which are obtained at different expansion thresholds and corresponding to sparse graphs at large thresholds, or fully connected reconstructed graphs at small thresholds, see Table~\ref{table:t}  {\em ii)} Regularized cross entropy vs. $t$, {\em iii)} number of total clusters $N_{cl}$ and  number of 2-site clusters $N_{2}$ included in the expansion vs. $t$, {\em iv)} maximal cluster size versus $t$. The spACE expansion stops at $N_{2}=200.$}
	\label{fig:ace-crossentropyvst}
\end{figure}

Fig.~\ref{fig:ace-crossentropyvst} shows the behavior of the ACE algorithm as a function of the inclusion threshold $t$ for one particular graph, hereafter called ER05, with $B=1000$ sampled configurations, analyzed with a color compression of $f_0=0.01$. This representative data set will be our reference case.
For each threshold $t$ used to select clusters in the ACE expansion, the model frequencies $\langle \delta(a_i,a) \rangle $ and $\langle \delta(a_i,a) \delta(b_i,b) \rangle$ calculated by Monte-Carlo simulation are compared to the data frequencies $f_i(a)$ and $f_{ij}(a,b)$ (see Eq.~\ref{eq:constraint}).

As detailed in \cite{cocco2012adaptive, barton2016adaptive}, to monitor the ability of the inferred model's ability to reproduce the measured frequencies and correlations while avoiding overfitting, 
we define a relative error that is the ratio between the deviations of the predicted observables from the data, 
$\Delta f_i(a)= \langle \delta(a_i,a)\rangle - f_{i}(a)$ and 
$\Delta f_{ij}(a,b)= \langle \delta(a_i,a) \delta(b_i,b) \rangle - f_{ij} (a,b)$,
and the expected statistical fluctuations due to finite sampling, $\sigma_{i} (a)=\sqrt{f_i(a) (1-f_i(a))/B}$ and $\sigma_{ij} (a,b)=\sqrt{f_{ij}(a,b)(1-f_{ij}(a,b))/B}$. The relative error on frequencies is
\begin{equation}
\epsilon_{\rm p}= \frac 1 {Nq}\sqrt{ \sum_{i,a}{\left ( \frac{\Delta f_i(a)}{\sigma_{i} (a)}\right)^2}} \,.
\label{eq:ep}
\end{equation}
The relative error on connected correlations, $c_{ij} (a,b)=\langle \delta(a_i,a) \delta(b_i,b) \rangle-\langle \delta(a_i,a)\rangle \langle\delta(b_i,b) \rangle$, is
\begin{equation}
\epsilon_{\rm c}= \frac 2 {N(N-1)q^2}\sqrt{ \sum_{i<j,a,b}{\left ( \frac{\Delta c_{i,j}(a,b)}{\sigma^c_{i,j} (a,b)}\right)^2}}\,,
\label{eq:ec}
\end{equation}
where we estimate the standard deviation in the connected correlations as $\sigma^c_{i,j} (a,b)= \sigma_{ij}(a,b)+f_j(b)\sigma_{i}(a)+f_i(a)\sigma_{j}(b)$. Finally, the maximum relative error is
\begin{equation}
\epsilon_{\rm max}= \max_{\{i,j,a,b\}} \frac{1}{\sqrt {2 \log{ ( {M })}}} \left ( \frac{|\Delta f_i(a)|}{\sigma_{i} (a)}, \frac{|\Delta f_{ij} (a,b)|}{\sigma_{ij} (ab)} \right) \,,
\label{eq:emax}
\end{equation}
where ${ M }=Nq+(N(N-1)/2)q^2$ is the total number of one- and two-point correlations. As shown in Fig.~\ref{fig:ace-crossentropyvst} (top panel) the relative errors defined above have a nonmonotonic behavior as a function of the threshold, reaching relative minima that successfully reconstruct the data ($\epsilon_{\rm max}<5$) at multiple values (marked by asterisks) of the expansion threshold $t$, see~Table~\ref{table:t}.
The regularized cross entropy, the total number of clusters included in the expansion, and their maximal size as a function of the cluster inclusion threshold $t$ are also
shown Fig.~\ref{fig:ace-crossentropyvst}.

The cluster inclusion threshold acts as an additional regularization.
There are 3 plateaus in the regularized cross entropy as a function of the cluster inclusion threshold $t$ of Fig.~\ref{fig:ace-crossentropyvst}:
the first plateau corresponds to an independent model, the second one to a sparse interaction network, and the third one to a fully connected network.
The number of edges present in the inferred graph of Fig.~\ref{fig:ace-crossentropyvst} is given by the number $N_2$ of 2-site clusters in the ACE expansion and is shown in Table~\ref{table:t} for the threshold corresponding to the minimal relative errors $\epsilon_{max}$.
In particular the two relative minima better reproducing the data correspond to 2 different inferred networks. The minimum with $\epsilon_{\rm max} =3.9$ is at high threshold ($t=0.036$) and is characterized by a numbers of edges $N_2=55$ smaller than the total number $N \,(N -1)/2=1225$ of possible pairs.
The inferred model is therefore a sparse graph, with a number of edges $N_2$ comparable with the number of edges of the model used to generate the data ($N_0=59$ for the model used to generate the data in Fig.~\ref{fig:ace-crossentropyvst}).
The second relative minimum with $\epsilon_{\rm max} =1$ is at low threshold $t=6.4 \times 10^{-5}$ where the expansion includes the maximal number of 2 site clusters $N_2=N\times (N-1)/2$, corresponding to a fully connected graph. As can be guessed by the difference in the connectivity between the original and inferred model, and as
we will better quantify in Section~\ref{sec:KL}, the fully connected solution is overfitting the data. 
 \begin{table}[]
\begin{tabular}{| c | c | c | c | c |}
\hline
t & $\epsilon_{max}$ & $N_2$ & $S^{\lambda}_c$ & $S_c$ \\ 
\hline
1 & 17 & 0 &56&56 \\ \hline
$9.6\times 10^{-2}$ & $4.9^*$ & 29 &51.7& 51.2 \\ \hline
$3.6 \times 10^{-2}$ & $3.9^*$ & 55 &50.6& 49.9 \\ \hline
$2.2 \times 10^{-2}$ & 34 & 90 &49.7& 48.8\\ \hline
$3.4\times 10^{-5}$ & $1^*$& 1225 & 42.3 & 39 \\ \hline
\end{tabular}
\caption{Inferred models obtained varying the expansion thresholds for the  reference data (ER05,B=1000 configurations), of Fig.~\ref{fig:ace-crossentropyvst} , with  
compression frequency $f_0=0.01$. The Table gives the cluster inclusion threshold $t$, the number $N_2$ of 2-site clusters, the maximal relative error $\epsilon_{max}$, the regularized cross entropy $S^{\lambda}_c$ and the cross entropy $S_c$ obtained with the cluster expansion at different thresholds $t$. The entropy of the model having generated the data is $S=50.3$. Possible  solutions $(\epsilon_{\rm max}<5) $ are indicated by asterisks. The optimal threshold determined by the spACE procedure is $3.6\times 10^{-2}$.}
\label{table:t}
\end{table}

\subsubsection{spACE, a variant of ACE for sparse interaction networks inference} 
\vspace{-0.4cm}
To force the ACE algorithm towards a sparse solution we introduced a new procedure in the cluster expansion (available at \url{https://github.com/johnbarton/ACE}),, which stops the algorithm at a maximal number $N_2^{max}$ of 2-site clusters, see Appendix~\ref{ace:pseudocode:appendix}. This procedures imposes a prior knowledge on the sparsity of the interaction graph by giving an upper bound for the number of edges.  The Erd\H{o}s-R\'enyi random graph models used here to generate the data have an average connectivity of 2.5 neighbors per site, so we can use this prior knowledge to fix a maximal number of edges to $N_2^{max}=N \times 2.5\approx 125$. In practice  to find the best sparse graph with a number of edges smaller than the prescribed value $N_2^{max}$ we spanned the threshold range in regular intervals \footnote{The first interval is $I/\tau < t < 1$, the second is for $1/\tau^2 < t < 1/\tau$ and so on. We have chosen here $\tau$ =3.4 (see
command -trec on the GitHub site)},  at which the local minima of the absolute error  and the corresponding t parameters are recorded (see Table~\ref{table:t}) 
The parameters corresponding to the global minimum (under the sparsity conditions) are then chosen ($t=3.6\times 10^{-2}$ for Table~\ref{table:t}).
The spACE procedure is stable when changing $N_2^{max}$ around its  prior fixed average value. 
 In the following we have stopped the algorithm for $N_2^{max}=200$ (shown on Fig.~\ref{fig:ace-crossentropyvst}), after checking that the value $N_2^{max}=100$ gave similar results. This procedure greatly reduces the computational time, which increases linearly with the number of computed clusters and grows exponentially, as $q^  {K_{max}}$, with their maximal size $K_{max}$ (see Sec.~\ref{sec:times} and Appendix~\ref{acetime:appendix}): as shown in  Fig.~\ref{fig:ace-crossentropyvst} $K_{max}=3$ at $N_2^{max}=200$ while $K_{max}=7$ for the fully connected graph. 
The number of inferred parameters
$M_{cc}$  in Fig.~\ref{fig:Nparam} is further reduced by  a factor $5\;10^{-2}$ for the best sparse model on ER data in Table~\ref{table:t}.
having couplings only on connected sites.

\section{Benchmarking of color compression and decompression on synthetic data}
\label{sec:results}

To carry out an extensive analysis of the effects of the color compression introduced in Section \ref{sec:compression} on the quality of the inference, we will apply it to artificial data generated by Potts models on Erd\H{o}s-R\'enyi (ER) random graphs. The model and the generation of the data are described in Section \ref{sec:ER}.

Once the data are obtained, we apply the compression schemes introduced in Section \ref{sec:compression} with no color compression and with frequency cut-off $f_0=[0, 10^{-4}, 10^{-3}, 3\cdot10^{-3}, 10^{-2}, 3\cdot10^{-2}, 10^{-1}]$. Note that all frequency thresholds $0<f_0\leqslant 1/B$ give the same color compression, so we infer the model only for the upper value in this range and thus the number of the tested frequency thresholds depends on $B$. Moreover, $f_0=0$ corresponds to removing from the inference only the unseen states.

Given the 10 realizations of the Erd\H{o}s-R\'enyi model, the 4 sample sizes and the 5 to 8 (depending on the sampling) values of the frequency threshold define 280 data sets. For each of them, we have inferred the corresponding Potts parameters, both with the ACE and the PLM algorithms.
 
\subsection{Probability distributions}
\label{sec:KL}

The Kullback-Leibler (KL) divergence measures how the inferred probability distribution of the possible configurations diverges from the empirical one (defined from the data samples), and can be computed as:
\begin{eqnarray}
 	D(P_{\boldsymbol{J}^{real}} || P_{\boldsymbol{J}^{inf}}) &=& \sum_{{\mathbf{a}}} P_{\boldsymbol{J}^{real}}({\mathbf{a}}) \log \frac{P_{\boldsymbol{J}^{real}}({\mathbf{a}})}{P_{\boldsymbol{J}^{inf}}({\mathbf{a}})} =  \log(Z_{inf})-\log (Z_{real})+\left\langle E_{inf}({\mathbf{a}})-E_{real}({\mathbf{a}})\right\rangle_{{\mathbf{a}}\:in\:real}\,, \nonumber
 	\label{eq:KL}
\end{eqnarray}
where ${\mathbf{a}}=\{a_1,\ldots a_N\}$ is a configuration and $\langle \cdot\rangle_{{\mathbf{a}}\:in\:real}$ indicates the average over the configurations generated by Markov Chain Monte Carlo (MCMC) from the real model. The first and the second lines are identical only when an infinite configuration sample is employed.
Here, we estimate the average over $P_{\boldsymbol{J}^{real}}$ using an ensemble of 50,000 MCMC configurations sampled from the model. 

As described before, the computation of the partition function $Z$ is far from being trivial, and was done in two ways. First, we used Annealed Importance Sampling (AIS) \cite{Salakhutdinov08}, starting from  the independent-site model: All initial couplings were set to zero, while initial fields were computed as $h^0_i(a)=\log(f_i(a)+\alpha)-\log(f_i(cons_i))$ where $cons_i$ is the most common state at site $i$ and $\alpha=1/B$ is the smallest observed frequency used as regularization. 
Then a chain of models with increasing couplings (up to the inferred values) are thermalized and the ratios of their partition functions may be estimated. Secondly, the Kullback-Leibler divergence and the logarithm of the partition function can also be directly estimated by the ACE procedure (Table~\ref{table:t}), see Appendix~\ref{ace:appendix}. The  KL divergences obtained directly from the ACE expansion and the ones obtained with importance sampling are very similar, as shown in Table~\ref{table::KLandZcomp}, for the reference case in Fig.~\ref{fig:ace-crossentropyvst} at the optimal cluster inclusion threshold corresponding to a sparse inferred network. The values of the logarithm of the partition function, and of the entropy are also consistent between the two methods.
In the following we will use the annealed importance sampling to calculate the KL divergence to compare results from PLM and ACE.

\begin{table}[]
\begin{tabular}{| c | c | c | c | c |c|c|}
\hline
cluster inclusion threshold t &	logZ(AIS) &	logZ(ACE) &	S(AIS) &	S(ACE) &	KL(AIS) &	KL(ACE)\\
\hline
1&	28.9&	28.6&	56.9&	56&	6.4&	6 \\
$9.6\times 10^{-2}$&	32&	31.9&	52.6&	51.4&	2.2&	2.3 \\
$3.6\times 10^{-2}$&	32.5&31.8&51.8&	50.7&	1.5&	1.5 \\
$2.2\times 10^{-2}$&	34.2& 32.8&	52.9&	50.9&	2.5&	3 \\
$3.4\times 10^{-5}$&	36&	26&	57.8& 49.1&	7.4&	5 \\
\hline
\end{tabular}
\caption{Comparison between importance sampling (AIS) and ACE methods to obtain the logarithm of the partition function $Z$, the entropy $S$, and the Kullback-Leibler divergence (KL), at the different sparsity threshold $t$ for the reference model, ER05, data sampling: B=1000 and color compression $f_o=0.01$, the optimal threshold determined by the spACE procedure is $3.6\times 10^{-2}$.  Fluctuations of the above values in AIS estimation over  repetition of MC sampling are of the order $5\,10^{-3}$ for the sparse models and $5\,10^{-2}$ for the fully connected model.}
\label{table::KLandZcomp}
\end{table}
\subsubsection{KL Divergence for ACE models at different inclusion thresholds $t$}
\begin{figure}[bth!]
	\subfloat [ACE]
	{ \includegraphics[width=0.4\linewidth]{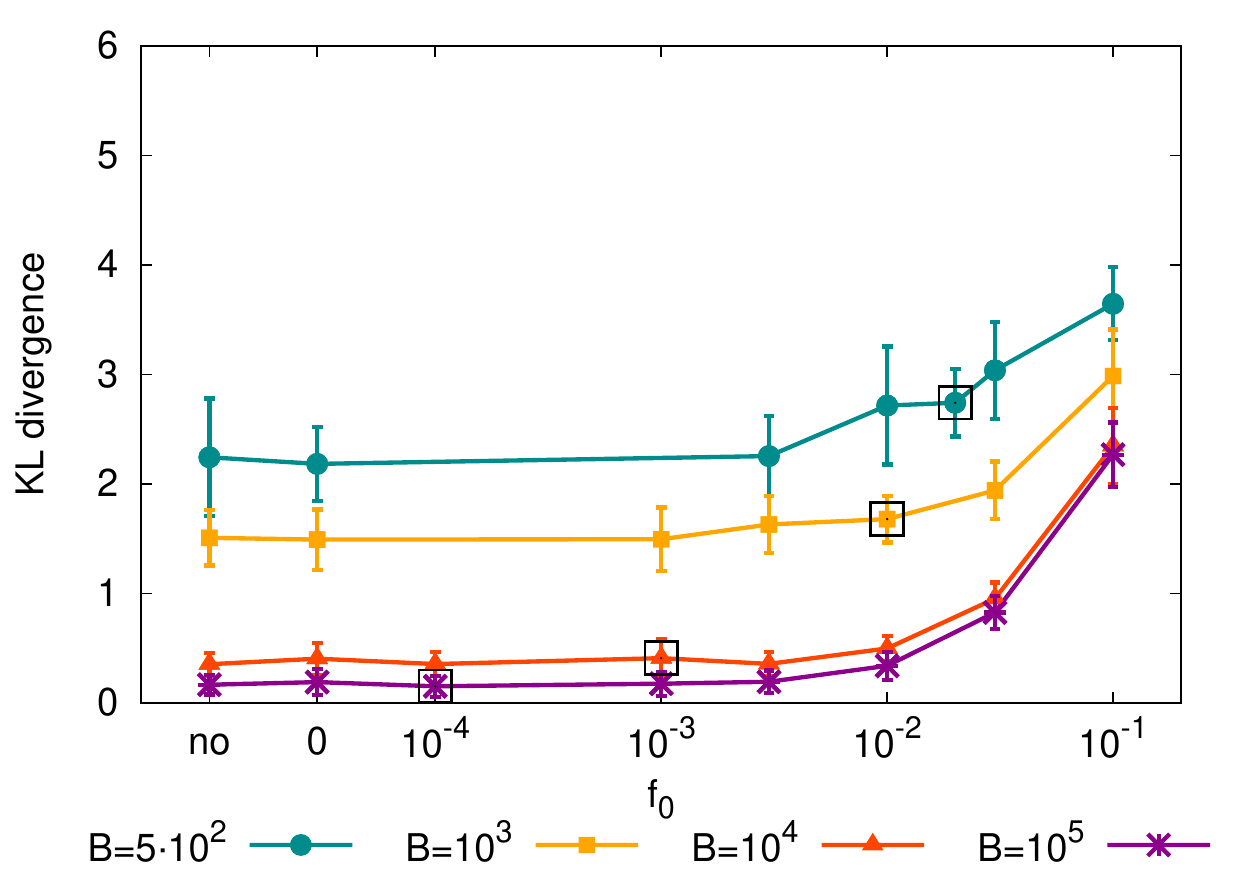} }
	\subfloat [PLM]
	{ \includegraphics[width=0.4\linewidth]{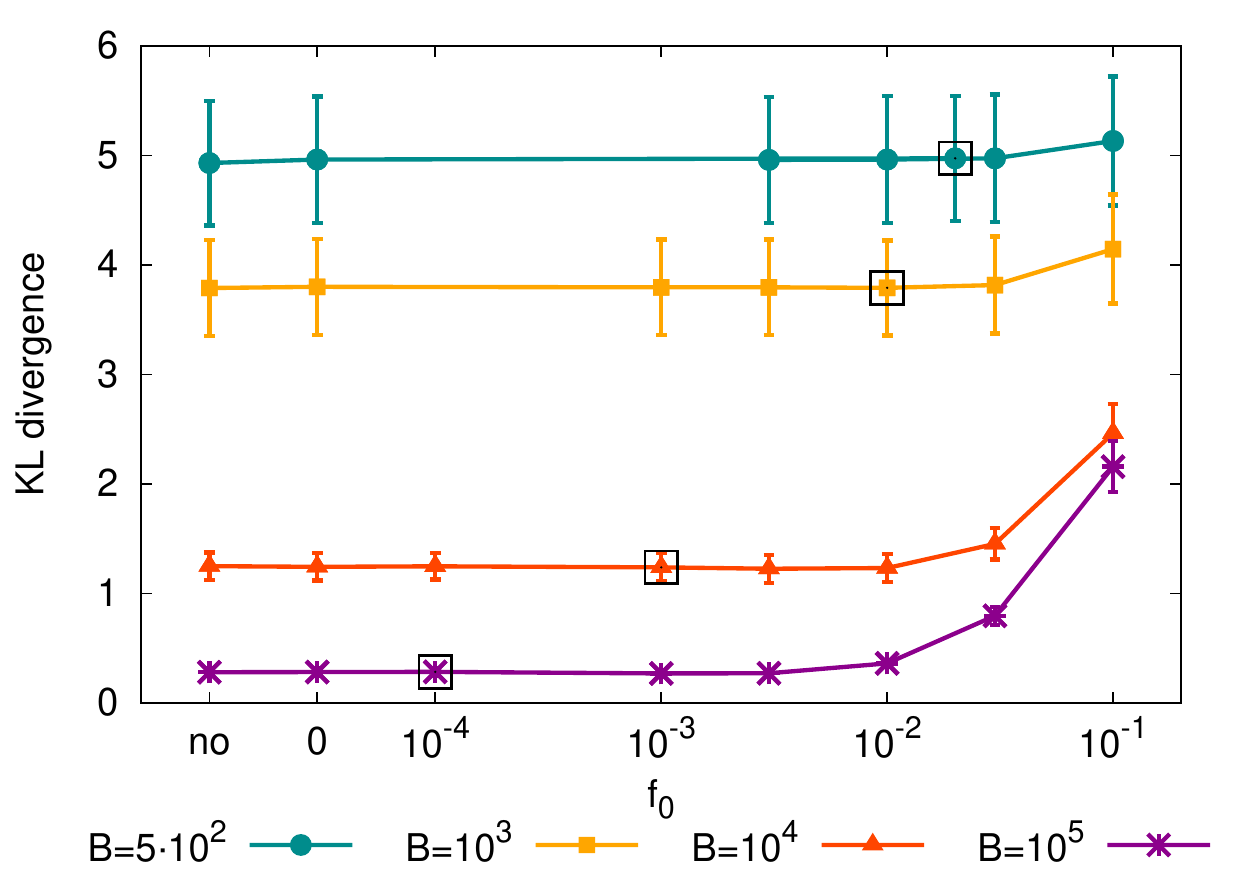} }
	\caption{Kullback-Leibler divergence between real and inferred probability distributions averaged over 10 realizations plotted as function of the compression parameter $f_0$ for different sample sizes. The left plot is for ACE, the right plot for PLM. Error bars are standard deviations over the 10 realizations. Black empty squares correspond to the reference cutoff frequency $f^{*}_0=10/B$. 
	\label{fig:KL}}
\end{figure}

\label{sec:acesparsitythreshold}
Table \ref{table::KLandZcomp} displays the KL divergences for the reference data set and different cluster inclusion thresholds of Fig.~\ref{fig:ace-crossentropyvst} and in table \ref{table:t} obtained both with importance sampling and the ACE expansion.

The fully-connected-graph model has a larger KL divergence and is therefore overfitting the data, while the sparse-graph model reproduces better the original model. All results for the ACE expansion presented in the following are obtained with the spACE procedure to infer a sparse graph. For the fully connected solution, due to overfitting, the cross entropy of Table~\ref{table:t} is not a good approximation to the entropy. Therefore, the estimate of the logarithm of the partition function and of the entropy given in table \ref{table::KLandZcomp} are significantly different from the ones obtained by the AIS method.

\subsubsection{KL Divergence as a function of the sampling depth B and the color compression threshold $f_0$}
\label{sec:plmlowreg}
Fig.~\ref{fig:KL} shows the mean over the ten ER realizations of the KL divergence between the real and the inferred distributions for various sampling depths and compression parameters for both ACE (left) and PLM (right). As expected, the KL divergence decreases for bigger samples, becoming very close to zero for $B=10^5$. The same happens for the standard deviations over the 10 realizations. ACE gives smaller KL divergences with respect to PLM, showing that the sparsity imposed in the spACE procedure gives a model reproducing better the original, sparse ER models. 

The black empty squares  in Fig.~\ref{fig:KL} indicate a reference compression threshold  
$f^{*}_0=10/B$, {\em i.e.} grouping symbols observed less than 10 times. We consider that frequencies larger than $f_0$ are reliably estimated \footnote{We estimate the standard deviation $\sigma$ of the osberved frequency based on a Binomial variable with probability $\hat f$ in a sample of size $B$: $\sigma=\sqrt {({\hat f} \, (1-{\hat f} ) /B}\simeq \sqrt {{\hat f} /B}$. Probabilities larger than $f_0$ correspond to signal-to-noise ratios ${\hat f}/\sigma > \sqrt{f^*_0\, B}> 3$. } For  $f_0>f^*_0$ the compression procedure  discards information on 'reliable' colors, and a loss in performance is expected.
  For both inference procedures the increase in KL divergence in Fig.~\ref{fig:KL}  becomes visible  only at a large value of the cut frequency $f_0 \gtrsim 0.001$, independently from the sampling depth.
We  indeed  expect that the increase of the KL divergence due to the color compression procedure be a  monotonic function of the cut frequency $f_0$, so it is  irrilevant at small cut frequency.
For high values of $f_0$  the increase  of the KL divergence  with respect to the uncompressed model is of course much more significant at high $B$.

Table~\ref{table::KLandZcomp} shows  that the strong regularization ($\gamma_J\approx N/B$) used in the fully connected PLM inference is essential to reduce overfitting:
a PLM  model inferred with a weak regularization ($\gamma_J\approx 1/B$) gives  indeed very large KL divergences, especially at low sampling depth.
   Color compression, acting as an additional regularization,  helps in this  latter case to reduce overfitting  
and to  lower the KL divergence. As shown in Appendix~\ref{appendix:PLM:regularization} and Fig.\ref{fig:KL_lowlambda}  the KL divergences for weak regularized PLM model
reach indeed a minimal value, at  intermediates or large compression thresholds $f_0$ depending on  sampling depth, 
   compatible, but slightly larger, than the one obtained for the usual strongly regularized PLM of Table~\ref{table::KLandZcomp}.




\begin{table}
\begin{tabular}{ l | c | c | c }
& PLM & \centering{PLM} & ACE \\
B & ($\gamma_J=\frac{1}{B}$,$\gamma_h=\frac{0.002}{B}$) & ($\gamma_J=\frac{50}{B}$,$\gamma_h=\frac{0.1}{B}$) & ($\gamma_J=\frac{1}{B}$,$\gamma_h=\frac{0.01}{B}$) \\
 \hline
$10^3$ & 12.97 & 3.80 & 1.35 \\
$10^4$ & 2.85 & 1.28 & 0.37 \\	
$10^5$ & 2.10 & 0.30 & 0.18 \\
\end{tabular}
\caption{KL divergences between inferred and empirical distributions
on the reference data sets at different $B$ and with no color compression.
The  value  obtained by  weakly regularized PLM  $\gamma_J=\frac{1}{B}$  is compared to the ones obtained by the standards strongly regularized  PLM inference and  weakly regularized spACE inference, used in Fig.~\ref{fig:KL}.}
\label{appendix:PLM:reg:table}
\end{table}

\begin{figure}[h!]
\hspace{-1cm}
	\subfloat
	{ \includegraphics[width=0.25\linewidth]{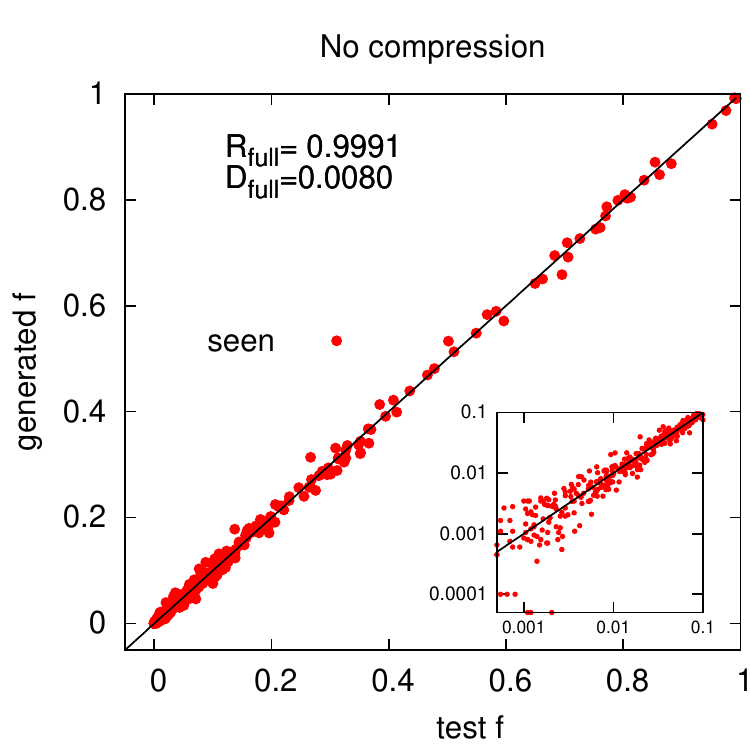}}
	\subfloat
	{ \includegraphics[width=0.25\linewidth]{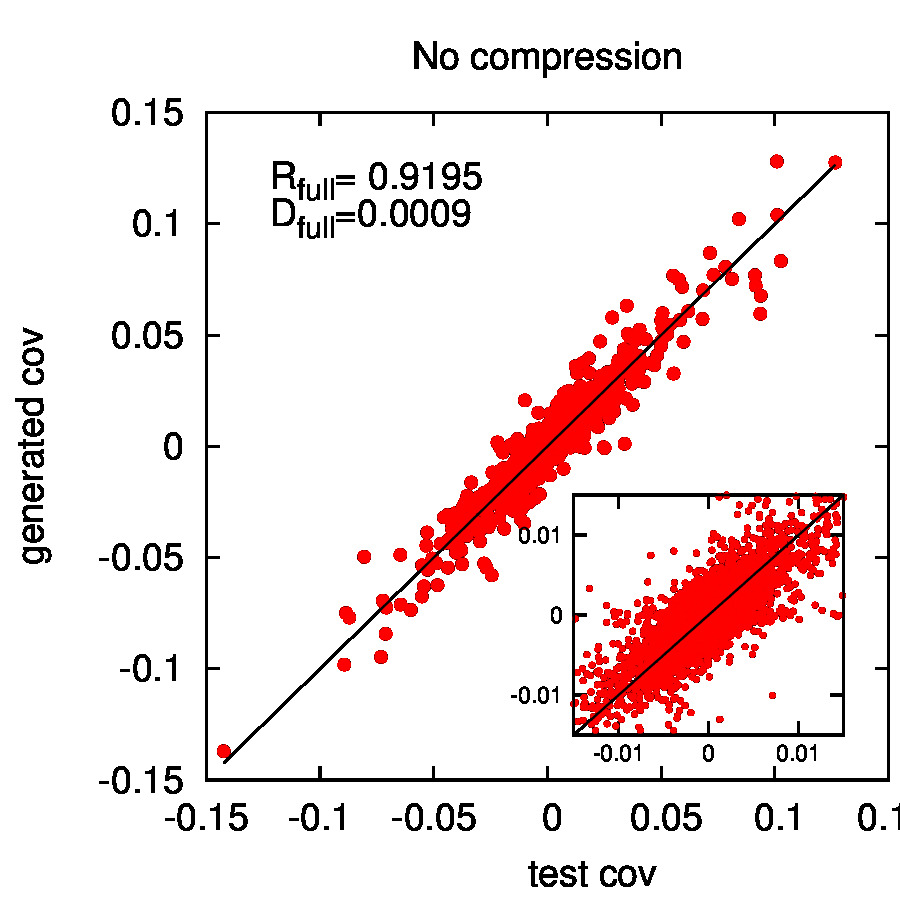}}
	\subfloat
	{ \includegraphics[width=0.25\linewidth]{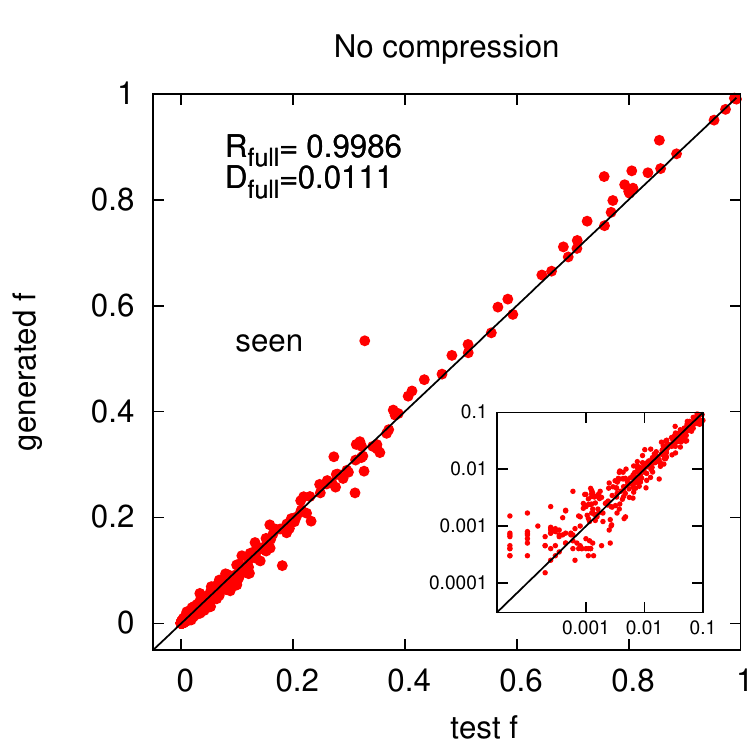}}
	\subfloat
	{ \includegraphics[width=0.25\linewidth]{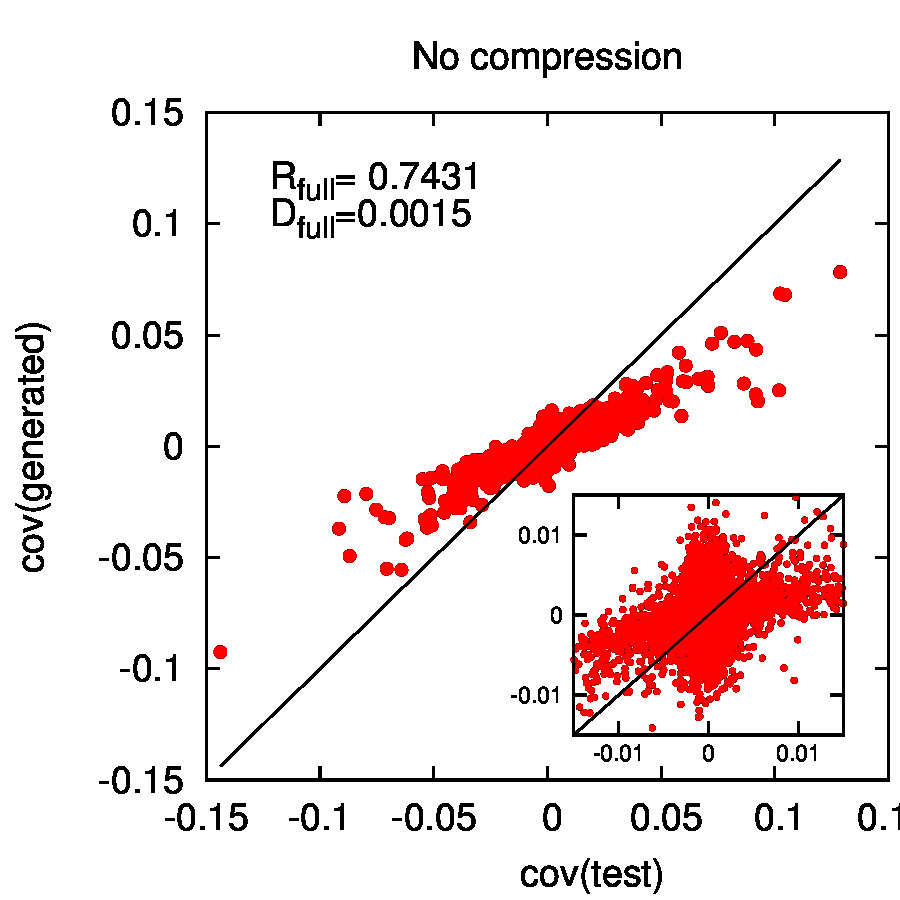}} 
	 \\
	 \hspace{-1cm}
	\subfloat
	{ \includegraphics[width=0.25\linewidth]{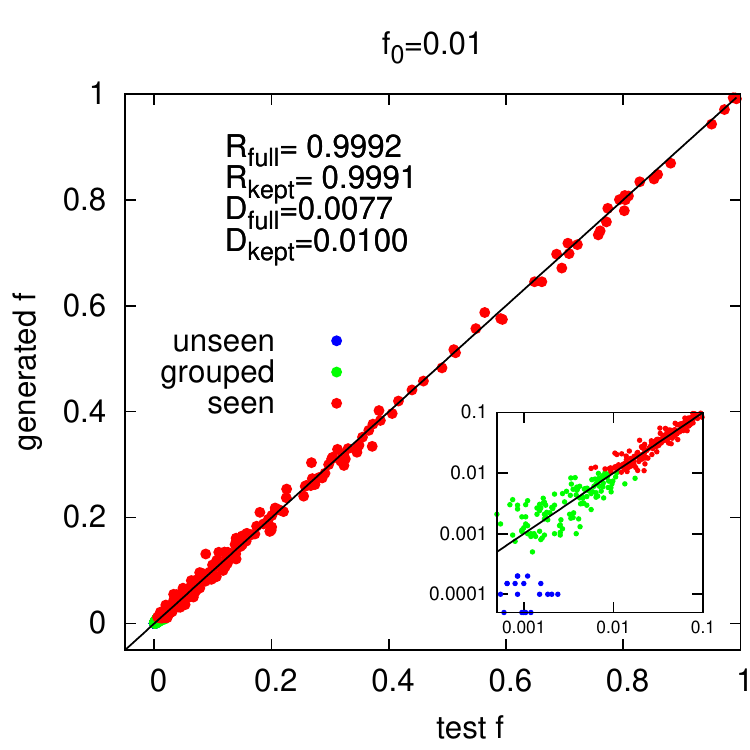}}
	\subfloat
	{ \includegraphics[width=0.25\linewidth]{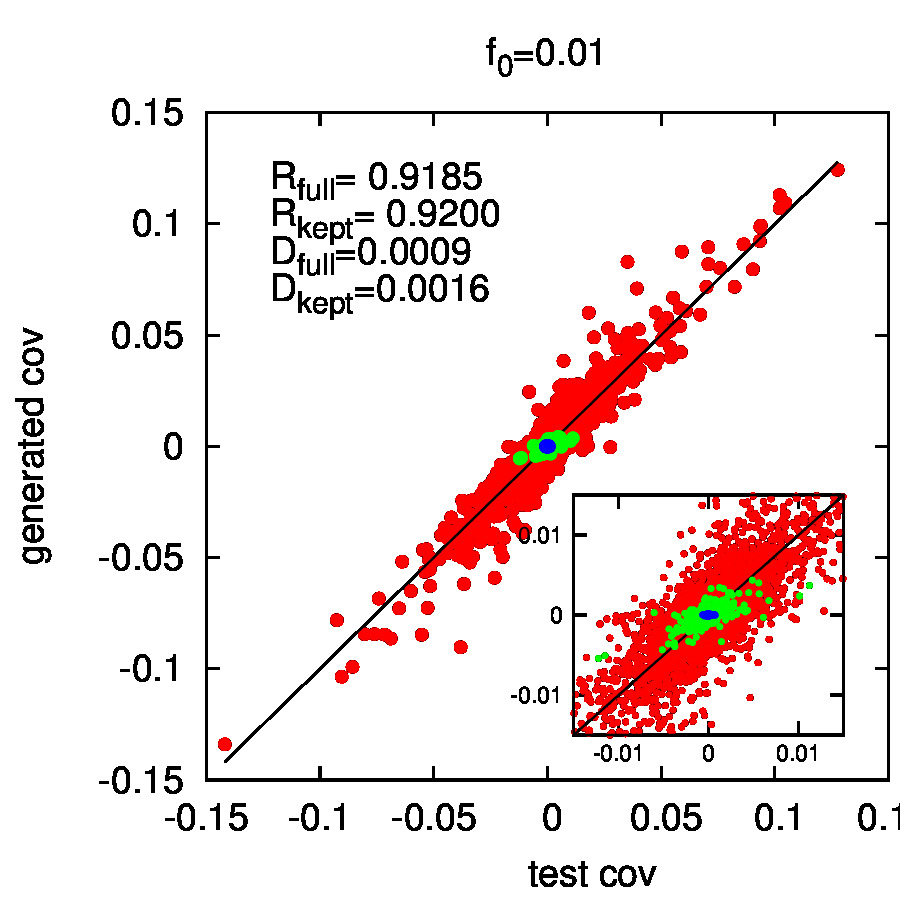}}
	\subfloat
	{ \includegraphics[width=0.25\linewidth]{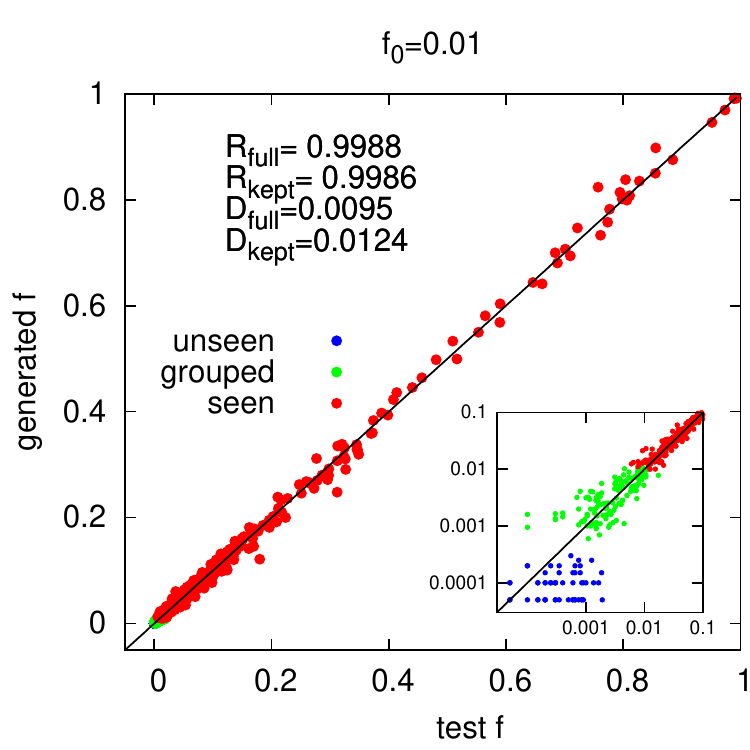}}
	\subfloat
	{ \includegraphics[width=0.25\linewidth]{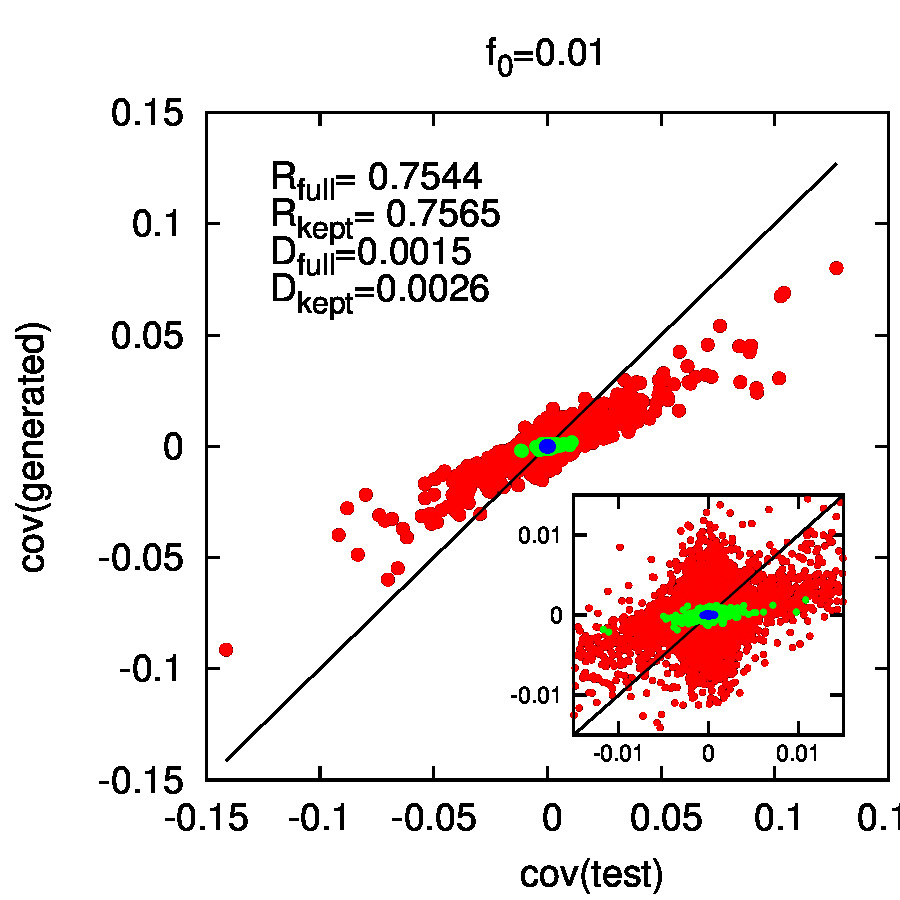}} \\
	  \hspace{-7cm} \subfloat   (a) ACE 
	   \hspace{7cm} \subfloat  (b) PLM
	\caption{Reconstruction of average frequencies and covariances for the reference case (ER05, B=1000). Comparison between generated data and test set for no color compression (top) and $f_0=0.01$ for ACE ( left,a) and PLM ( right, b). The Pearson correlation coefficient ($R$) and the absolute error ($\Delta$, Eq.~\ref{eq:mse}) are marked on top of the plots both for the full model and the for the reduced one (only explicitly modeled states). }
	\label{fig:genACEandPLM}
\end{figure}

\subsection{Low-order Statistics} \label{sec:generated}

In this Section we discuss the generative 
properties of the inferred models, in particular its ability to reproduce the low order statistics of the original model : the site frequencies $f_i(a)=\displaystyle{\sum_{generated\: {\mathbf{a}}} \delta(a_i,a)/B_{gen}}$ and covariances $cov_{ij}(a,b)=\displaystyle{\sum_{generated \:{\mathbf{a}}} \left[\delta(a_i,a)\delta(a_j,b)/B_{gen}\right]-f_i(a)f_j(b)}$.
To benchmark the generative power of the inferred model as a function of the color compression two sets of 20000 configurations are generated by Markov Chain Monte Carlo, respectively with the real and with the inferred model for each $B$, $f_0$, and graph realization, and their low order statistics are compared.

Figure~\ref{fig:genACEandPLM}  shows, for the models inferred from the reference data set, the comparison of the frequencies and covariances computed from the configurations generated by the real model (test sequences) and by the models inferred with ACE and PLM, without color compression (top panels) and with $f_0=0.01$ (bottom panels). As shown in the figures PLM covariances are downscaled, due to the strong regularization, as happens for the couplings (Section \ref{sec:HJ}). 
Moreover PLM assigns smaller frequencies to the unseen Potts states (left panels of Fig.~\ref{fig:genACEandPLM} in log-log scale). This is probably due to the fact that the pseudocount used during decompression seems to be well fixed for the weak regularization used in ACE but not for the large regularization used in PLM.
The insets in Fig~\ref{fig:genACEandPLM} show that, contrary to spACE,  zero covariances are set to non-zero values with PLM due to  overfitting.

\begin{figure}[h!]
	\subfloat
	{\includegraphics[width=0.4\linewidth]{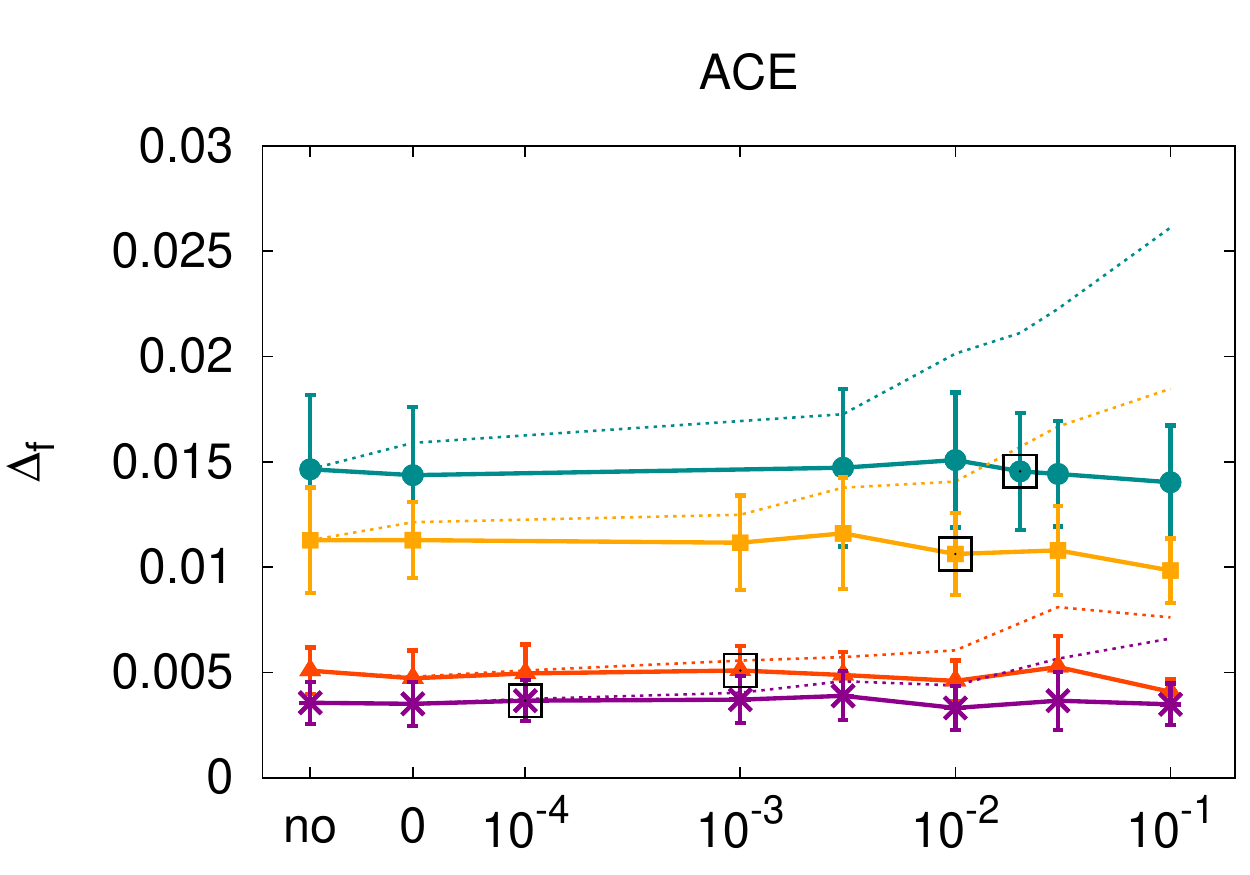}}
	\subfloat
	{\includegraphics[width=0.4\linewidth]{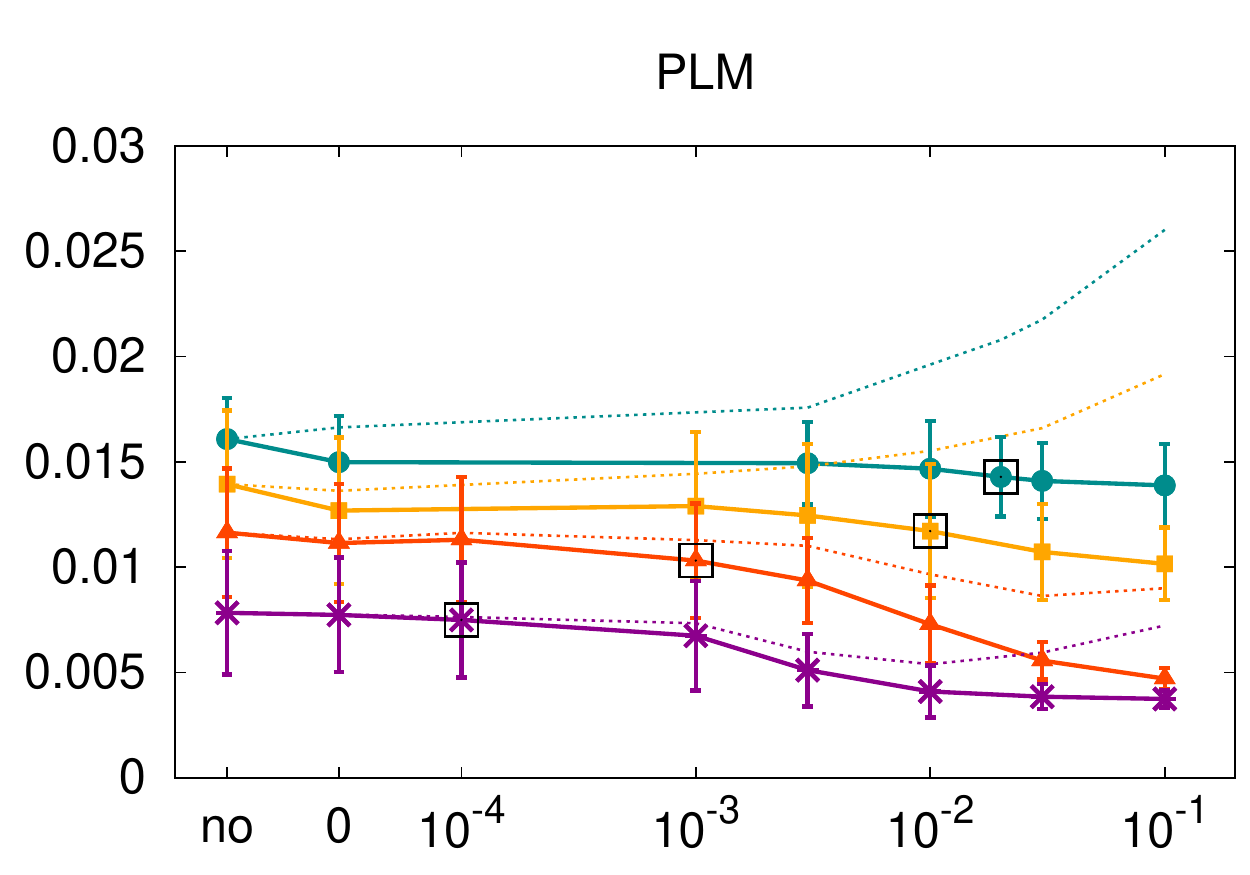}}\\
	\subfloat
	{\includegraphics[width=0.4\linewidth]{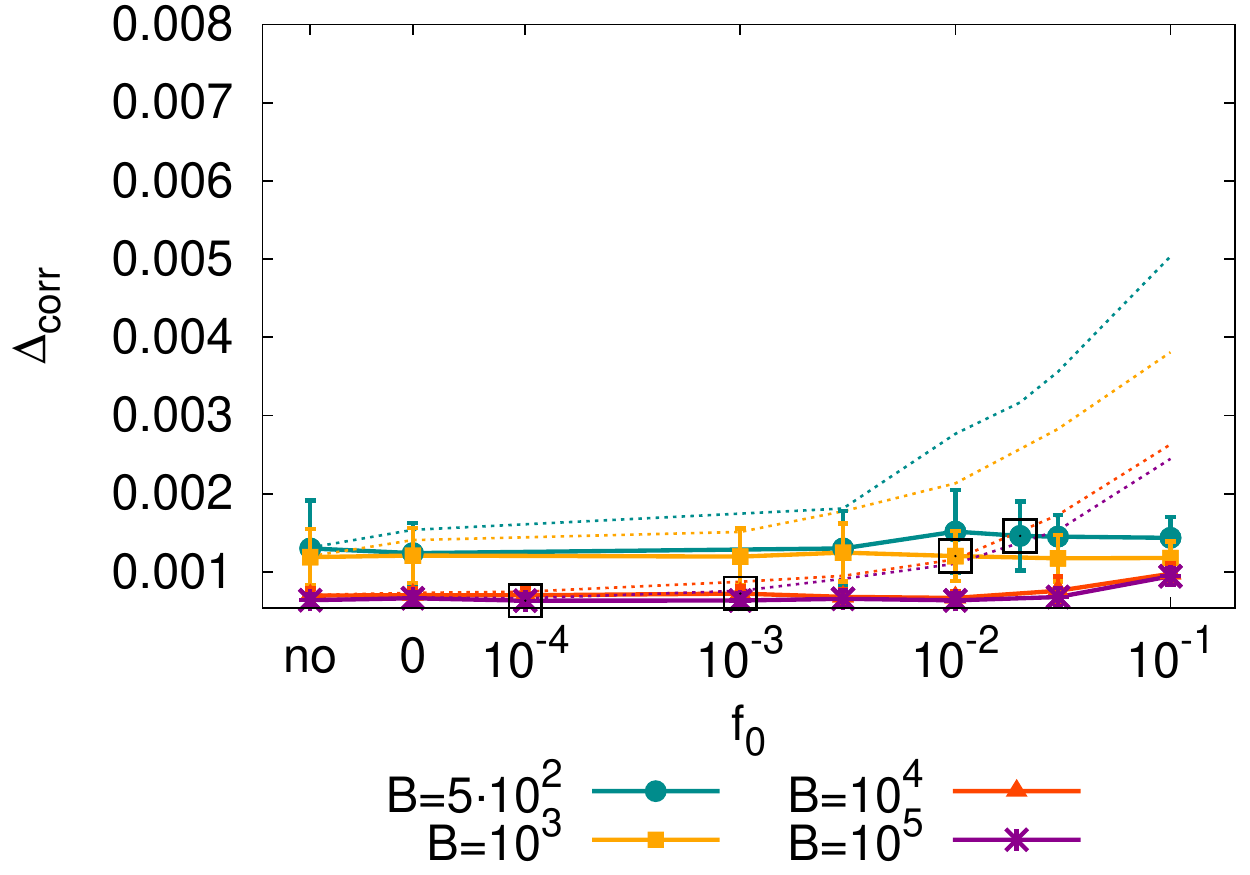}}
	\subfloat
	{\includegraphics[width=0.4\linewidth]{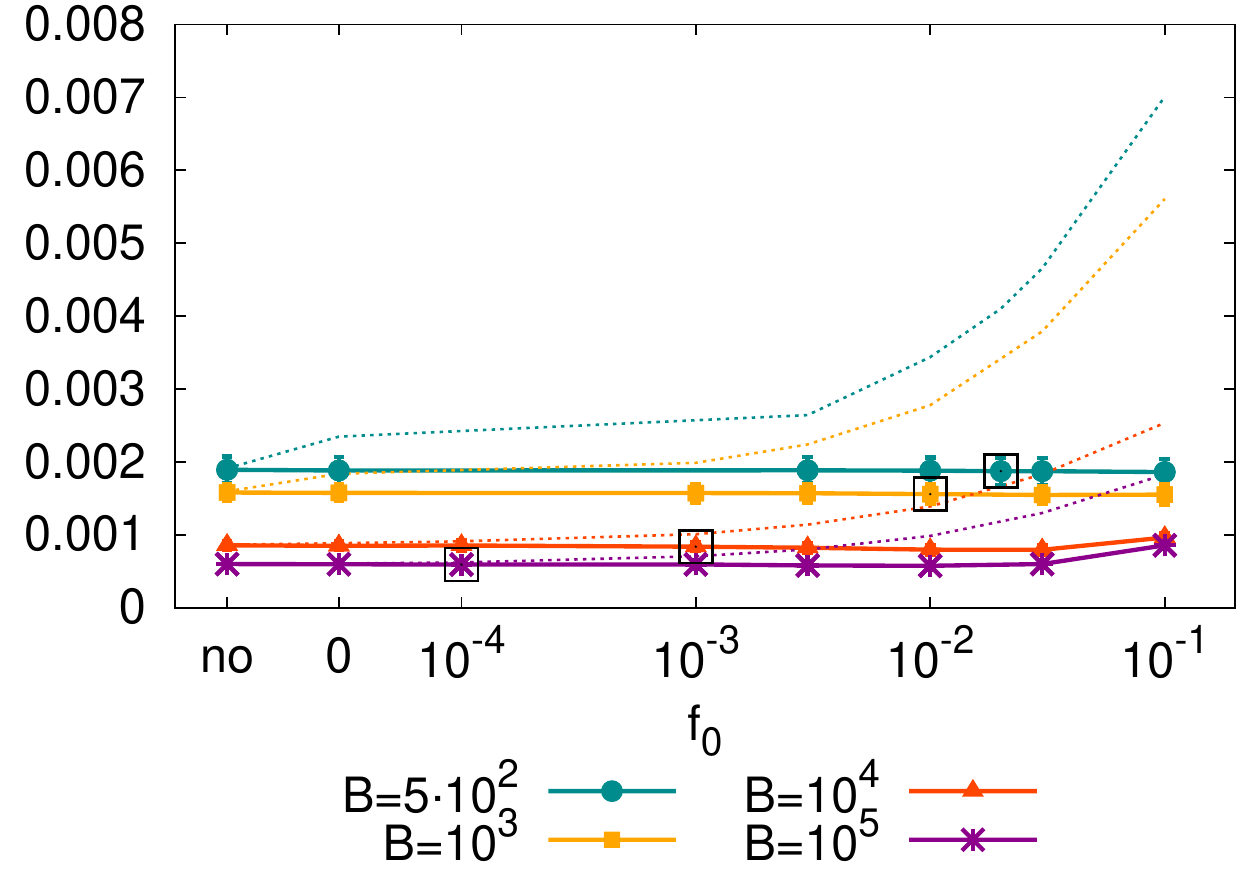}}
	\caption{Absolute error (Eq.~\ref{eq:mse}) of frequencies ($\Delta f$, top panels) and covariances ($\Delta\text{corr}$, bottom panels) averaged over 10 ER realizations, as a function of the color compression for several sample sizes.
	Dashed lines: error on explicitly modeled Potts states only. Full lines: error on all parameters.
	Error-bars are standard deviations computed over the 10 realizations. Inference is performed respectively by ACE (left) and PLM (right).
	}
	\label{fig:delta_testgen_ace_plm}
\end{figure}

To have a more systematic comparison, we analyzed their absolute mean square error defined as:
\begin{equation}
\label{eq:mse}
\Delta f = \sqrt{ \frac{\sum_{i}\sum_{a} \left(f^{gen}_i(a) - f^{test}_i(a)\right)^2 }{\sum_{i} q_i} } \ , \quad
\Delta \text{corr}= \sqrt{ \frac{\sum_{ij} \sum_{ab} (cov_{ij}^{test}(a,b)-cov_{ij}^{gen}(a,b))^2}{\sum_{ij} q_i \cdot q_j }} \ .
\end{equation}
These quantities are then computed for different $B$ and $f_0$ and averaged over 10 graph realizations.

 Figure~\ref{fig:delta_testgen_ace_plm}  shows the  Mean Square Errors (MSE) for the frequencies and covariances.
spACE  has again better performances than PLM \footnote{Similar results are obtained when considering the Pearson correlations between the real and inferred frequencies or correlations rather than the absolute errors (not shown).}.
 The MSE for the full Potts model (full lines) shows a little dependence on the compression frequency $f_0$; meanwhile 
the MSE restricted to explicitly modeled colors (dotted lines) generally increases at large compression performances. The only exception is that frequencies
 are better reproduced with PLM at strong color compression because the decompression procedure 
(Eq.~\ref{eq:decompress}),  acting as an independent model, correctly assigns fields to grouped states.
  The  above observations are simply explained by the fact that  the contribution to the MSE  of a color increases with its frequency.
 The MSEs  on  the full Potts model, after  decompression (full lines in Fig.~\ref{fig:delta_testgen_ace_plm} ) are  quite independent from $f_0$,  as the averages in the MSE are
dominated by the  large amount of colors with small frequencies, see Fig.~\ref{fig:genACEandPLM}. 
On the contrary the MSE on explicitly modeled colors (dot lines  in Fig.~\ref{fig:delta_testgen_ace_plm})  shows a large increase with the  compression threshold because 
 the mean is restricted to more and more  frequent colors.



\begin{figure}[h!]
	\subfloat[ACE's $PPV(n_{min})$ ]{ \includegraphics[width=0.4\linewidth]{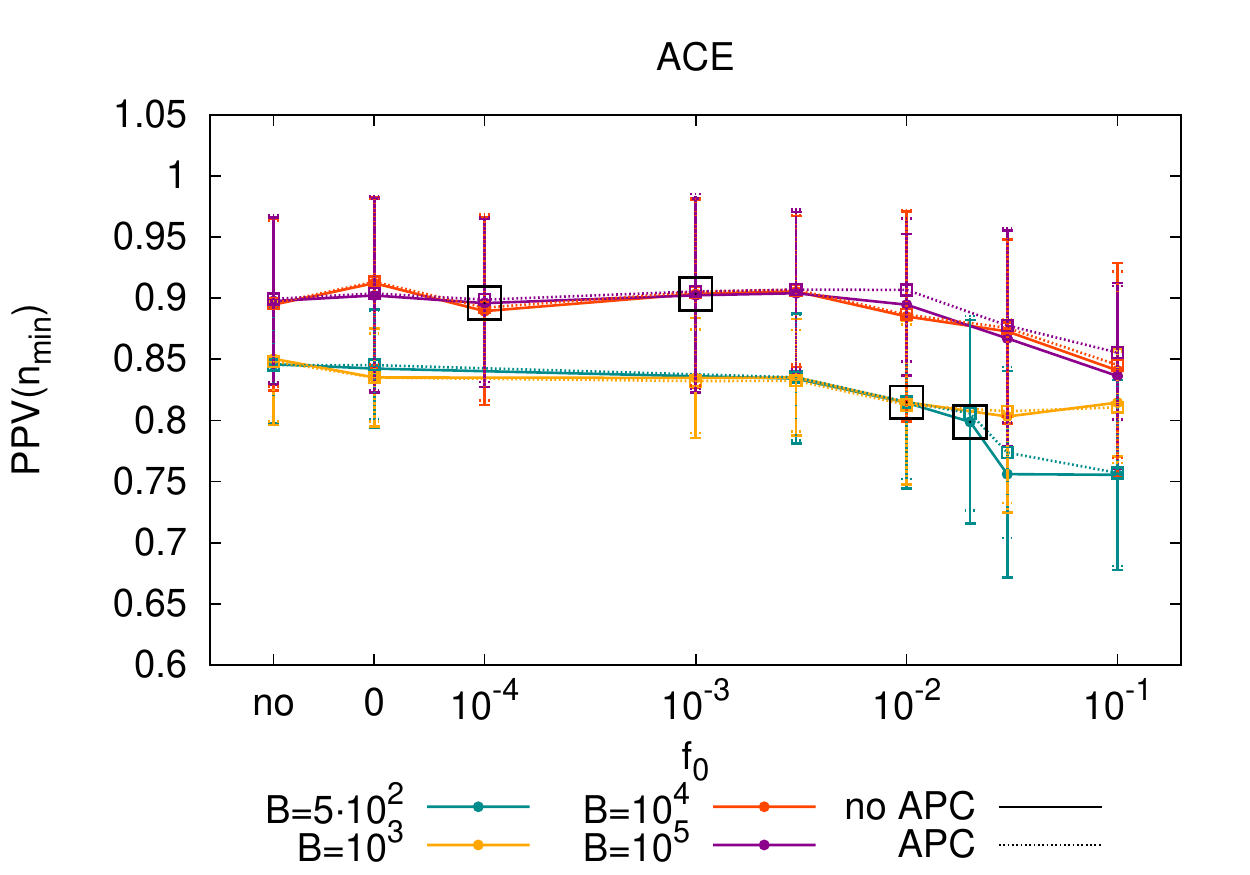}} 
\subfloat[PLM $PPV(N_0)$ ]{
\includegraphics[width=0.4\linewidth]{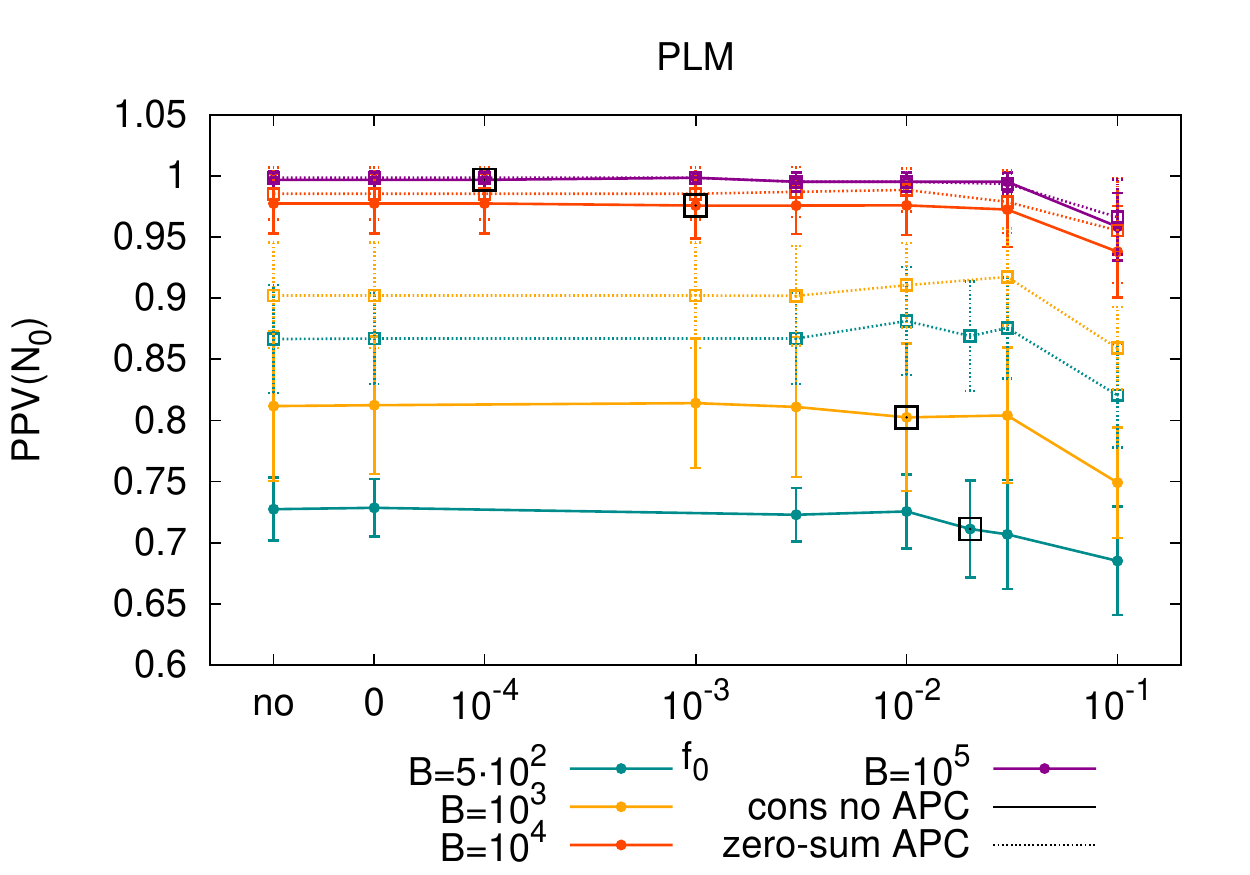}}
	\caption{ACE and PLM interaction graph reconstruction as a function of the color compression $f_0$. 
Positive Predicted value $PPV(n=min \left\lbrace N_0,N_{pred}\right\rbrace)$ for ACE and PLM inference. Points and error bars are averages and standard deviations obtained on 10 ER realizations. For ACE all quantities are computed in the consensus gauge with (dotted line) and without (full line) APC. (Left and middle plot). Right Plot $PPV(N_0)$ with PLM either in the consensus gauge without APC (full line), or in the zero-sum gauge with APC (dotted line).}
	\label{fig:PPVACEandPLM}
\end{figure}

\subsection{Interaction networks\label{sec:contacts}}

In this section we focus on the reconstruction of the interaction network and the prediction of pairs of sites that are connected, or ``in contact'', in the interaction graph. The original ER graph is sparse, with an average connectivity of about 2.5. We can predict contacting sites as those site pairs with large couplings, as traditionally done for protein structures
\cite{sulkowska2012genomics, nugent2012accurate, hopf2012three}.
To this end, we compute the Frobenius norm of the (10$\times$10) inferred and decompressed coupling matrix between each pair of sites $i, j$, $F_{ij}=\sqrt{\sum_{a,b} J_{ij}(a,b)^2}\,.$


The heat map of the  Frobenius norms of the couplings inferred by ACE and PLM at different color compression gives very similar results and allow us to identify   the largely coupled sites  by ACE and PLM (see Figs.~\ref{fig:cmapACEandPLM} in Appendix).
The only difference between ACE and PLM is that  spACE infer a sparse network with zero Frobenius norms on the majority of links,
while PLM with the $L_2$ norm regularization described in Eq.~\ref{eq:L2} infers  a dense Frobenius norm matrix,
There is therefore no straightforward separation between pairs of sites predicted to be in interaction or not.

To gain more insight into these predictions, as done for protein structure \cite{morcos2011direct,ekeberg2014fast,cocco2018inverse}, we can sort links by decreasing Frobenius norm and follow the precision obtained progressively including the corresponding links in the so called Positive Predicted Value (PPV) curve (see Appendix~ \ref{sec:graphdef}).This is shown in  Figs.~\ref{fig:PPVACEandPLM} for a number of links up to the last  inferred one for ACE or to $N_0=50$ for PLM, for the reference case.

The Frobenius norm is computed in the consensus gauge, which will be used to compare the couplings and fields in the next section and in the  zero sum gauge  used on protein structure prediction \cite{dunn2008mutual,morcos2014coevolutionary,cocco2018inverse}. Performance can be further improved with the Average Product Correction (APC) (defined in Appendix ~\ref{sec:graphdef}) when using zero sum gauge.
When inferring sparse networks, APC only corrects the ranking of the predictions in the PPV curve, but it does not change the overall number of site pairs predicted to be in interaction, nor the global precision,
on the contrary APC on the zero sum gauge largely improve PLM precision, as expected.
 Figure~\ref{fig:PPVACEandPLM} (bottom)  shows the average PPV, over all the data realization and as a function of $f_0$,  at the number of real links $PPV(N_0=50)$  for PLM and at the 
 lesser between the predicted  $N_{pred}$ or real  $N_0$ number of links with ACE.

As expected, the plots show that the contact prediction improves with sampling, and that the APC significantly improves the results for PLM in zero sum gauge. PLM generally gives higher PPV, especially at high sampling depth $B=10^4$ and $B=10^5$ 
where the reconstruction error for spACE  are due to the sparsity threshold. We have verified that, at large sample and when inferring fully connected netwoeks ACE has the same PPV as PLM.
We see that, for both algorithms, the performance is usually stable against the introduction of color compression. 

 \begin{figure} [h!]
	\subfloat[ACE: no color compression]	{\includegraphics[width=0.4\linewidth]{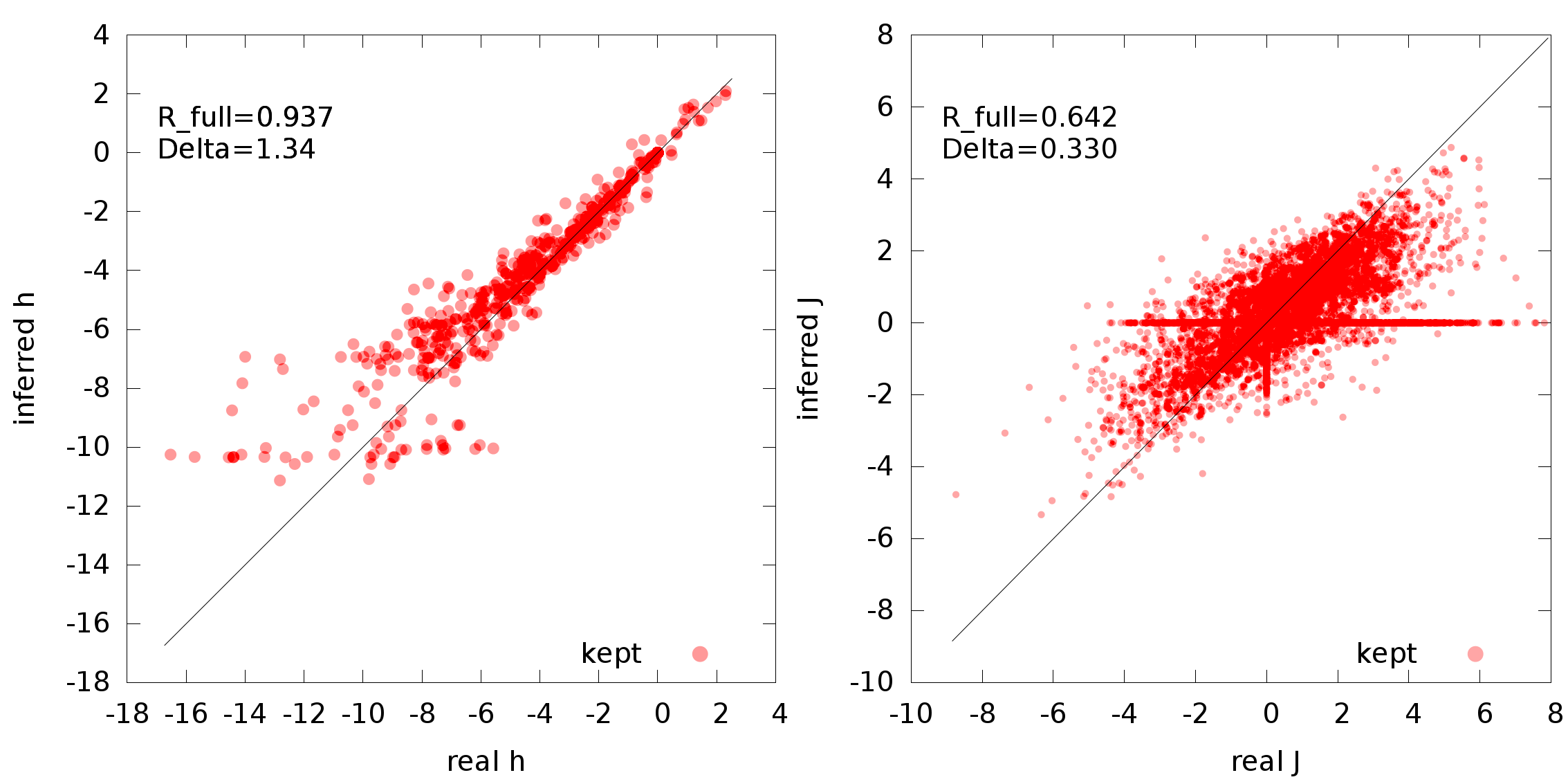}} 
	\subfloat[PLM: no color compression]{\includegraphics[width=0.4\linewidth]{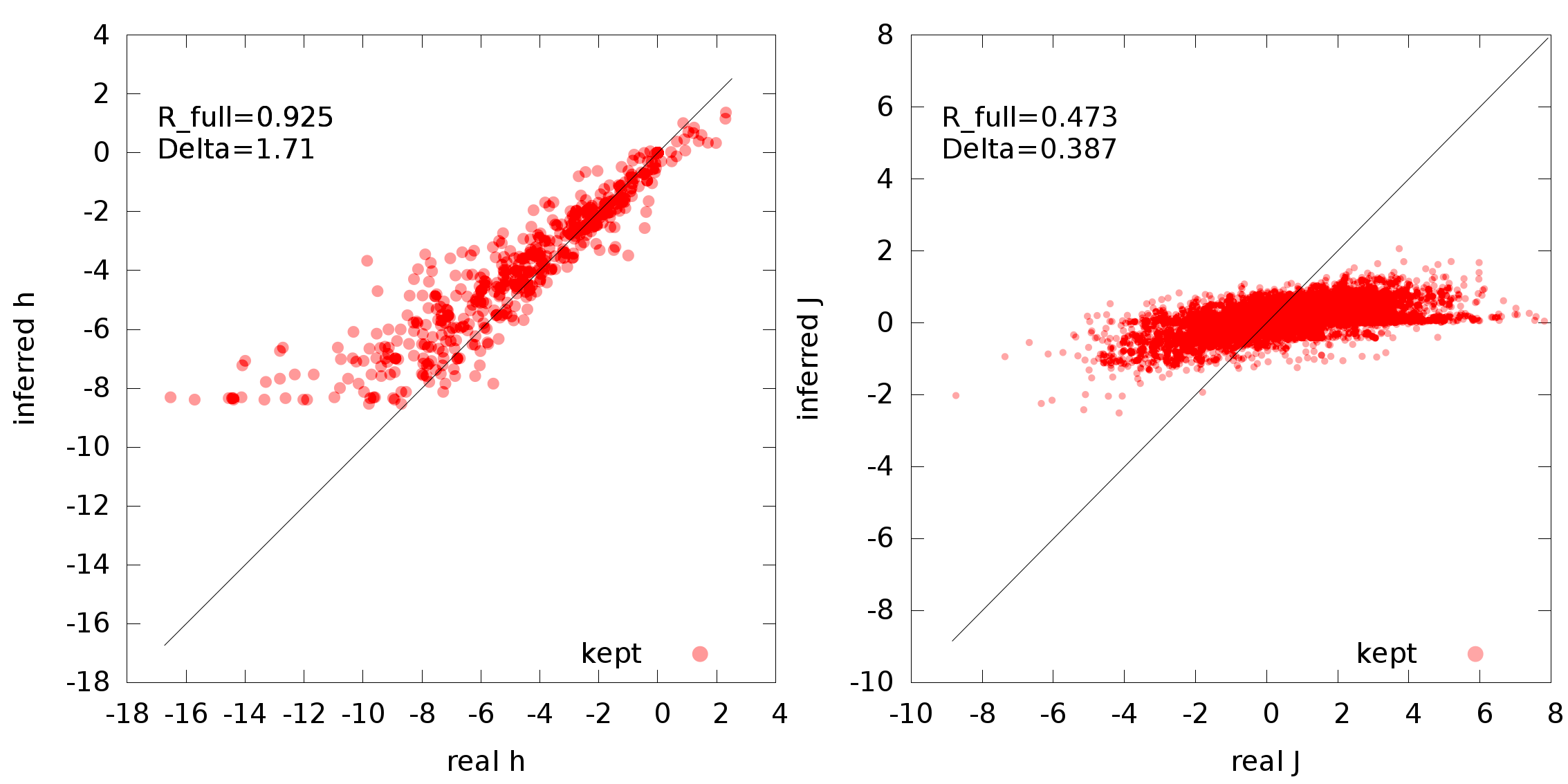}} \\
	 \subfloat[ACE: $f_0=0.01=10/B$]
	{\includegraphics[width=0.4\linewidth]{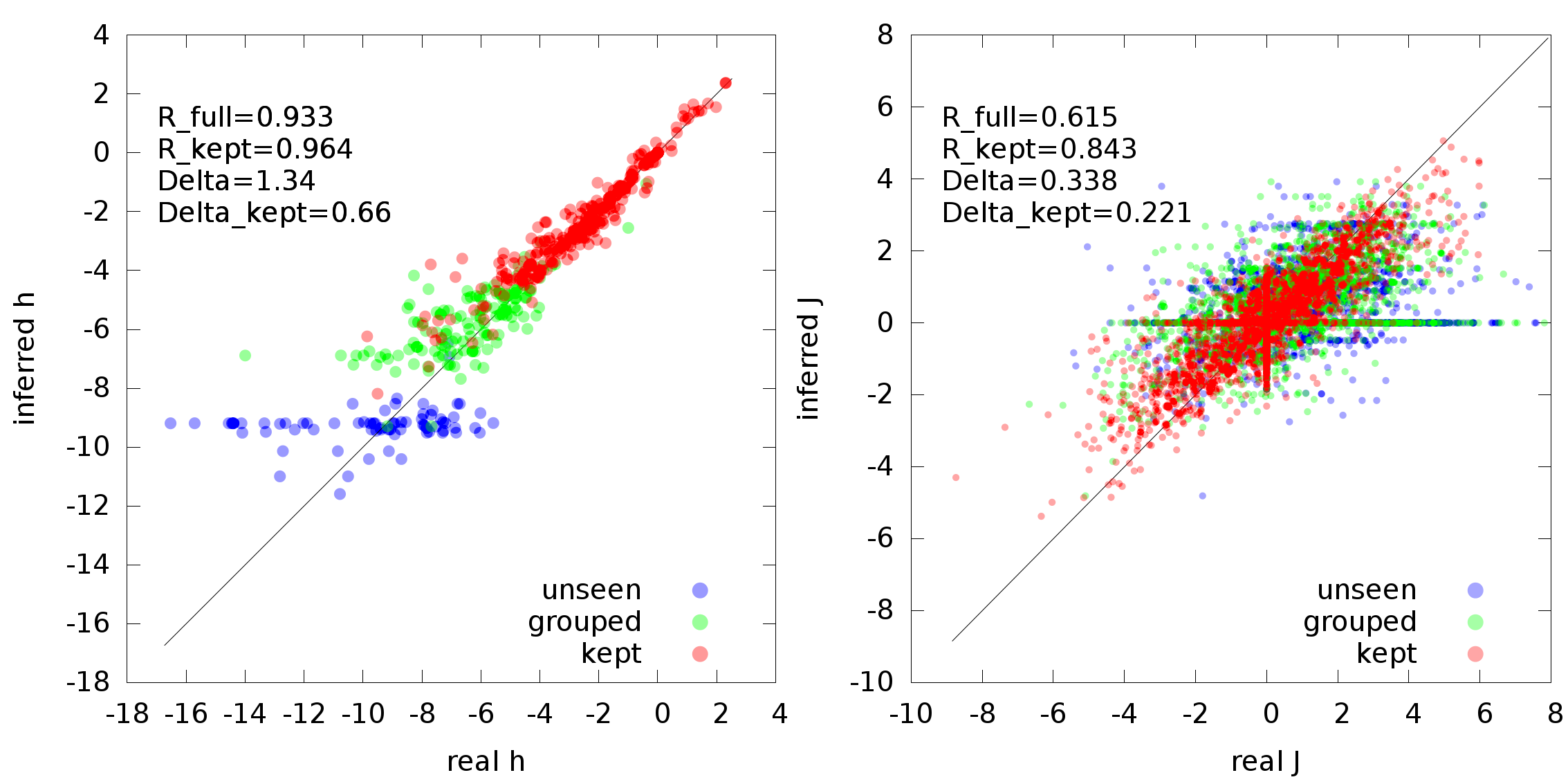}}
	\subfloat[PLM: $f_0=0.01=10/B$]
	{\includegraphics[width=0.4\linewidth]{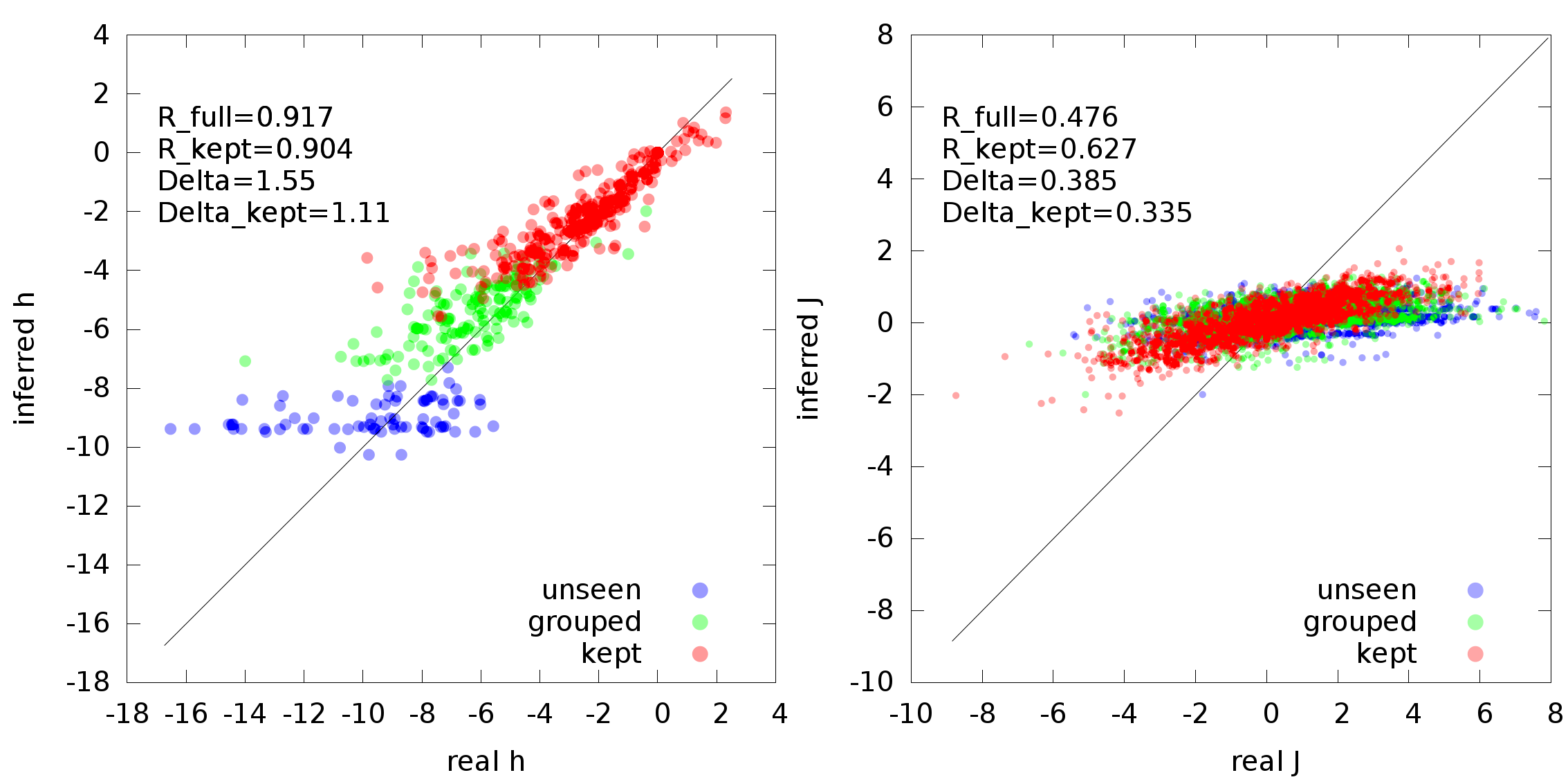}}
\caption{Comparison of inferred and real fields and couplings with ACE  and PLM for one realization of ER graph with no color compression (top) and $f_0=0.01$ (bottom), for B=1000 sampled configurations. Parameters on explicitly modeled (kept), grouped, and unseen Potts states are colored differently. Left: field comparison. Right: coupling comparison. On top of each plot, the Pearson correlation coefficient ($R$) and the absolute error ($\Delta$, as in Eq.~\ref{eq:deltaJ}) are indicated. }
\label{fig:Jh_ACEandPLM}
\end{figure}

\subsection{Couplings and Fields\label{sec:HJ}}

In Fig.~\ref{fig:Jh_ACEandPLM} we compare the fields and the couplings of the real model (x-axis) and the inferred Potts model (y-axis) obtained for the reference data of the graph ER05 sampled at $B=1000$ without color compression (top panels) and with $f_0=0.01=10/B$ (bottom panels), respectively, with ACE and PLM. Different colors in Fig.~\ref{fig:Jh_ACEandPLM} 
show Potts states (or Potts states pairs) occurring at different frequencies and therefore treated in the color compression procedure as explicitly modeled, grouped, or unseen in the configuration sample. For couplings, if at least one of the two Potts state is unseen, the pair is considered as unseen; if at least one site is grouped, the pair is considered as grouped;  if both sites are explicitly modeled, the pair is considered as explicitly modeled. The comparison is performed in the consensus gauge. 
\color{black}

Figure~\ref{fig:Jh_ACEandPLM} shows that, as observed in Section ~\ref{sec:contacts} the sparse procedure, spACE,
misses some edges and the corresponding couplings are fixed to zero. PLM couplings are systematically smaller in amplitude than real ones, ending up in a tilted entry-by-entry comparison. This is due to the large regularization introduced to avoid overfitting. 

\begin{figure}[h!]
	\subfloat
	{\includegraphics[width=0.4\linewidth]{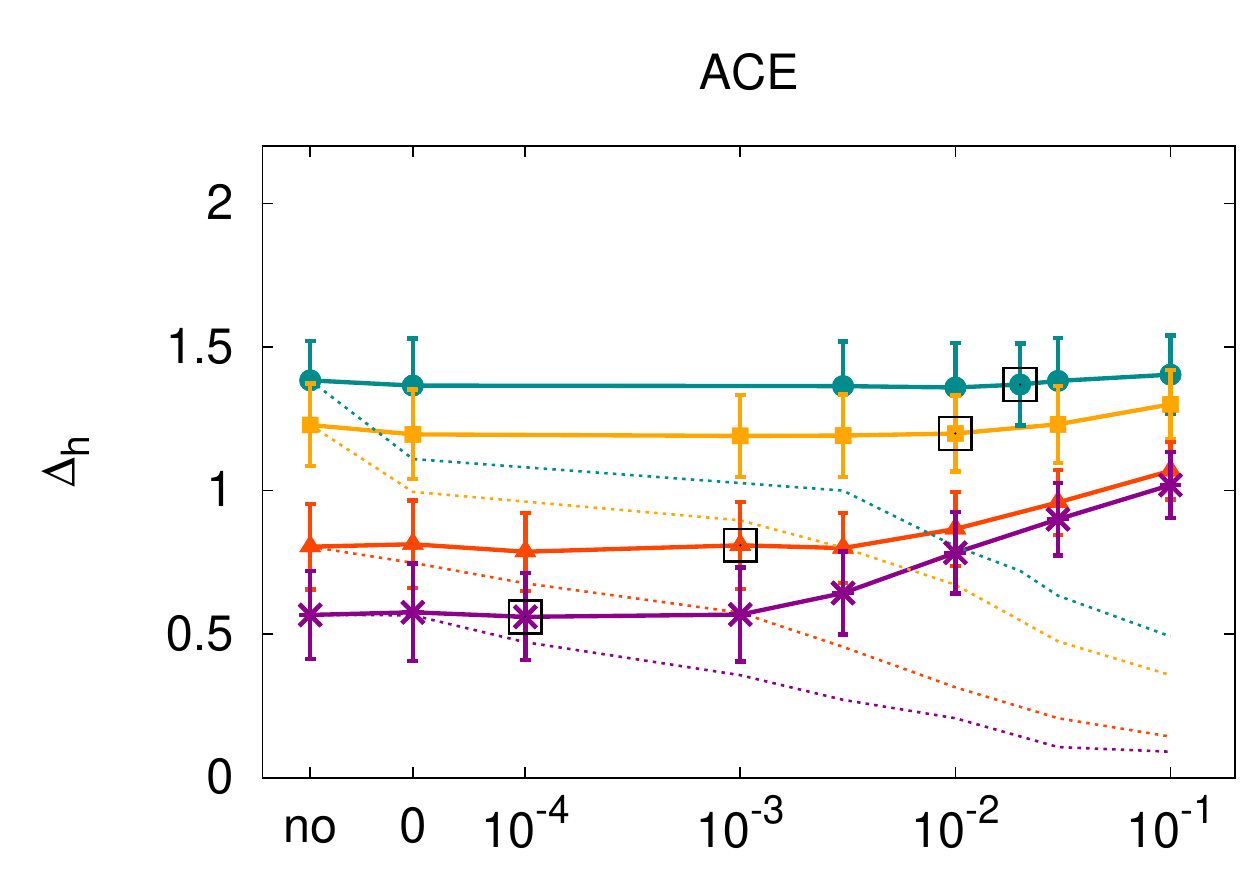}}
	\subfloat
	{\includegraphics[width=0.4\linewidth]{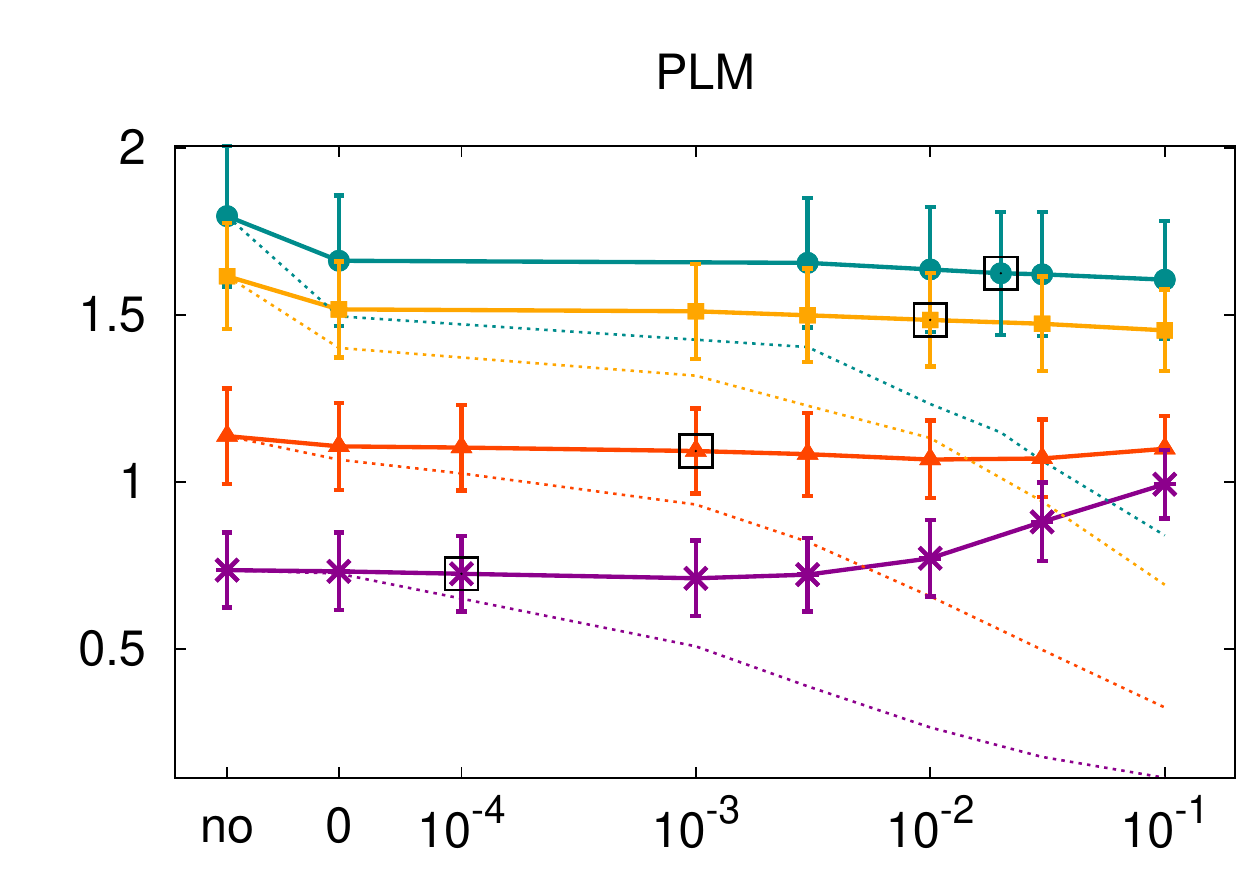}}\\
	\subfloat
	{\includegraphics[width=0.4\linewidth]{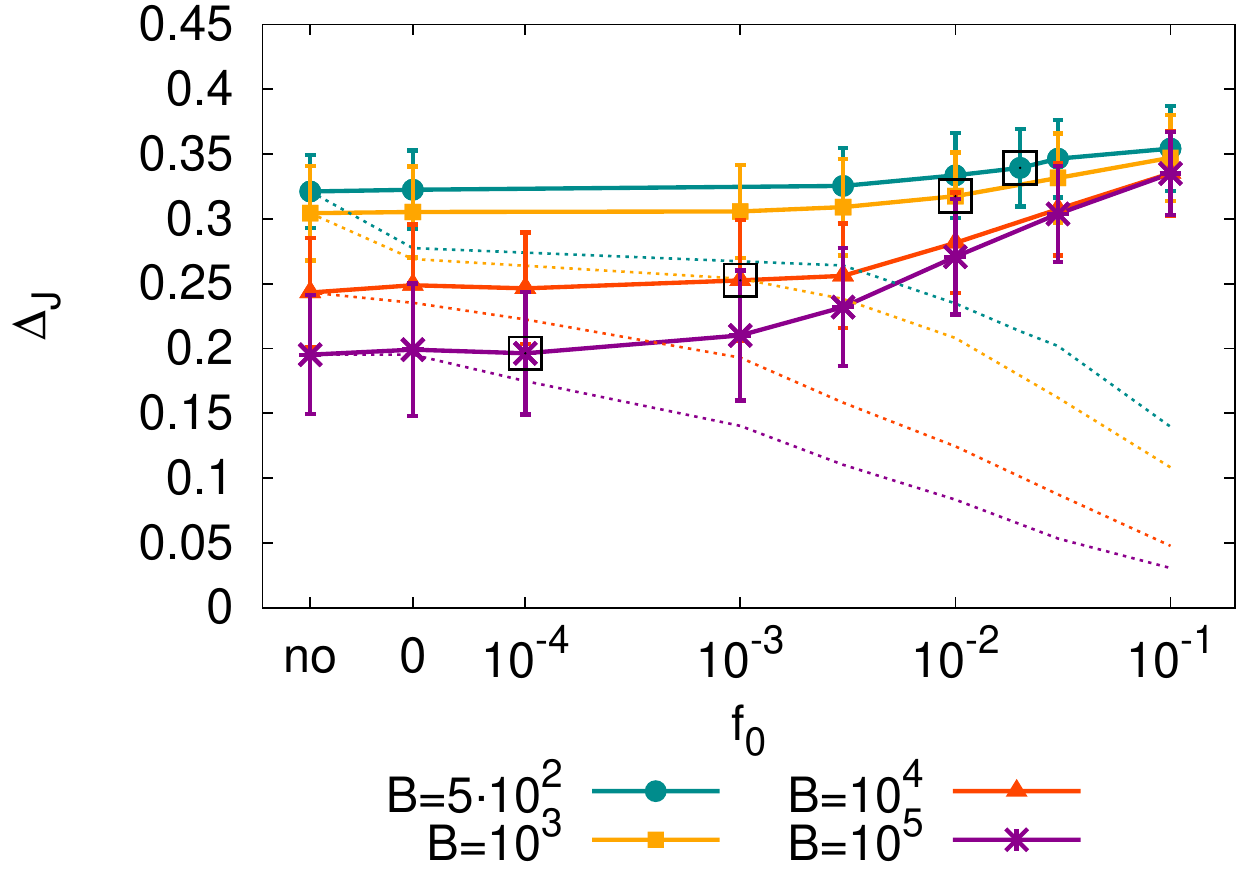}}
	\subfloat
	{\includegraphics[width=0.4\linewidth]{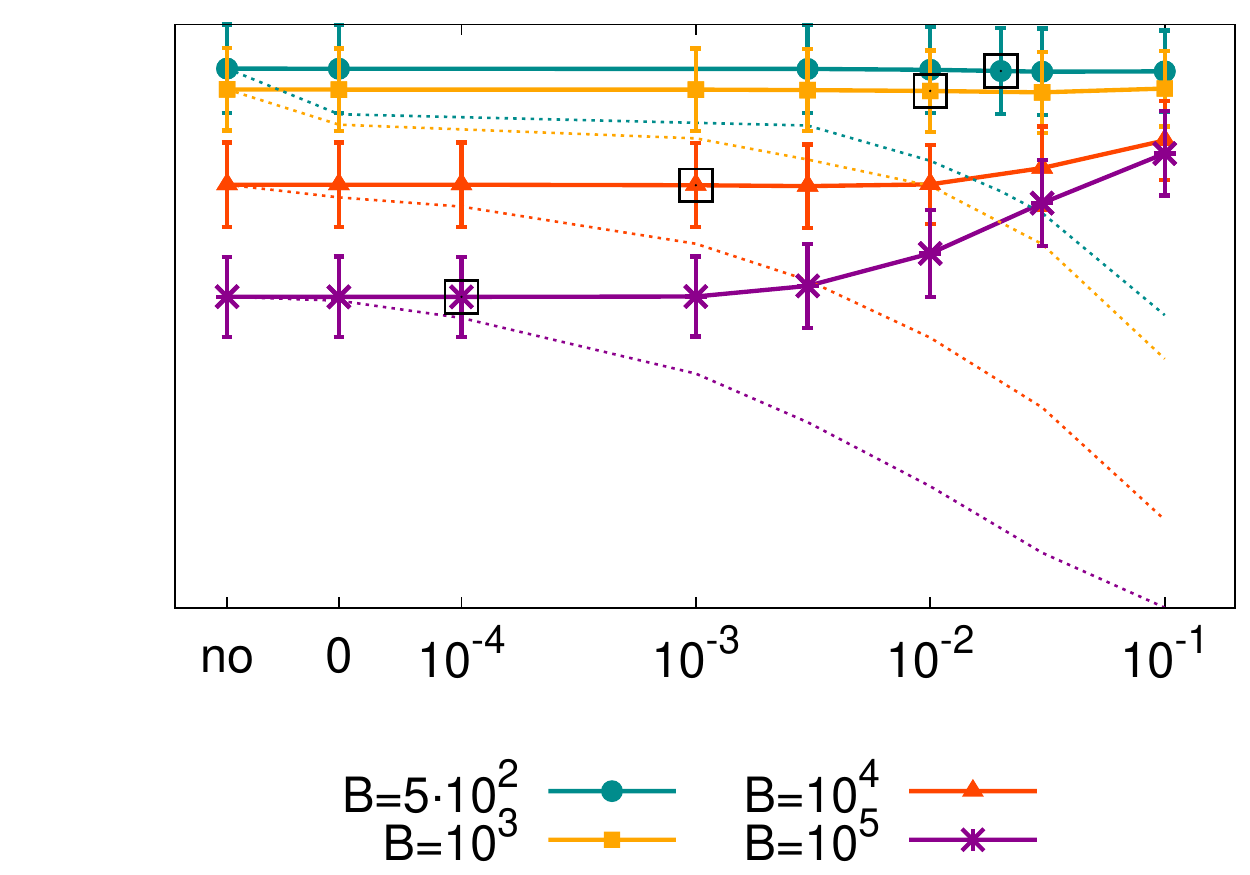}}
	\caption{Absolute errors (Eq.~\ref{eq:deltaJ}) on fields ($\Delta h$, top panels) and couplings ($\Delta J$, bottom panels) averaged over 10 ER realizations, as a function of the color compression for several sample sizes.
	Dashed lines: error on parameters related to explicitly modeled Potts states. Full lines: error on all parameters, after decompression of unseen and grouped Potts states (see Sec.~\ref{sec:compression}).
	Error-bars are standard deviations computed over the 10 realizations. Inference is performed respectively by ACE (left) and PLM (right).
	}
	\label{fig:delta_ace_plm}
\end{figure}

 Figure~\ref{fig:delta_ace_plm} shows the absolute errors on the fields and couplings defined through
\begin{eqnarray}
\Delta h = \sqrt{ \frac{\sum_{i}\sum_{a} \left(h^{inf}_i(a) - h^{real}_i(a)\right)^2 }{\sum_{i} q_i} } \ , \quad 
\Delta J = \sqrt{ \frac{\sum_{ij} \sum_{ab} (J_{ij}^{real}(a,b)-J_{ij}^{inf}(a,b))^2}{\sum_{ij} q_i \cdot q_j }}\ ;
\label{eq:deltaJ}
\end{eqnarray}
These errors measure the average distances from the diagonal of the points in the scatter plots of Fig.~\ref{fig:Jh_ACEandPLM}.
We observe that the couplings and fields reconstruction performances are stable as a function of the color compression up to the reference value $f^{*}_0\simeq {10\over B}$, where they drop because the compression become too strong. Such drop is perfectly clear for the couplings parameters, especially  within the small regularization used in spACE. Large compression threshold,  as well as  large coupling regularizations,  degrades indeed the informations on the correlations of grouped, badly sampled, variables.
The increase of  field reconstruction error  at all compressions is negligible for small sampling depth $B=500,B=1000$  and for both algorithm, showing that  the decompression 
procedure introduced  in Sec.~\ref{sec:compression},  correctly assign, in the limit of the available information, the large and negative fields for the grouped and unseen Potts symbols,  as done by using prior information with the $L_2$ regularization  and shown in Fig.~\ref{fig:Jh_ACEandPLM}.
The dashed lines in Fig.~\ref{fig:delta_ace_plm}, in agreement with  Fig.~\ref{fig:Jh_ACEandPLM},   show that by restricting the coupling comparison to  the explicitly modeled symbols,  the better and better sampled ones at larger and larger $f_0$, the reconstruction indicators are better and better.  In other words, even in the largely undersampled regime, parameters for well sampled colors are correctly inferred and are not affected by the poorly sampled states. This underlines the difference between sites and states in a Potts model. In the standard renormalization procedure \cite{cardy1996scaling} when the number of sites are reduced in an effective ``renormalized'' Potts model the parameter values of the retained sites change. In contrast, in the space of Potts states, grouping some of them, and keeping the probabilities conserved, does not affect the others \footnote{Similar results are obtained when considering the Pearson correlations between real and inferred parameters, rather than their absolute differences. ACE gives better results than PLM for parameter reconstruction, especially on couplings (not shown). This is due to the fact that spACE avoid overfitting of data and setting many non-zero couplings for non interacting sites in the real interaction graph.}.

\section{Application to sequence-based models of protein families }

\label{sec:real}

We now apply our inference approach to protein sequence data. Input samples are multiple sequence alignments of protein families, the nodes of the graph are the protein sites, and states are the 20 amino acids plus the insertion-deletion symbol ($q=21$). In this context, we aim at reconstructing the contact map \cite{morcos2011direct, hopf2012three} or the fitness landscape \cite{ferguson2013translating, mann2014fitness, figliuzzi2016coevolutionary, hopf2017mutation}. In particular, we would like to compare the change of energy corresponding to single point mutations with respect to a wild-type protein sequence to the experimentally measured changes of fitness of the protein. 
Differently from ER data, sequences are not uniformly sampled. A reweighting procedure, described in Appendix~\ref{app:w},  is usually introduced  to  reduce the initial number $B$
of sequences in the sample to an effective number of independent ones $B_{eff}.$

\begin{figure}[h!]
\subfloat[ACE]
	{\includegraphics[width=0.45\linewidth]{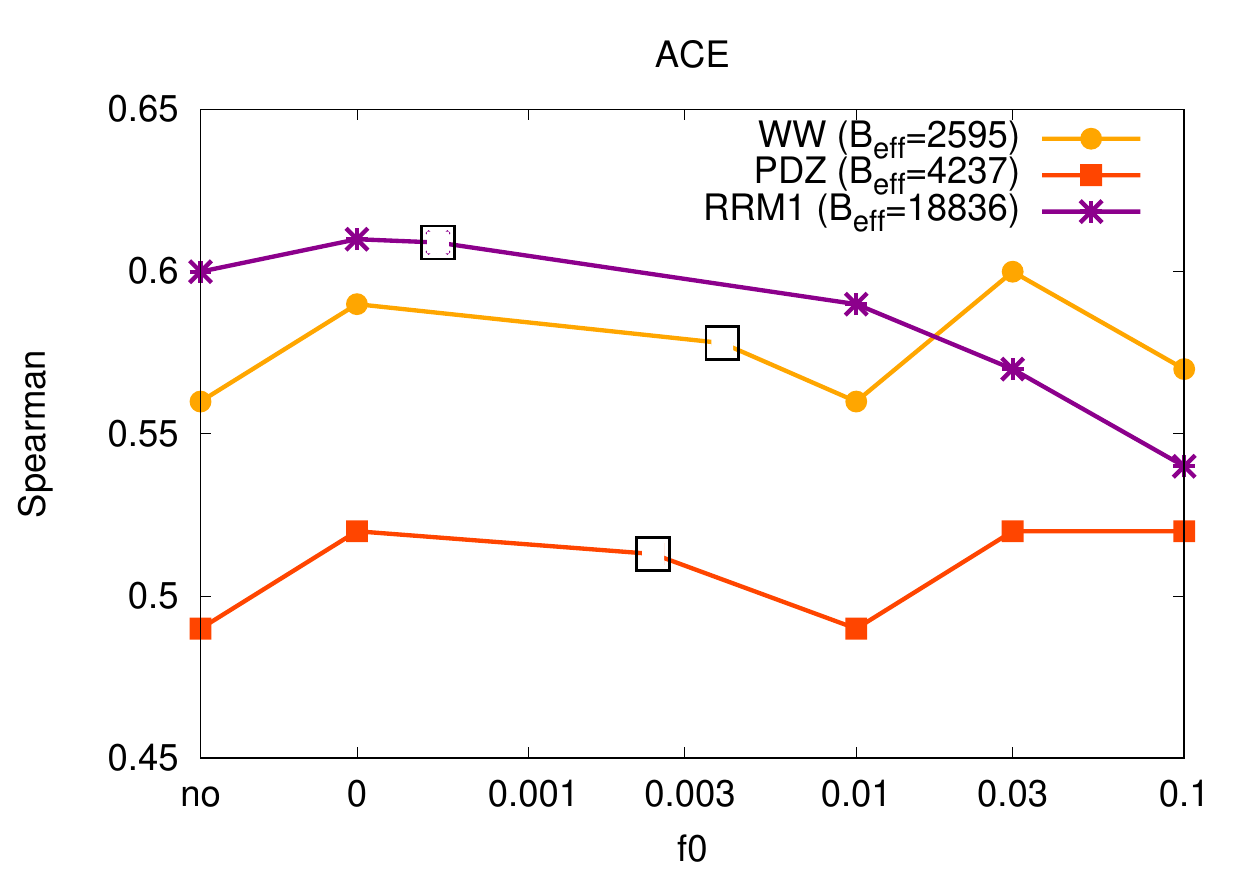}}
	\subfloat[PLM]
	{\includegraphics[width=0.45\linewidth]{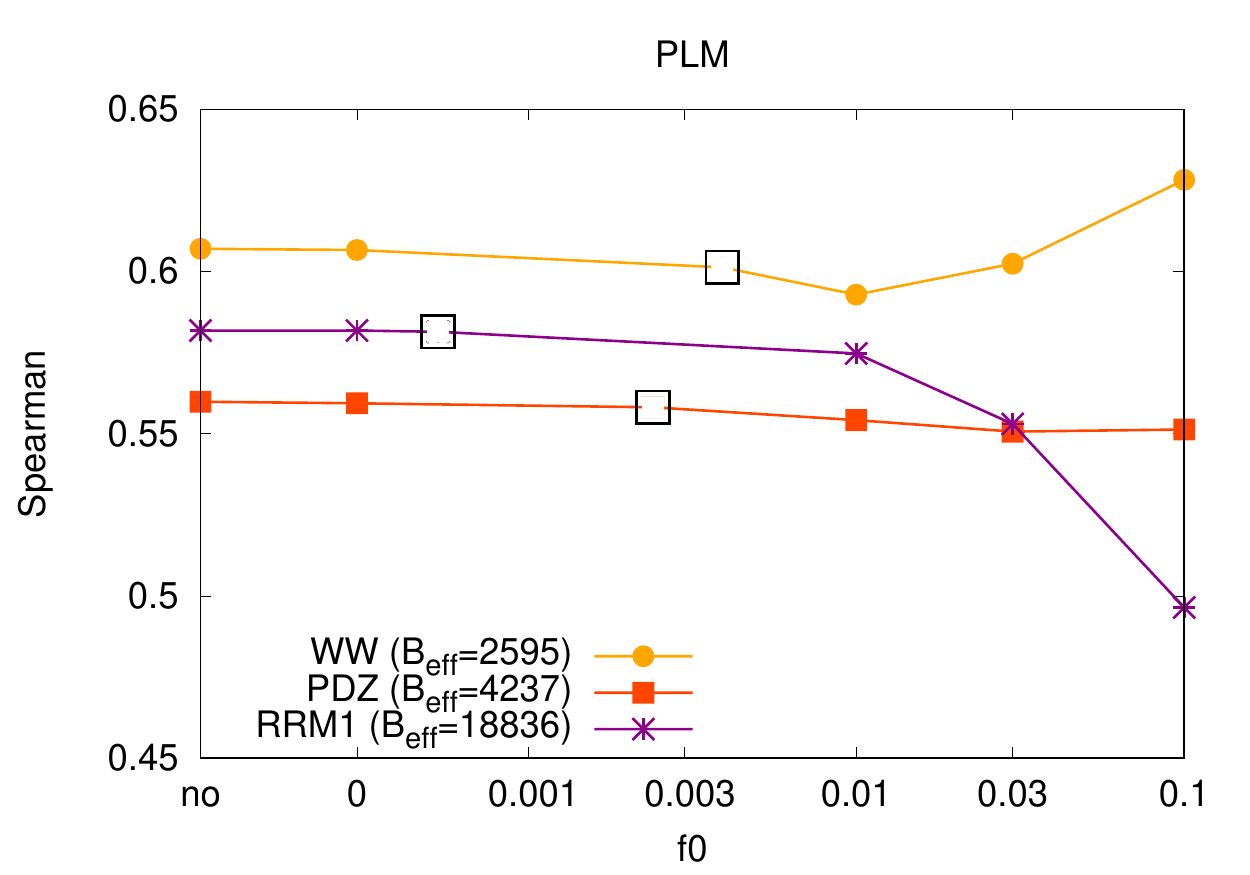}}
		\caption{Spearman correlation coefficient between the fitness predictions and the experimental measures found in literature, as a function of the color compression  for ACE  and PLM.  The protein families used here are: WW (PF00397, in orange), PDZ (PF00595, in green) and RRM1 (PF00076 in light blue)}
	\label{fig:realprot}
\end{figure}

We here consider three protein families, whose fitnesses have been systematically assessed against single-point mutations. We specify for each of them the number of sites in the alignment $N$,
the number of sequences $B$ and the number of effective sequences $B_{eff}$:
\begin{itemize}
\item WW  ( $N=31$,  $B=8251$, $B_{eff}=2590$) is a  protein domain that mediates specific interactions with protein ligands.  Here fitness has been measured in terms of the capability to bind a certain ligand \cite{araya2012fundamental}.
\item PDZ  ($N=84$,  $B=24795$, $B_{eff}=4240$) is a protein domain present in signaling proteins. Here fitness has been measured in terms of binding affinity \cite{mclaughlin2012spatial};
\item RRM  ($N=82$,  $B=70780$, $B_{eff}=18800$)is an RNA recognition motif; fitness was estimated through growth  rate measurements in \cite{melamed2013deep}.
\end{itemize}
Alignments and experimental fitness measures used in this section were taken from \cite{hopf2017mutation}. The PLM procedure was applied with the same  large regularization used for the artificial data,  $\gamma_J=N/B$, $\gamma_h=0.1/B$. The SpACE procedure was applied with a threshold cutoff $N_2^{max}=3N$, and  a smaller regularization $\gamma_J=10/B$, $\gamma_h=0.1/B$ ; however the relative error $\epsilon_{max}$ (Eq.\ref{eq:emax}) was generally too large even in its local minima, indicating that the procedure had not converged. 
 As shown in \cite{barton2016adaptive}  a Boltzmann Machine Learning (BML) procedure was further used, starting from the spACE inferred parameters as an initial guess, to better reproduce the low order statistics of the data and therefore the quality of the inferred model.
Mutational cost  of single mutations is predicted  by their energetic cost, corresponding to the value of the field of the mutated amino acid, after having transformed the field parameters in the gauge of the wild type sequence through Eq.~\ref{eq:consensus_transform} \cite{cocco2018inverse}.   
 
Contrary to what happens for synthetic data, where the true model is known, the relationship between inferred energies and experimental fitness values may be nonlinear, so we use as a quality measure of the inference the Spearman correlation coefficient between them rather than the Pearson.
 The Spearman values of Fig.~\ref{fig:realprot}  for the full model (no compression)  and PLM are in agreement to the ones previously obtained  \cite{hopf2017mutation} with PLM algorithm and a slightly different procedure, giving Spearman values of 0.6, 0.57, 0.5 for RRM1, WW and PDZ respectively.  ACE + BML performances are compatible with PLM  results, better for families with a large number of sequences such as RRM, and slightly  worse in the PDZ case. 
  The impact of color compression on the prediction is then investigated and shown in Fig.~\ref{fig:realprot}  and has to be compared to the quality of  field reconstruction  in Fig.~\ref{fig:delta_ace_plm}, after decompression (full lines).
 For the WW and PDZ families, as for  PLM with the ER data with  $B=1000$ and  $B=500$, the predictions are not affected by large compression thresholds.
Very large compressions may even  improve performances,  as for the  to WW family,  where inference with almost binary Potts variables (at $f_0=0.1$  $<q_i>=2.5$ see 
Fig.~\ref{fig:Nparam}) give better predictions  than with a full $q=21$-state Potts model.
 A similar effect was observed also for HIV fitness predictions in \cite{mann2014fitness}.
The fact that  large color-compressions can improve fitness predictions may be related to the   experimental uncertainties and  limited resolution on the fitness measures and also  to the difficulty in estimating  the effective number of independent sequences in the data. 
  For the better sampled  $RRM$  family, predictions worsen at large compression  $f_0> 10/B_{eff}$, as expected and observed in Fig.~\ref{fig:delta_ace_plm} for the well sampled  artificial data (at  $B=10^{4}$ and $B=10^{5}$). 

\section{Gain in computational time for synthetic  and protein sequence data}
\label{sec:times}

\begin{figure}[h!]
	\includegraphics[width=0.2\linewidth]{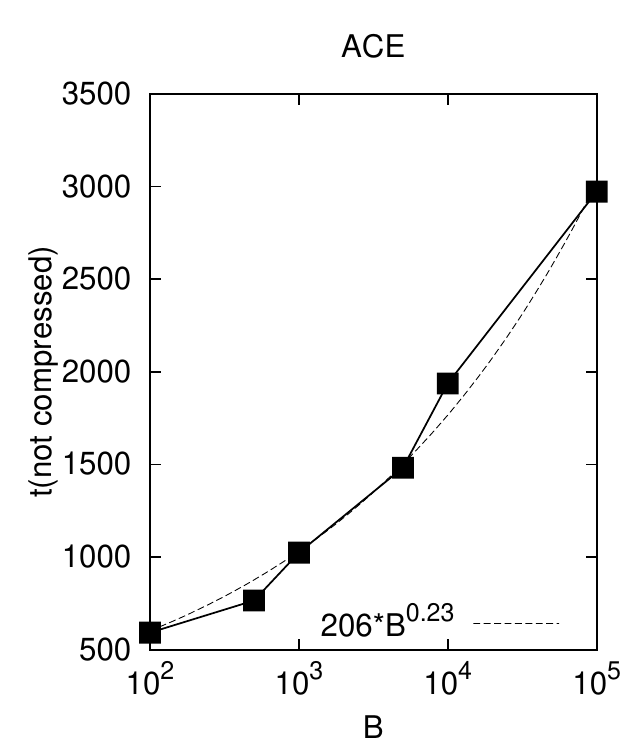} 	
	\includegraphics[width=0.2\linewidth]{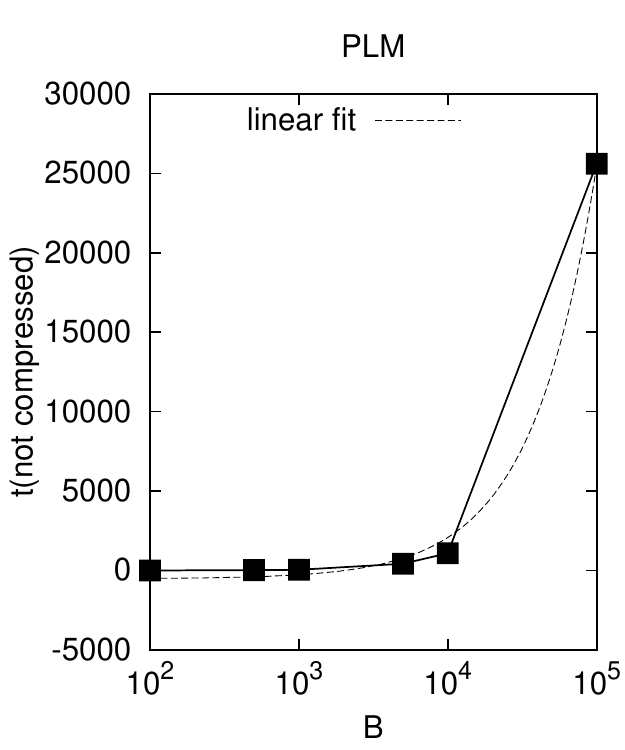}
	\includegraphics[width=0.2\linewidth]{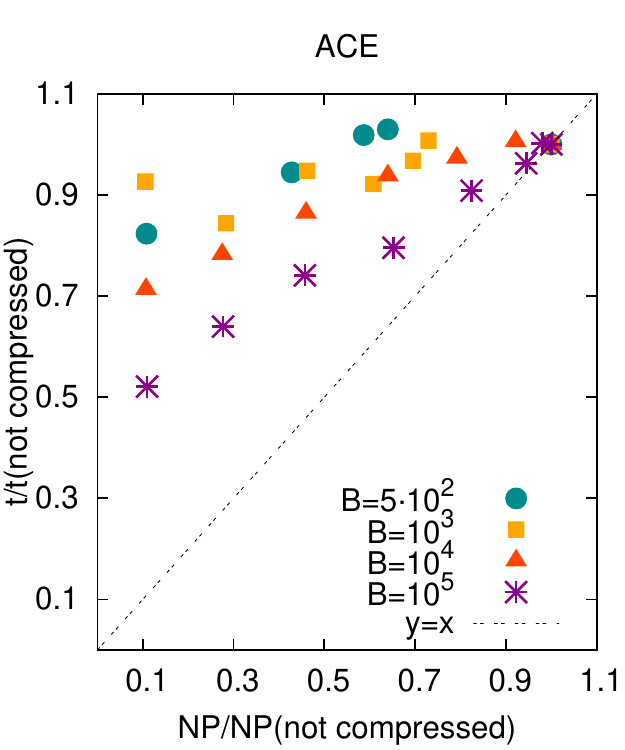}
	\includegraphics[width=0.2\linewidth]{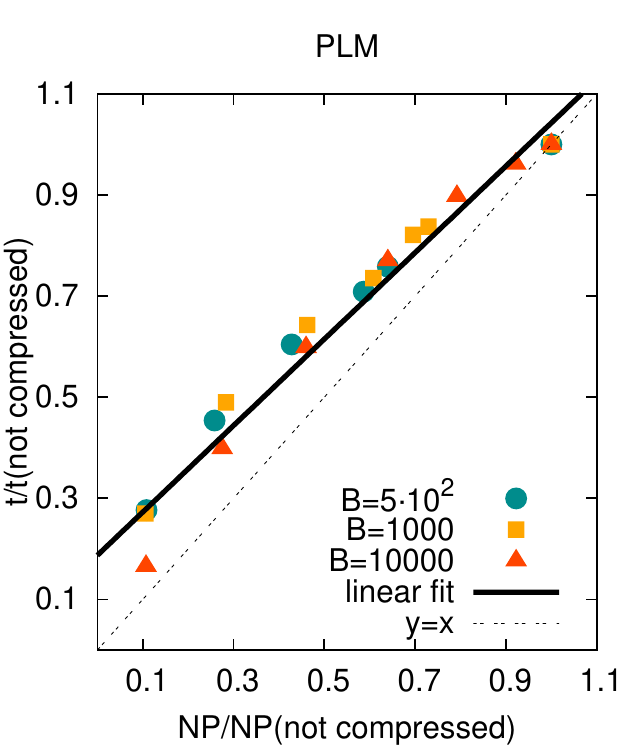}
	\caption{Computational times  averaged on all ER data for ACE and PLM for different B and color compression. Left: computational time, in seconds, using only 1 CPU, for ACE  and PLM  as a function  $B$; a power law/ linear fits are added  to ACE /PLM (dashed line). Right: computational time ratio between the compressed and decompressed inference for ACE (1 CPU) and PLM (25 parallel CPU) as a function of fraction of parameters to infer;  dotted line: x=y; full line: linear PLM fit. }
	\label{fig:ERtimes}
\end{figure}

\begin{figure}[h!]	
	{\includegraphics[width=0.4\linewidth]{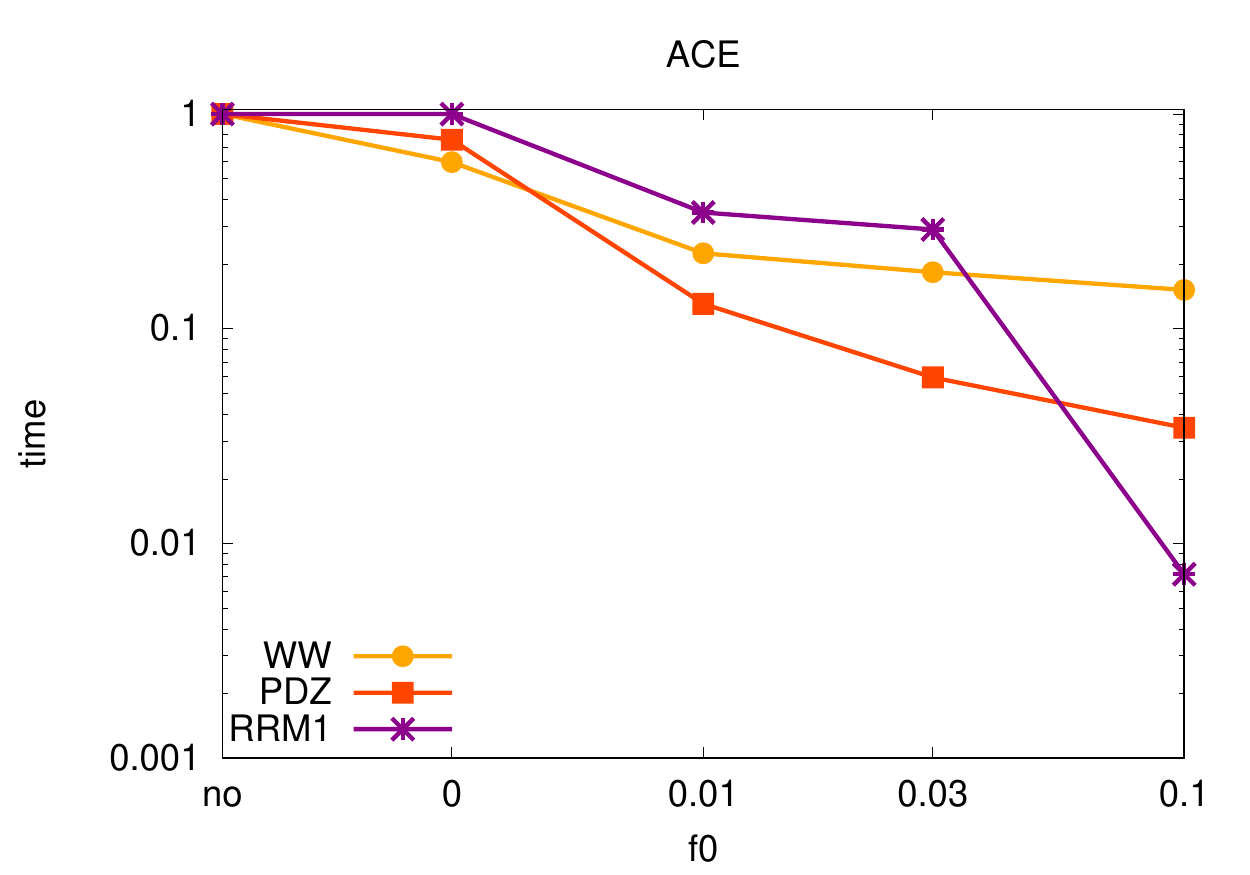}}
	{\includegraphics[width=0.4\linewidth]{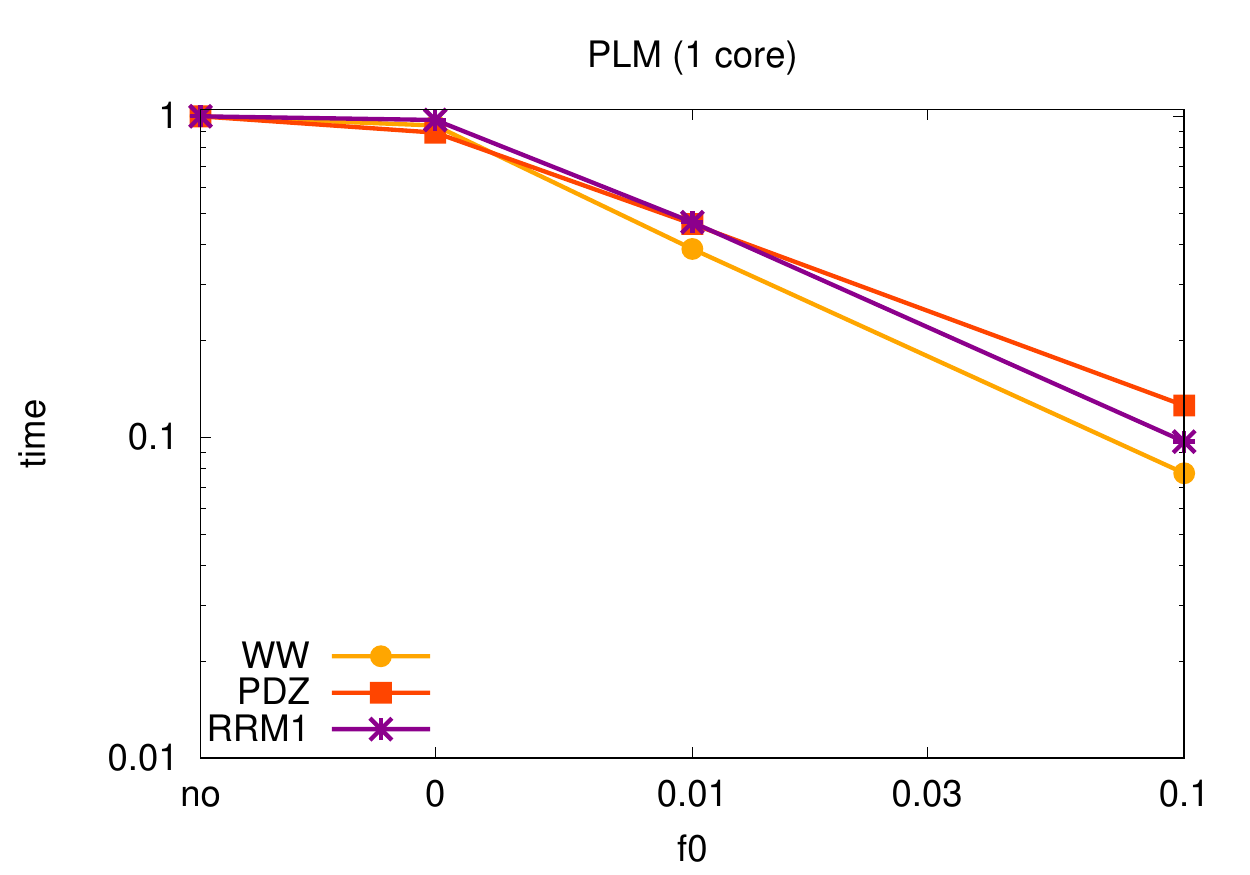}}
\caption{ Time gain due to color compression on protein sequences data with ACE (Left)  and PLM (Right). The protein families used here are: WW (PF00397), PDZ (PF00595) and RRM1 (PF00076).  }
	\label{fig:RPtimes}
\end{figure}
 \begin{table}  [h!]
$B=10^3$ \hspace{4cm} $B=10^5$ \\
\begin{tabular}{| c|c | c | c | c| c|c|| c|c | c | c | c| c|c}
\hline
 $f0$ & $t_{ACE}$  & $\hat N_{cl}$&$k_{max}$& $<q_{i}>$& $t^{1}_{PLM}$ & $t^{20}_{PLM}$& $t_{ACE}$  & $\hat N_{cl}$&$k_{max}$& $<q_{i}>$& $t^{1}_{PLM}$ & $t^{20}_{PLM}$\\
 \hline
 no & $10^3$  & 2511 &4& 10 & 39 & 8&  $9\, 10^2$ & 1306 &4& 10 & $1.8\,10^4$ & $1.5 \,10^3$\\
 0 &  $10^3$& 2320 &4&8.6 & 32 & 8& $9\, 10^2$&1306 &4& 9.9& $1.8 \,10^4$ & $1.5 \,10^3$\\
 0.01 & $10^3$ &2065 &4&5.8&22 & 7&   $9\, 10^2$ &1305 &4&5.8& $ 10^4$  & $10^3$\\
 0.1 & $8\,10^2$ & 1350&4&2.2 &11 & 7&   $6\, 10^2$ &1314&4 & 2.3& $2\,10^3$ & $3 \,10^2$ \\ 
 \hline
 \end{tabular}
  \caption{Running times, in seconds, for spACE ($N_2^{max}=4N$) and PLM on 1 and 20 cores, for the data from the reference graph ER05, at 2 different sampling depth ER 1000 $(N=50 B=1000)$, ER100000 (N=50,B=100000),
 for different color compression thresholds $f_0$. The number of clusters processed by the algorithm $\hat N_{cl}$ and the average number of colors explicitly modeled are also indicated  $\langle q_{kept} \rangle.$ }
 \label{tab:ERtimes}
 \end{table}
 \begin{table} [h!]
WW \hspace{4cm} PDZ  \hspace{4cm} RRM1\\
\hspace{-1.1cm}
\begin{tabular}{ | c|c | c | c | c| c|c||c | c | c | c| c|c||c | c | c | c| c|c|}
\hline
 $f0$ & $t_{ACE}$  & $\hat N_{cl}$&$k_{max}$& $<q_{i}>$& $t_{PLM}^{1}$ & $t^{20}_{PLM}$ & $t_{ACE}$  & $\hat N_{cl}$&$k_{max}$& $<q_{i}>$& $t_{PLM}^{1}$ & $t^{20}_{PLM}$ & $t_{ACE}$  & $\hat N_{cl}$&$k_{max}$& $<q_{i}>$& $t_{PLM}^{1}$ & $t^{20}_{PLM}$ \\
 \hline
  no & $1.5 \,10^3$ & 702&4& 21 &$2\,10^2$ & 25 &  $4 \,10^4$ &4355 & 5&21&  $10^4$ &  838 & $2\,10^5$ & 4127& 6&21&  $6 \,10^4$ &  $4 \,10^3$  \\
  0 & $10^3$ &702&4&18.6 & $2\,10^2$ & 24 &  $3 \,10^4$ & 4355& 5&19.8 &$10^4$   & 728 & $2\,10^5$   & 4127&6&20.9 &  $6 \,10^4$ &$4 \,10^3$  \\
   0.01 & $4\,10^2$ &702  &4&9.4&85 & 15 &   $5 \,10^3$& 4548&6&11.2 & $5\,10^3$  & 428 & $6\,10^4$ & 4992&7&12&  $3 \,10^4$ & $2 \,10^3$ \\
 0.1 & $2.5 \,10^2$  &714&4 &2.2 &17 & 8 &$10^3$ & 4146&5 &2.5& $1.5\,10^3$& 123 & $10^3$ & 3724& 5&2.5& $5 \,10^3$  &$5\,10^2$ \\ 
 \hline
 \end{tabular}
\caption{ Running times (in seconds) for ACE and PLM for the three studied protein families: WW (N=31 B=8251), PDZ (N=84,B=24795), RRM1 (N=82,B=70780)   for different color compression thresholds $f_0$.   spACE has been stopped at $N_2^{max}=93,252,246$ (WW, PDZ, RRM1); the number of clusters processed by the algorithm $\hat N_{cl}$ and the average number of colors explicitly modeled are also indicated  $<q_{kept}>$. PLM times are given for both on 1 and   20 cores.}\label{tab:RPtimes}.
\end{table}

In this section we study how the computational time scales with the sample size and the color compression frequency threshold  for spACE and PLM on synthetic and protein sequence data. Times have been obtained on a processor Intel$\textsuperscript{\textregistered}$ Xeon(R) CPU E5-2690 v4 $@ $ 2.60GHz x 56. 
The  limiting factors for computational time with the ACE expansion are three: the number of colors $<q>$ in the inferred model, due to the computation of the partition function which grows as $<q>^K$ on clusters of size $K$, the overall number of clusters which enters in the  construction rule to 
  and the number of configurations sampled by Monte-Carlo to calculate  the relative errors in the reconstruction of the statistics of the data.\\
For  PLM the  average over the sampled configurations  for the pseudo-likelihood and moments calculation determines  a linear dependence on the sample size $B$, and on the number of parameters.   
Fig.~\ref{fig:ERtimes} and  Table~\ref{tab:ERtimes} show the computational times for  the reference ER graph realization as
a function of $f_0$ and B.  
On such data spACE  (Fig.~\ref{fig:ERtimes})   weakly depends on $K$ and $\langle q\rangle$,which take small values,  the limiting step  is therefore  the number of Monte Carlo  sampled configurations. Having a large number of MC steps (here fixed to 500000) is  important to correctly estimate the relative errors at very large sample size, given the small value of the sampling variances. The computational time show therefore a weak linear dependence on the compression,    as shown in 
Fig.~\ref{fig:timevsfo-r5b1000full} (Appendix).
 The  computational times for spACE and PLM  on real proteins
 are compared in Tables ~\ref{tab:RPtimes} and  in Fig.~\ref{fig:RPtimes}.
  As shown  in  Table~\ref{tab:RPtimes}  for real proteins  a   larger  numbers and  larger  sizes of clusters
  are summed on in the  ACE expansion,  as compared to the artificial data in Table~\ref{tab:ERtimes}.  The total number of processed clusters  ($\hat{N}_{cl}$) and their average  size $\langle q\rangle$  are clearly the limiting factor for the running  time,
  making ACE  slower than PLM. PLM further benefits from parallel computing, here results obtained on 1 and 20 are compared, while for spACE the computation has also been followed by time consuming MC-learning.
  Fig.~\ref{fig:RPtimes} shows the large reduction in computational time within  spACE  by color compression, 
  such compression makes often feasible the numerical computation, especially for non sparse ACE inference (see Fig.~\ref{fig:timevsfo-r5b1000full})  which  would take prohibitively long times without significant compression.

\section{Conclusion}
\label{sec:conclusion}
The present work reports an extensive numerical  benchmarking of inferred  color compressed Potts model with 2 algorithms (ACE and PLM ) on synthetic data,
as well as on  protein sequence data.
Knowing the ground-truth model that generated the data, we can  assess the inference performance at different compression strengths, by (1) computing the Kullback-Leibler divergence between the real and inferred models; (2) checking  the reconstruction of low-order statistics; (3) testing the reconstruction of the structure of the interaction network, and of the couplings and field parameters. \\
We will in the following resume and discuss  the comparison between PLM and ACE inference methods before discussing  the results on color compression.\\
A first important advantage shown by  ACE with respect to PLM  inference  is that
ACE can be easily stopped at large value of the cluster inclusion threshold to reconstruct a good sparse model for  the artificial data, 
which have been generated from sparse random graphs.
Imposing sparsity  regularization naturally reduces the  number of coupling parameters in the inferred model, leaving the possibility to  choose a small value for the $L_2$  regularization parameter on the left non-zero couplings.  Such double regularization scheme allows a very good reconstruction of the overall model, the model parameters and the statistics. 
On the contrary for  PLM , the inferred model is fully connected  and therefore a large $L_2$  regularization is needed  to avoid overfitting, 
in the optimal  model reconstruction, with the consequences that all the amplitudes of coupling parameters  are systematically underestimated, and the statistics of the data less well reproduced. 
Further theoretical investigations
are needed to obtain the optimal  regularization value found for the inference of a fully connected model, as a function of the sparsity of the original graph, the number of sites and color and the number of data and will be carried out on a forthcoming paper \cite{fantome2019} . It would be  also interesting  to introduce a  double regularization scheme for PLM algorithm.
The situation is different on protein sequence data because  the spACE procedure  with the simple stopping  criterium implemented here
does not converge to small statistical reconstruction errors, without additional MC learning algorithm. The  resulting inferred network is no more sparse 
 and no gain of performances of ACE with respect to PLM are achieved at larger computational costs. 
spACE procedure can be improved to better impose sparsity constraint in a more general case \cite{franz2019fast,cocco2019}.
 A question which  deserves further investigation is if the inference of sparse connectivity graph is  more appropriate for parameter reconstruction in the large undersampling regime,  even for data generated with models, such as protein sequence data, which are not necessary sparse.  Preliminary results for fitness prediction on  protein sequence data seems to indicate that sparse models gives  very good performances \cite{cocco2019}.
Next we will resume conclusion and discussion on color compression. \\
The central finding of the present work is that color compression does not degrade  the studied performances in a very large range of  frequency compression cut off, while  largely reducing  the dimensionality of the inferred model and its computational time.  The reduction of computational time obtained thanks to color compression  often become essential for solving the inverse problem in reasonable times.
 For example, ACE inference of models for many HIV proteins   \cite{barton2015scaling, barton2016relative, barton2016entropy, louie2018fitness} has been possible thanks to color compression. The color compressed version of the PLM algorithm introduced  adapted here from the routine of the group of Aurell and collaborators \cite{ekeberg2013improved, ekeberg2014fast}  can be crucially important  when dealing with large protein sequences  or with whole genome inference \cite{gao2018correlation, pensar2019high} with a much larger number of variables than single protein sequence data.
The color compression and decompression procedures introduced here are not restricted to pairwise graphical models. They could be used in other machine learning approaches, on protein sequence data, such as restricted Botzmann machines \cite{tubiana2019learning},  or variational auto-encoders \cite{riesselman2018deep} .

\vskip .3cm \noindent {\bf Acknowledgements.} We thank Lorenzo Posani for useful discussions. This work benefited from the financial support of  the   grants PSL ProTheoMicS and  RBMPro ANR-17-CE30-0021-01.

\section{Appendix}

\subsection{Reminder about  Adaptive Cluster Expansion and the inclusion threshold}
\label{ace:appendix}
In the ACE inference procedure the cross entropy is expanded as the sum of cluster contributions.
 Defining a cluster as a sub-set of variables:
 $\Gamma = \{i_1, \ldots, i_k\}, k\leq N$,
we can formally write the cross-entropy as the sum of cluster contributions:
\begin{equation} \label{eq:S}
S({\bf J}| {\bf f })= \sum_{\Gamma} \Delta S_{ \Gamma},
\end{equation}
where the sum is over all nonempty clusters of the $N$ variables. The cluster cross-entropy contributions $\Delta S_{\Gamma}$ are recursively defined through
\begin{equation} \label{eq:dS}
\Delta S_{\Gamma} = S_{\Gamma}-\sum_{\Gamma' \subset \Gamma} \Delta S_{\Gamma'}\,.
\end{equation}
Here $S_\Gamma$ denotes the minimum of the cross entropy (\ref{eq:crossentropy}) restricted only to the variables in $\Gamma$. Thus, $S_\Gamma$ depends only on the frequencies $p_i(a)$, $p_{ij}(a,b)$ with $i,j\in\Gamma$. Provided that the number of variables in $\Gamma$ is small (typically $\lesssim 10$ for $q=10$ Potts state as in the present work) numerical maximization of the likelihood restricted to $\Gamma$ is tractable. The definition of $\Delta S_\Gamma$ ensures that the sum over all clusters $\Gamma$ in 
(\ref{eq:S}) yields the cross entropy for the entire system of $N$ variables.  As detailed in \cite{cocco2012adaptive,barton2016adaptive}, a recursive construction rule is used to avoid, before selection, the computation of all cluster entropies. Such rule consists in building up clusters of size $k$
 by combining selected clusters selected of size $k-1$. The ACE expansion consists in truncating the expansion in Eq.~(\ref{eq:S}) by fixing a cluster inclusion threshold $t$ and summing up in Eq.(\ref{eq:dS}) only cluster contribution with 
 $|\Delta S_{\Gamma} |>t$. 
 
 \subsection{ ACE and SpACE pseudocodes}
 \label{ace:pseudocode:appendix}
  The  ACE algorithm, described in detail in \cite{barton2016adaptive}, is based on the selection and summation of individual cluster contributions to the Cross Entropy.
 It is built by the recursion of the following routine.  Given a list $L_k$ of clusters of size $k$, beginning with the list of all  the $N\;(N-1)/2$ possible clusters of size $k=2$ :
\begin{enumerate}
\item For each cluster $\Gamma\in L_k$
\begin{enumerate}
\item Compute $S_\Gamma$ by numerical minimization of  Eq.~(\ref{eq:crossentropy}) restricted to $\Gamma$.
\item Record the parameters minimizing Eq.~(\ref{eq:crossentropy}), called ${\bf J}_{\Gamma}$.
\item Compute $\Delta S_\Gamma$ using Eq.~(\ref{eq:dS}).
\end{enumerate}
\item Add all clusters $\Gamma\in L_k$ with $|\Delta S_\Gamma |>t$ to a new list $L^\prime_k(t)$.
\item Construct a list $L_{k+1}$ of clusters of size $k+1$ from overlapping clusters in $L^\prime_k(t)$.
\end{enumerate}
The rule  used by default for constructing new clusters of size $k+1$ from selected clusters of size $k$ is the so called strict rule:  a new cluster is  added only if all of its $k+1$ subclusters of size $k$ belong to $L^\prime_k(t)$ . The above process is then repeated until no new clusters can be constructed  (for ACE) or  also until the  maximal number of 2-site clusters $N_2^{max}$ is in the selected list $L^\prime_k(t)$ (for SpACE ). 
After the summation of clusters terminates, the approximate value of the parameters minimizing the cross-entropy, given the current value of the threshold, is computed by
\begin{equation} \label{eq:dJ}
{\bf J}(t) = \sum_k \sum_{\Gamma \in L^\prime_{k}(t)} \Delta{\bf J}_\Gamma, \quad \Delta {\bf J}_\Gamma = {\bf J}_\Gamma - \sum_{\Gamma^\prime \subset \Gamma} \Delta {\bf J}_{\Gamma'}\,.
\end{equation}
Then a Monte Carlo simulation is run to estimate the model one and two point correlations functions, which are compared to the empirical ones, taking into account their expected  statistical fluctuations, by  the computations of relative errors defined in Eq.~(\ref{eq:ep}, \ref{eq:ec}, \ref{eq:emax}) see also Fig.~\ref{fig:ace-crossentropyvst} in main text. When the maximal error is smaller than one the algorithm stops. Within the SpACE  approximation even if a
relative error  of 1 is not reached, the smaller relative errors among the ones obtained for logarithmically spaced  threshold intervals is chosen. 
%


 \begin{figure}
	\includegraphics[width=0.6\linewidth]{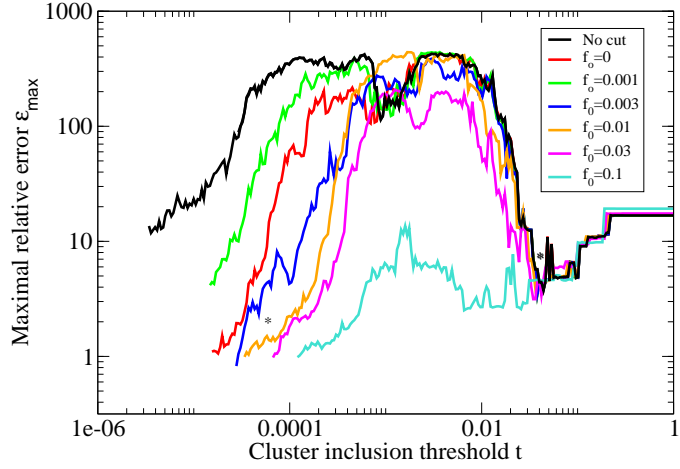}
	\caption{ Maximum relative error as a function of the expansion threshold for a particular graph realization (same used in Fig 5: ER05, sampled with B=1000),for different color compression$f_0$ . }
	\label{fig:ace-crossentropyvstvsfo}
\end{figure}

 \begin{figure}
	\includegraphics[width=0.45\linewidth]{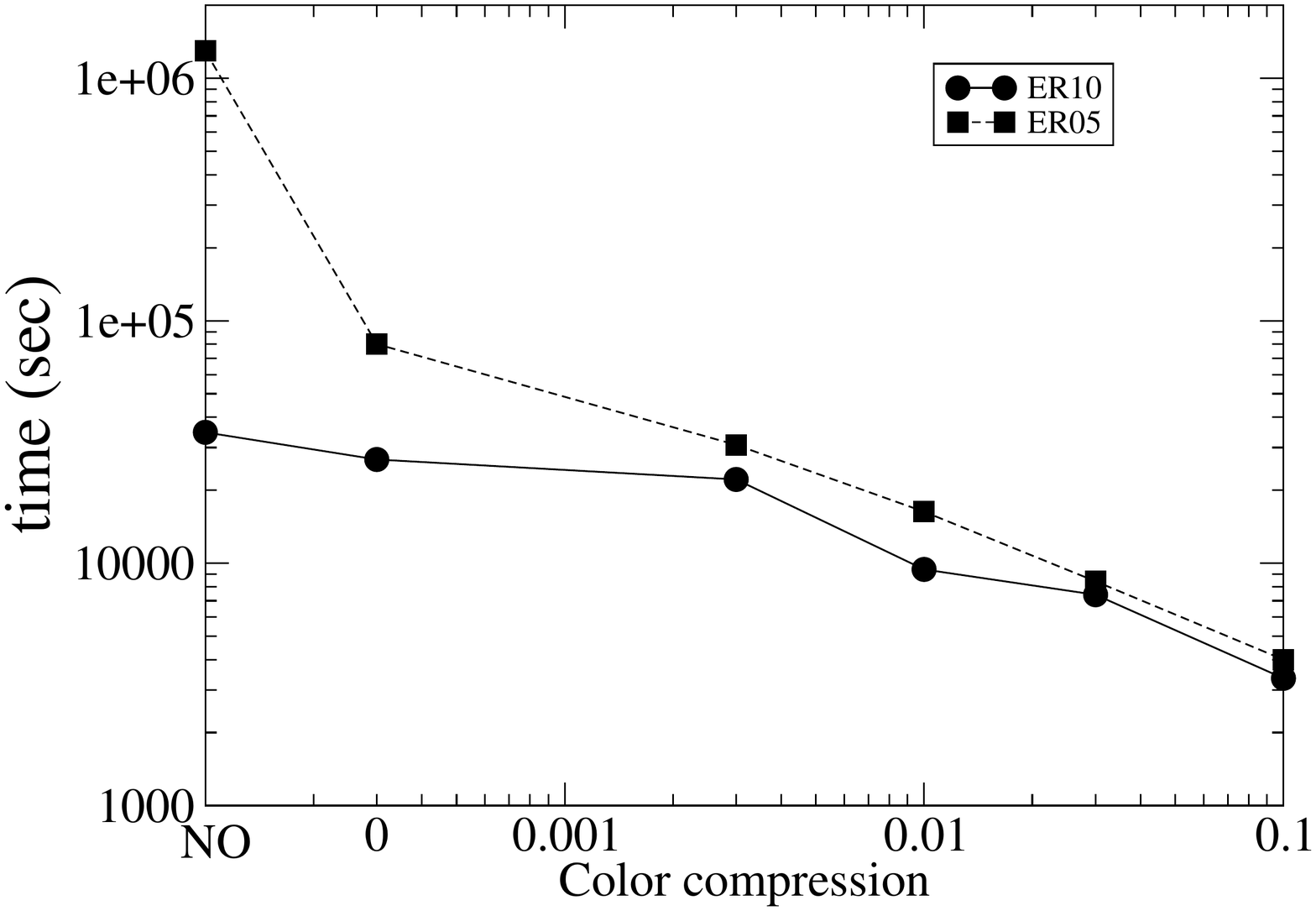}
	\includegraphics[width=0.45\linewidth]{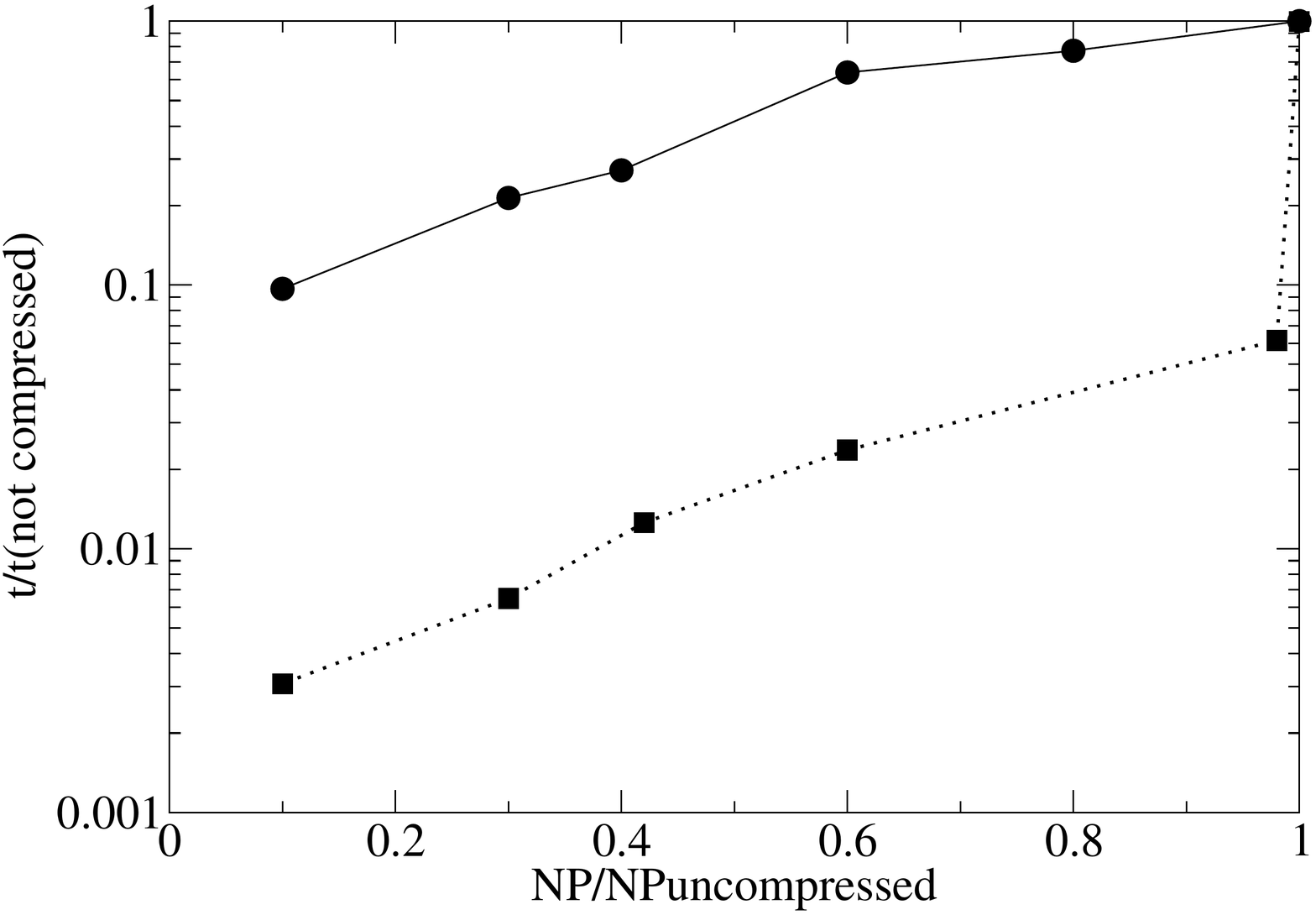}
	
	\caption{Reduction in computational time due to the color compression for fully connected ACE inference  on 2 data sets obtained by
	sampling B=1000 configurations from two Erd\H{o}s-R\'enyi random graph models.
	Left: Computational time at the  fully connected, low-threshold minimum, as a function of the color compression threshold $f_0$.
	Right: Computational time relative to the one with no color compression as a function of the number of parameters. }
	\label{fig:timevsfo-r5b1000full}
\end{figure}


\subsection{ Cluster expansion and computational time as a function of the color compression $f_0$ for fully connected graphs.} 
\label{acetime:appendix}
Fig.~\ref{fig:ace-crossentropyvstvsfo} shows that the behavior of the maximal relative reconstruction error $\epsilon_{max}$
 as a function of the cluster inclusion threshold $t$ when changing the color compression threshold $f_0$.
The presence of two relative minima corresponding to a sparse and a fully connected models fitting the data is observed for all the values of $f_0$, see the two stars in Fig.~\ref{fig:ace-crossentropyvstvsfo}. Moreover the threshold $t$ corresponding to the sparse inferred graph is largely independent of the level of color compression. 

Fig.~\ref{fig:ERtimes}  in the main text shows a mild computational gain as a function of the color compression when inferring a sparse interaction network (large-threshold minima). Such gain is
generally huge when the expansion converges only at low threshold values and sums up clusters of larger and larger sizes $K$. The numerical computation of the cross entropy 
requires indeed the sums over a number of $q^K$ configurations for $K$ Potts variables with $q$ states each. To illustrate this effect in Fig.~\ref{fig:timevsfo-r5b1000full} we show the reduction in computational time
when the cluster expansion is stopped at the small threshold value corresponding to a fully connected inferred graph. One can reach a 1000-fold computational time reduction with large color compressions. 
As shown in Fig.~\ref{fig:timevsfo-r5b1000full} the expansion was stopped to maximal relative error of order 10 at small threshold $t$ after 11 days while it took 50 minutes
to infer a good quality, fully connected model for the maximal color compression $f_0=0.1$. Note that the computational time to reach the sparse good model shown in Fig.~\ref{fig:ERtimes}
is smaller due to the reduced number of clusters. 
For large interconnected models color compression can therefore be essential to reach
convergence in a reasonable amount of time and infer a model that reproduces the statistics of the data. 

 \subsection{Kullback-Leibler Divergence from the ACE expansion}
 \label{appendix:KLACE}
 	The computation is done in the Ising case for the simplicity of the notations, the generalization to the Potts case being straightforward. We denote $\boldsymbol{J}^B=\{J_{ij}^B,h_i^B\}$ the inferred parameters at sample size $B$, and $\boldsymbol{J}^{true}=\{J_{ij}^{true},h_i^{true}\}$ the true underlying model parameters. The inferred cross-entropy at sampling $B$ writes
 	\begin{equation}
 	S_B = -\sum_{\boldsymbol\sigma} P_{\boldsymbol{J}^B}(\boldsymbol\sigma) \log P_{\boldsymbol{J}^B}(\boldsymbol\sigma) \ ,
 	\label{eq:sb}
 	\end{equation}
 	where the sum is over all possible configurations $\boldsymbol\sigma=\{\sigma_1,...,\sigma_N\}$. The inferred probability distribution at finite sampling $B$ is
 	\begin{equation}
 	P_{\boldsymbol{J}^B}(\boldsymbol \sigma)=\frac{\exp \left(\sum_{i=1}^N h_i^B\sigma_i+\sum_{k<l}^N J^B_{kl}\sigma_k \sigma_l \right)}{\mathcal{Z}_B} \ .
 	\label{eq:proba}
 	\end{equation}
 	The Kullback-Leibler (KL) divergence between the true and the inferred distributions writes

 	$$\begin{aligned}
 	D(P_{\boldsymbol{J}^{true}} || P_{\boldsymbol{J}^B}) &= \sum_{\boldsymbol\sigma} P_{\boldsymbol{J}^{true}}(\boldsymbol\sigma) \log \frac{P_{\boldsymbol{J}^{true}}(\boldsymbol\sigma)}{P_{\boldsymbol{J}^B}(\boldsymbol\sigma)} \\ 
 	& = -S_{true} - \sum_{\boldsymbol\sigma} P_{\boldsymbol{J}^{true}}(\boldsymbol\sigma)\left\{ \sum_i h_i^B\sigma_i + \sum_{k<l} J_{kl}^B\sigma_k \sigma_l -\log \mathcal{Z}^B \right \} \\
 	& = -S_{true} + \log \mathcal{Z}^B - \sum_{\boldsymbol\sigma} P_{\boldsymbol{J}^{true}}(\boldsymbol\sigma) \left\{ \sum_i h_i^B\sigma_i + \sum_{k<l} J_{kl}^B\sigma_k \sigma_l \right \} \ .
 	\end{aligned}$$
 	However, Eqs. (\ref{eq:sb}) \& (\ref{eq:proba}) give
 	$$\log \mathcal{Z}^B = S_B +\sum_{\boldsymbol\sigma} P_{\boldsymbol{J}^B} (\boldsymbol\sigma) \left\{ \sum_i h_i^B\sigma_i + \sum_{k<l} J_{kl}^B\sigma_k \sigma_l \right \}\ .$$
 	The KL divergence between the true and the inferred distributions then writes
 	$$\begin{aligned}
 	D(P_{\boldsymbol{J}^{true}} || P_{\boldsymbol{J}^B}) =& (S_B-S_{true}) - \sum_{\boldsymbol\sigma} P_{\boldsymbol{J}^{true}}(\boldsymbol\sigma) \left\{ \sum_i h_i^B\sigma_i + \sum_{k<l} J_{kl}^B\sigma_k \sigma_l \right \} \\
 	&+ \sum_{\boldsymbol\sigma} P_{\boldsymbol{J}^B} (\boldsymbol\sigma) \left\{ \sum_i h_i^B\sigma_i + \sum_{k<l} J_{kl}^B\sigma_k \sigma_l \right \} \ .
 	\end{aligned}$$
Moreover, a reasonable approximation is 
 	\begin{equation}
 	S_{true}=-\sum_{\boldsymbol\sigma} P_{\boldsymbol{J}^{true}} (\boldsymbol \sigma) \log P_{\boldsymbol{J}^{true}}(\boldsymbol\sigma) \approx S_{B \rightarrow \infty } = -\sum_{\boldsymbol \sigma} P_{\boldsymbol{J}^{B\rightarrow \infty}}(\boldsymbol \sigma) \log P_{\boldsymbol{J}^{B\rightarrow \infty}}(\boldsymbol \sigma) \ ,
 	\label{eq:compapprox}
 	\end{equation}
 	because the true underlying parameters are recovered by the inference method in the perfect sampling case: $P_{\boldsymbol{J}^{B\rightarrow \infty}}(\boldsymbol \sigma) \rightarrow P_{\boldsymbol{J}^{true}}(\boldsymbol \sigma)$. Therefore,
 	\begin{equation}
 	D(P_{\boldsymbol{J}^{true}} || P_{\boldsymbol{J}^B}) = (S_B-S_{\infty}) + \sum_i h_i^B (\left\langle \sigma_i\right\rangle^B - \left\langle \sigma_i\right\rangle^{\infty}) + \sum_{k<l} J_{kl}^B (\left\langle \sigma_k \sigma_l\right\rangle^B-\left\langle \sigma_k \sigma_l\right\rangle^{\infty}) \ ,
 	\end{equation}
 	where $\left\langle \cdot\right\rangle^B=\sum_{\boldsymbol\sigma} \cdot P_{\boldsymbol{J}^B}(\boldsymbol\sigma)$, and $\left\langle \cdot\right\rangle^{\infty}=\sum_{\sigma} \cdot P_{\boldsymbol{J}^{B\rightarrow \infty}} (\boldsymbol\sigma) \approx \sum_{\sigma} \cdot P_{\boldsymbol{J}^{true}} (\boldsymbol\sigma)$. 
It naturally generalizes to the $q$-states Potts case:
 	\begin{equation}
 	\begin{aligned}
 	D(P_{\boldsymbol{J}^{true}} || P_{\boldsymbol{J}^B}) =& (S_B-S_{\infty}) + \sum_{i=1}^N\sum_{a=1}^q h_i^B(a) (\left\langle \sigma_{ia}\right\rangle^B - \left\langle \sigma_{ia}\right\rangle^{\infty}) \\
 	&+ \sum_{\substack{k,l=1 \\ k<l}}^N \sum_{c,d=1}^q J_{kl}^B(c,d) (\left\langle \sigma_{kc} \sigma_{ld}\right\rangle^B-\left\langle \sigma_{kc} \sigma_{ld}\right\rangle^{\infty}) \ .
 	\end{aligned}
 	\label{eq:KLpotts}
 	\end{equation}
The artificial data are in a compressed representation (\textit{cf.} Section~\ref{sec:compression}). The complete inferred parameters are recovered as explained in Eq.~(\ref{eq:decompress}).

\subsection{Assignment of fields to zero-frequency states after inference}
\label{unseen:appendix}
In section \ref{sec:compression} we have discussed the decompression method used in the paper. In particular, we have seen that a pseudo-count is associated to the unseen states to assign them
a field with respect to the reference of the grouped/compressed state or the least probable state and, in principle, this is different to what implicitly done when the model is inferred without color compression.
To better understand the difference between the two approaches let us consider a simplified example of an independent-site model.
Without color compression, the fields are obtained as the minimum of:
\begin{equation}
S_{ind}=\log \sum_{a=1}^q \; e^{h_i(a)}-\sum_{a=1}^q h_i(a) \;p_i(a)+\gamma_h \cdot \sum_{a=1}^{q} h_i(a)^2 \label{eq:DeltaS}
\end{equation}
which, for the unseen colors in the gauge of $Z=\sum_{a=1}^q \;e^{h_i(a)}=1$, gives:
\begin{equation}
h_u^{\Gamma}=-Lw(\frac{1}{2\gamma_h})
\end{equation}
 where $Lw(y)$ is the Lambert function, solution of $xe^{x}=y.$
On the other hand, the field which we assign to these symbols during color decompression is, in the same gauge,
 \begin{equation}
h_u^p= \log(\frac{\alpha}{B})
\end{equation}
where we set, as in the rest of the paper, $\alpha=0.1$.
In this independent-site approximation there is then a shift between the two procedures given by $\Delta h= h_i(a)^{\Gamma}-h_i(a)^p $, that depends from the pseudocount $\alpha$ and from the value of the regularization $\gamma_h$.
In table \ref{tab:unseen} we give these shifts for the two values of $\gamma_h$ used respectively by ACE ($\gamma_h=0.01/B$) and PLM ($\gamma_h=0.1/B$).

\begin{figure} [h!]
	{\includegraphics[width=0.5\linewidth]{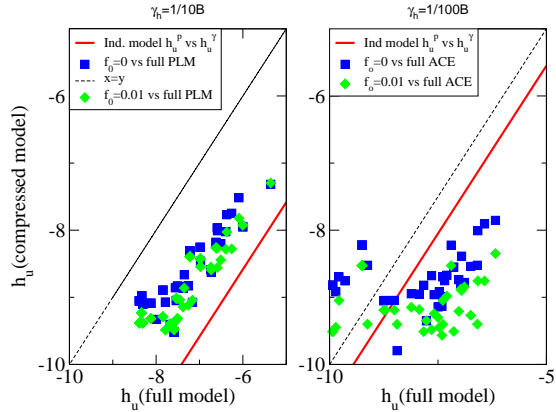} }
	\caption{Fields for the unseen Potts symbols for ER05 and $B=1000$ in the non compressed model and the reference model with $f_0=0.01$ of Fig.~\ref{fig:Jh_ACEandPLM}. Dotted line x=y. Red Line: the shift given in the independent model for $\gamma_h=1/10B$ as the one used with PLM. }
	\label{fig:comppseudogamma}
\end{figure}

If, in the approximation of independent-sites, the difference $\Delta h$ is the same in all gauges, the specific values of $h_i(a)^{\Gamma}$ and $h_i(a)^p$ of table \ref{tab:unseen} are specific of the $Z=1$ gauge.
To have a comparison in the consensus gauge as done in the rest of the paper one has to subtract $h_i(a)^{\Gamma}$ and $h_i(a)^p$ the field of the most common color $c_i$ on the considered site. In the independent-site approximation this is just $h_i(c_i)=log(p_i(c_i))$, and makes that unseen colors of different sites are found to have different fields.
In Fig.~\ref{fig:comppseudogamma} we plot the fields for the unseen symbols with a color compression $f_0=0.01$ (green diamonds) and $f_0=0$ (blue squares) versus the one for no color-compression for PLM and ACE (same fields as in Fig.~\ref{fig:Jh_ACEandPLM} of the main paper), in the consensus gauge. We can see a systematic shift towards lower values (at least for PLM, to be checked for ACE). We can compare this shift with the theoretical shift obtained with the independent model as described above. Even if we neglect the terms due to the couplings we can well reproduce such shift as the difference between the field obtained with the pseudo-count with respect to the one obtained with the regularization, there is a good agreement between what observed and the theoretical shift for independent variables. In particular the shift is smaller for the regularization chosen by the ACE procedure.

 \begin{table} [h!]
\begin{tabular}{ l | c | c |c}
$B$& $h_u^{\Gamma(ACE)}$& $h_u^p$&$ \Delta h_u$ \\ \hline
$10^2$ &-6.6 & -6.9& 0.3\\
$10^3$ & -8.6 &- 9.2&0.5\\
$10^4$ & -10.7& -11.5&0.8\\
$10^5$ & -12.9 & -13.8&0.9\\
\end{tabular}
\;\;\;\;
\begin{tabular}{ l | c | c |c}
$B$& $h_u^{\Gamma(PLM)}$& $h_u^p$&$ \Delta h_u$ \\ \hline
$10^2$ &-4.7 & -6.9& 1.5\\
$10^3$ & -6.6 & -9.2 &2.5\\
$10^4$ & -8.7 &-11.5 &2.8 \\
$10^5$ & -10.7 &-13.8 &3.1\\
\end{tabular}
\caption{Difference between the fields fixed by regularization and the one computed with the pseudo-count for the unseen Potts variables in the approximation of independent sites respectively for the fields regularization $\gamma_h=0.01/B$ used in ACE and $\gamma_h=0.1/B$ used in PLM } 
\label{tab:unseen}
\end{table}

\subsubsection{KL divergence for PLM at low $L_2$ regularization as a function of the sampling depth $B$ and of the color compression threshold $f_0$ }
\label{appendix:lowPLM}
\label{appendix:PLM:regularization}
In this section we study what happens with PLM at lower $L_2$ regularization, {\em e.g.} $\gamma_J=1/B$ as the one used for ACE, when the threshold for color compression is varied. Without color compression, the performance obtained for $\gamma_J=1/B$ becomes significantly worse, see Table~\ref{appendix:PLM:reg:table}. 

\begin{figure}[h!]
{\includegraphics[width=0.5\linewidth]{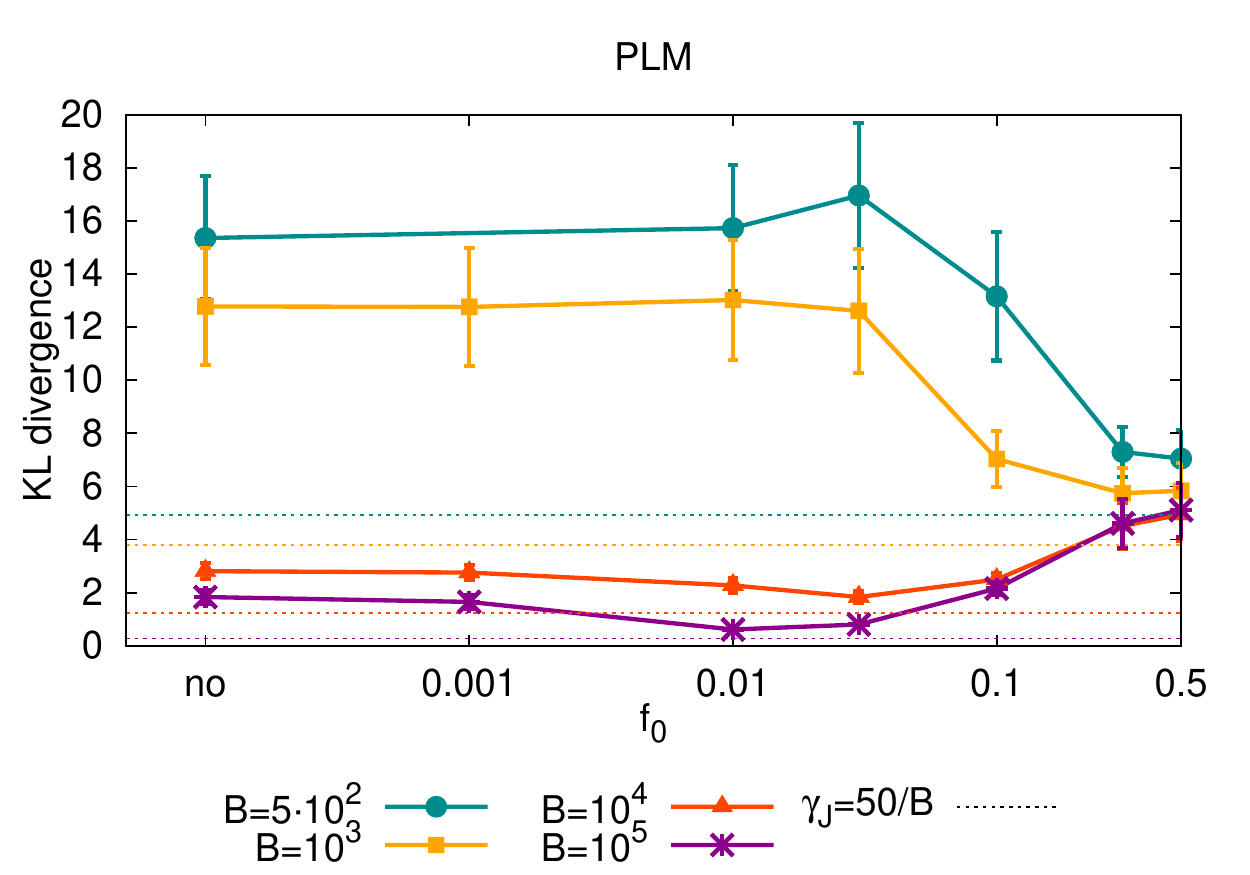}}
	 \caption{KL divergence between the true model and the one inferred with $\gamma_J=1/B$ averaged over 10 realizations for several sample sizes $B$ at different color compression thresholds $f_0$ (full line). Error bars are standard deviations over the 10 realizations. Horizontal dashed lines are there for comparison and correspond to the KL values obtained with $\gamma_J=50/B$ without color compression. }
 	\label{fig:KL_lowlambda}
\end{figure}

Figure~\ref{fig:KL_lowlambda} shows the average KL divergence between the true model and the inferred one for several color compression frequencies at low $L_2$ regularization ($\gamma_J=1/B$). 
For all the sample sizes, we observe the existence of an optimal value of $f_0$. Especially at small sampling depth $B\leq 1000$, 
large color compression leads to a very significant decrease in the KL divergence.
However, in spite of the improvement due to color compression, the KL divergences do not reach the minimal values obtained with strong $L_2$ regularizations (dashed lines in Fig.~\ref{fig:KL_lowlambda}), showing that a good choice of the regularization is always essential.

\subsection{Graphical reconstruction: Contact maps and $Fscore$ for spACE graph reconstruction as a function of color compression}

\begin{figure}[h!]
	\subfloat[ACE: no color compression]
	{\includegraphics[width=0.5\linewidth]{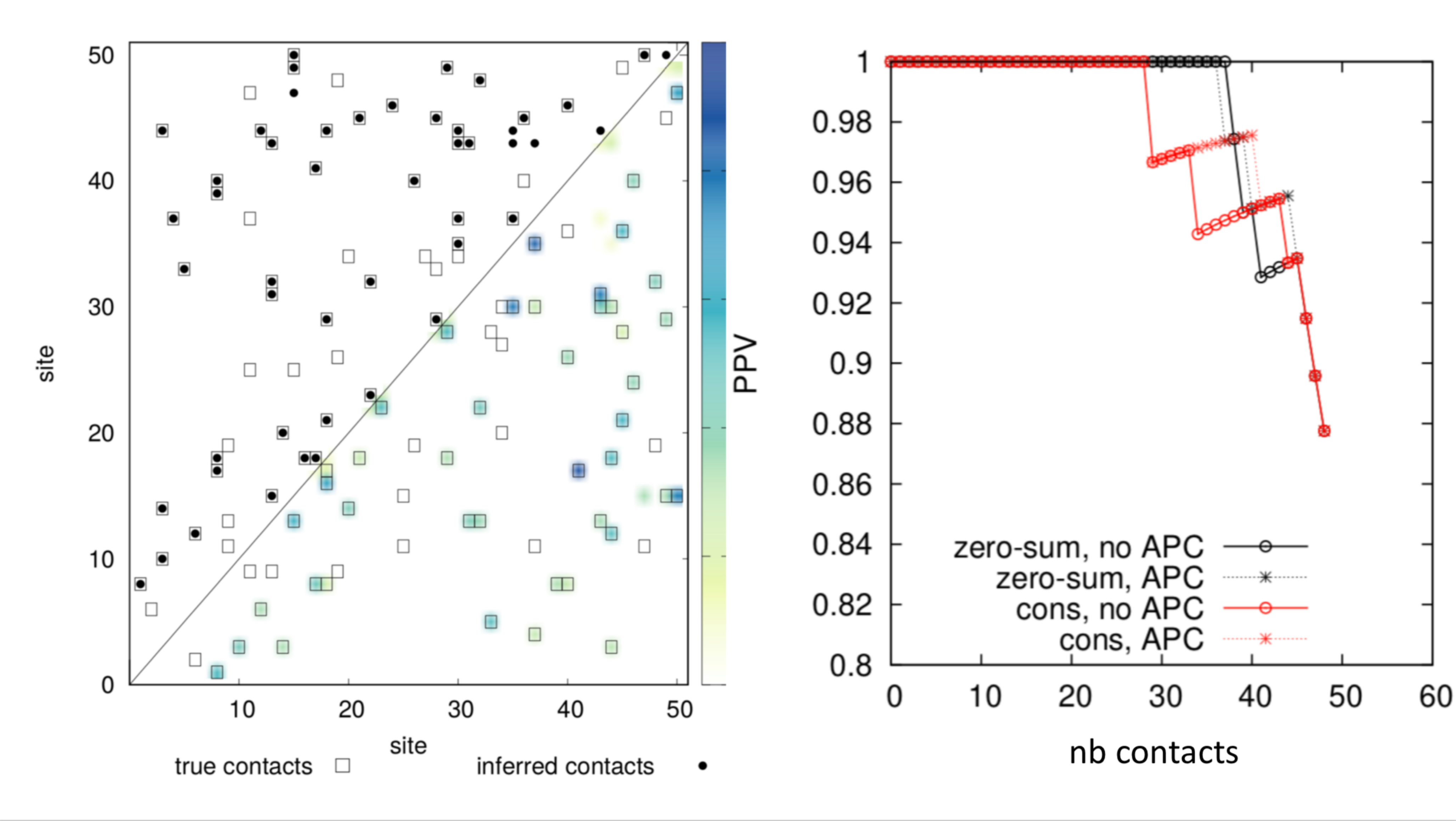}}
	  \subfloat[PLM: no color compression]
	{ \includegraphics[width=0.5\linewidth]{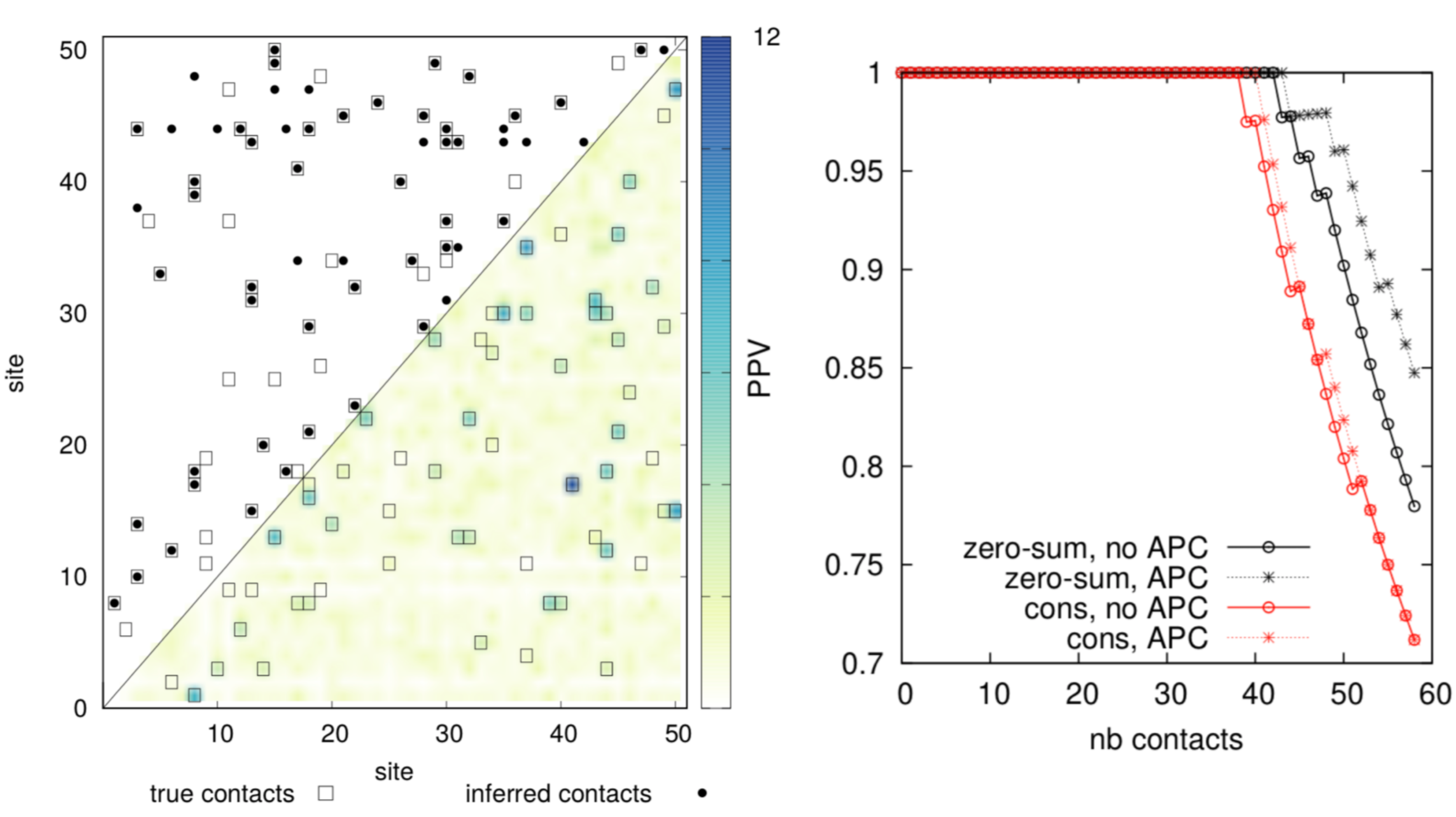} } \\
	\subfloat[ACE: $f_0=0.01$]
	{ \includegraphics[width=0.5\linewidth]{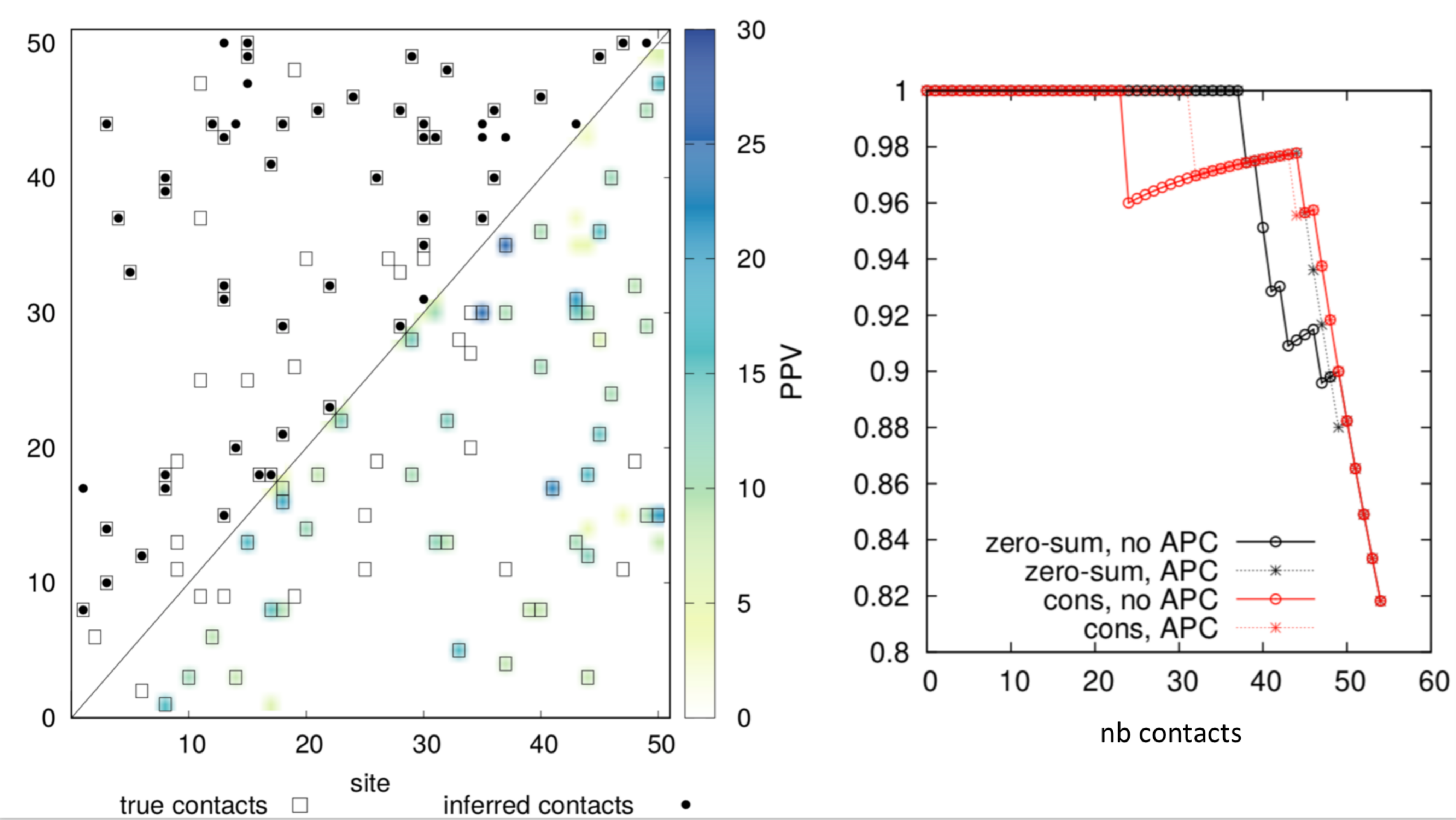}}
	\subfloat[PLM: $f_0=0.01$]
	{ \includegraphics[width=0.5\linewidth]{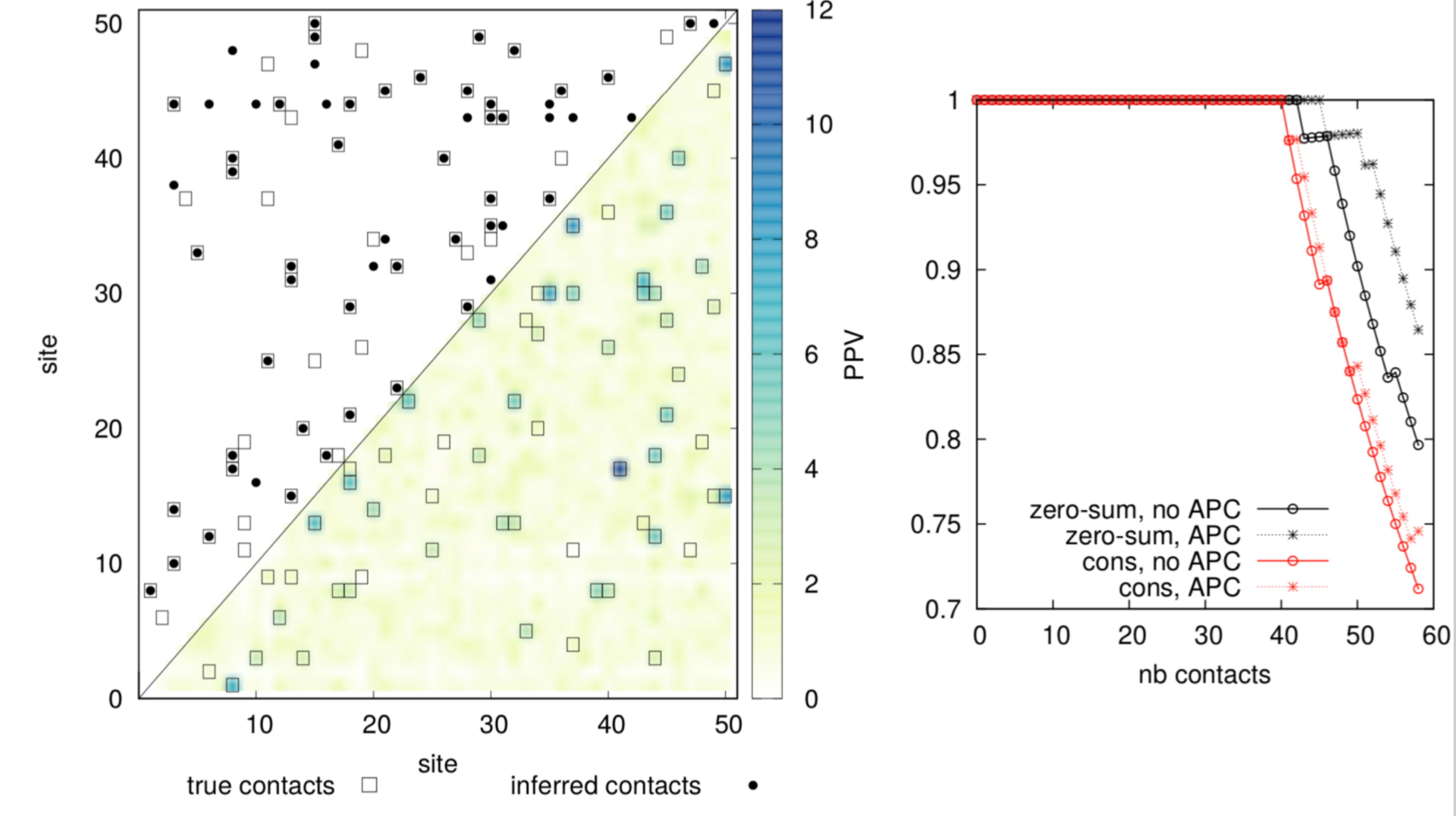}} \\
	\caption{ACE and PLM interaction graph reconstruction and PPV curve for one realization of ER graph. 
	Left: contact maps. Upper triangular: contact map with real contacts (black empty squares), inferred true positive (full circles in empty squares), inferred false positive (full circles) in consensus gauge without Average Product Correction (APC). Lower triangular: Frobenius norm of the inferred parameters in consensus gauge with color-scale on the right. 
	Right: Positive Predicted Value (PPV) curve in consensus and zero-sum gauge with and without APC.
	Top: no color compression. Bottom: $f_0=0.01$.}
	\label{fig:cmapACEandPLM}
\end{figure}

In Figs.~\ref{fig:cmapACEandPLM}  we compare the real contact maps with the heat map of the  Frobenius norms of the couplings inferred by ACE and PLM with no color compression  and $f_0=0.01$ for the reference data (ER05, $B=1000$).
Results at different color compressions are very similar, such as  the identification of largely coupled sites  by ACE and PLM.

\begin{figure}
	{ \includegraphics[width=0.6\linewidth]{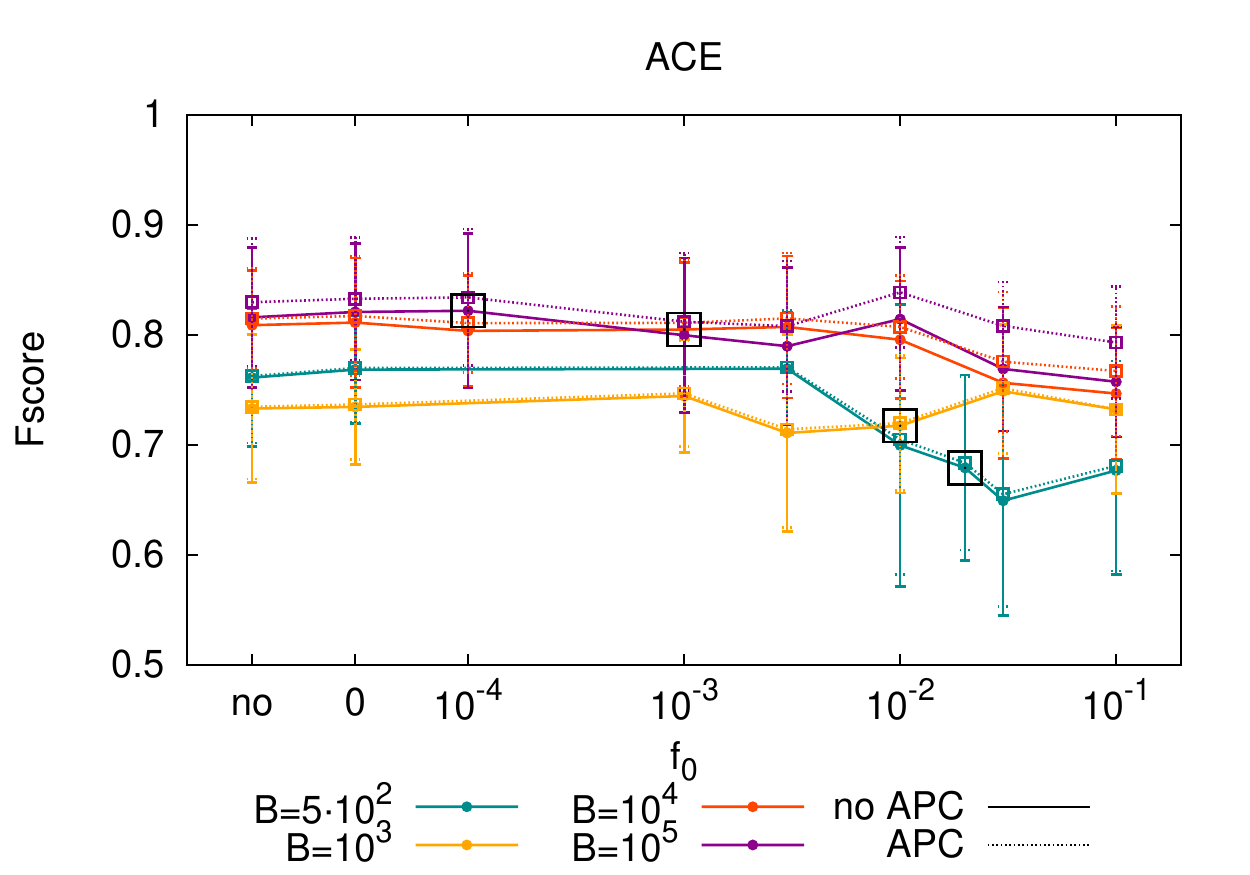}	}
	\caption{Fscore for ACE  inference. Points and error bars are averages and standard deviations obtained on 10 ER realizations. All quantities are computed in the consensus gauge with (dotted line) and without (full line) APC.}
	\label{fig:Fscore_ACE} 
	 \end{figure}

To  encompass both the precision and the recall in a single measure it is possible to use the Fscore, which is the harmonic mean between the two and gives
\begin{equation}
{\rm Fscore} = 2\frac{TP(N_{pred})}{N_{pred}+N_0}
\end{equation}
where $TP$ is the number of true predicted contacts, $N_{pred}$ is the number of predicted contacts and $N_0$ is the number real contacts. The Fscore for ACE inference are plotted in Fig.\ref{fig:Fscore_ACE}.

\subsection{Definition of Positive Predicted Value and Average Product Correction}
\label{sec:graphdef} 
The Positive Predicted Value (PPV) curve is defined as:
\begin{equation}
PPV(n)=\frac{TP(n)}n\\.
\end{equation}
Where $TP(n)$ is the number of true predicted edges in the top $n$ pairs.
 the Average Product Correction (APC) \cite{dunn2008mutual,morcos2014coevolutionary,cocco2018inverse},
\begin{equation}
F^{APC}_{ij}= F_{ij}-\frac{F_{i.}\,F_{.j} }{F_{..}}\,.
\end{equation}
Here the dot indicates the average over the corresponding variables, e.g.~$F_{i,.}$ is the average of $F_{ij}$ over the second index $j$.
\subsection{ Reweighting procedure}
\label{app:w}
  To reduce sampling bias, we decrease 
  the statistical weight of sequences having many similar ones in the 
  MSA. More precisely, the weight of each sequence is defined as the 
  inverse number of sequences within Hamming distance $d_H<xL$, with 
  an arbitrary but fixed $x\in (0,1)$:
  \begin{equation}
    \label{eq:reweighting}
    w_m = \frac 1{||\{ n | 1\leq n\leq M; 
    d_H[(a_1^n,...,a_L^n),(a_1^m,...,a_L^m)]\leq xL \}||} \ 
  \end{equation}
  for all $m=1,...,M$. The weight equals one for isolated sequences, and 
  becomes smaller the denser the sampling around a sequence is. Note that 
  $x=0$ would account to removing double counts from the MSA. The total
  weight
  \begin{equation}
    \label{eq:Meff}
    M_{eff} = \sum_{m=1}^M w_m
  \end{equation}
  can be interpreted as the effective number of independent sequences.

\newpage

\bibliographystyle{ieeetr}
\bibliography{mybib}
\end{document}